\documentclass[12pt]{article}
\usepackage{amsmath,amssymb}
\usepackage[dvipdfmx]{graphicx}
\usepackage{bm}
\usepackage{color}
\usepackage{comment}
\usepackage{braket}

\usepackage{fancybox}  

\def\mydate{23 June 2016}
\def\ignore#1{{}}

\makeatletter
\@addtoreset{equation}{section}
\makeatother

\newcommand{\siml}{%
\hspace{0.3em}\raisebox{0.4ex}{$<$}\hspace{-0.75em}\raisebox{-.7ex}{$\sim$}\hspace{0.3em}}

\newcommand{\beeq}{\begin{equation}}
\newcommand{\eneq}{\end{equation}}
\newcommand{\beqn}{\begin{eqnarray}}
\newcommand{\eeqn}{\end{eqnarray}}

\def\mybig{\displaystyle \strut }
\def\mbig{\displaystyle }
\def\dd{\partial}
\def\la{\raise.16ex\hbox{$\langle$}\lower.16ex\hbox{}  }
\def\ra{\raise.16ex\hbox{$\rangle$}\lower.16ex\hbox{} }
\def\go{\rightarrow}

\def\onehalf{ \hbox{$\frac{1}{2}$} }

\def\onethird{ \hbox{$\frac{1}{3}$} }
\def\twothird{ \hbox{$\frac{2}{3}$} }

\def\fivehalves{ \hbox{$\frac{5}{2}$} }
\def\fourfifths{ \hbox{$\frac{4}{5}$} }
\def\Tr{{\rm Tr \,}}
\def\tr{{\rm tr \,}}
\def\eff{{\rm eff}}

\def\cL{{\cal L}}

\def\SM{{\rm SM}}
\def\EM{{\rm EM}}
\def\diag{{\rm diag ~}}

\def\KK{{\rm KK}}

\def\ep{\epsilon}
\def\psibar{ \psi \kern-.65em\raise.6em\hbox{$-$} }
\def\psibarl{ \psi \kern-.65em\raise.6em\hbox{$-$} \lower.6em\hbox{} }

\def\myfrac#1#2{\frac{\mybig #1}{\mybig #2}}
\def\mfrac#1#2{\frac{\mbig #1}{\mbig #2}}

\begin{document}

\thispagestyle{empty}

{\small \noindent \mydate    \hfill OU-HET 893}

{\small      \hfill MISC-2016-06}

\vskip 3.5cm

\baselineskip=22pt plus 1pt minus 1pt

\begin{center}
{\Large \bf  Toward Realistic Gauge-Higgs Grand Unification}
\end{center}

\baselineskip=22pt plus 1pt minus 1pt

\vskip 1.5cm

\begin{center}
{\bf
Atsushi Furui$^1$, Yutaka Hosotani$^1$ and Naoki Yamatsu$^2$
}

\vskip 5pt

{\small \it $^1$Department of Physics, Osaka University, 
Toyonaka, Osaka 560-0043, Japan} \\
{\small \it $^2$Maskawa Institute for Science and Culture, 
Kyoto Sangyo University, Kyoto 603-8555, Japan} \\

\end{center}

\vskip 3.5cm


\baselineskip=16pt plus 1pt minus 1pt

\begin{abstract}
The $SO(11)$ gauge-Higgs grand unification in the Randall-Sundrum warped space is presented.
The 4D Higgs field is identified as the zero mode of the fifth dimensional component of the
gauge potentials, or as the fluctuation mode of the Aharonov-Bohm phase $\theta_H$ along
the fifth dimension.  Fermions are introduced in the bulk in the spinor and vector representations
of $SO(11)$.   $SO(11)$ is broken to $SO(4) \times SO(6)$ by the orbifold boundary conditions,
which is broken to $SU(2)_L \times U(1)_Y \times SU(3)_C$ by a brane scalar.
Evaluating the effective potential $V_\eff (\theta_H)$, we show 
that the electroweak symmetry is dynamically broken to $U(1)_\EM$.
The quark-lepton masses are generated by the Hosotani mechanism and brane interactions, with which
the observed mass spectrum is reproduced.
The proton decay is forbidden thanks to the new fermion number conservation.
It is pointed out that there appear light exotic fermions.  
The Higgs boson mass is determined with the quark-lepton masses given, 
which, however, turns out  smaller than the observed value.

\end{abstract}



\newpage

\baselineskip=20pt plus 1pt minus 1pt
\parskip=0pt

\section{Introduction}
Up to now almost all observational data at low energies are consistent with the standard model (SM) of 
electroweak interactions.  Yet it is not clear whether the Higgs boson discovered in 2012 
at LHC is precisely what is introduced in the SM.  Detailed study of interactions of the Higgs boson 
is necessary to pin down its nature.   From the theory viewpoint the Higgs boson sector 
of the SM lacks a principle which governs and regulates the Higgs interactions with itself and other fields,
quite in contrast to the gauge sector in which the gauge principle of $SU(3)_C \times SU(2)_L \times U(1)_Y$ 
completely fixes gauge interactions among gauge fields, quarks and leptons.  
Further the mass of the Higgs scalar boson $m_H$ generally acquires large quantum corrections
from much higher energy scales, which have to be canceled by fine-tuning bare masses in a theory.
It is called the gauge-hierarchy problem.

Many proposals have been made to overcome these problems.
Supersymmetric generalization of the SM 
is among them.  There is an alternative scenario of the gauge-Higgs unification in which the 4D Higgs
boson is identified with a part of the extra-dimensional component of gauge fields defined
in higher dimensional spacetime.\cite{YH1}-\cite{Hatanaka1998}
  The Higgs boson, which is massless at the tree level, acquires 
a finite mass at the quantum level, independent of a cutoff scale and regularization scheme.
The $SO(5) \times U(1)_X$ gauge-Higgs electroweak (EW) unification in the five-dimensional 
Randall-Sundrum (RS)  warped space has been formulated.\cite{ACP2005}-\cite{FHHOS2013} 
 It gives almost the same phenomenology at low energies 
as the SM, provided that the Aharonov-Bohm (AB) phase  $\theta_H$ in the fifth dimension is 
$\theta_H \siml 0.1$.
In particular, cubic couplings of the Higgs boson with other fields, $W$, $Z$, quarks and leptons, 
are approximately given by the SM couplings multiplied 
by $\cos \theta_H$.\cite{HS2007}\cite{HS2007a}-\cite{Kurahashi2014}
The corrections to the decay rates  $H \go \gamma \gamma, Z \gamma$, which take place 
through one-loop diagrams, turn out finite and small.\cite{FHHOS2013, FHH2015}
Although  infinitely many Kaluza-Klein (KK) excited states of $W$ and top quark contribute, 
there appears miraculous cancellation among their contributions.
In the gauge-Higgs unification the production rate of the Higgs boson at LHC is approximately
that in the SM times $\cos^2 \theta_H$, and the branching fractions of various Higgs decay modes 
are nearly the same as in the SM.  The cubic and quartic self-interactions of the Higgs boson show
deviations from those in the SM, which should be checked in future LHC and ILC experiments.
Further the gauge-Higgs unification predicts the $Z'$ bosons, namely the first KK modes of $\gamma$, 
$Z$ and $Z_R$, around 6 to 8 TeV range with broad widths for $\theta_H = 0.11$ to $0.07$,
which awaits confirmation at 14 TeV LHC in the near future.\cite{HTU2011, LHCsignal2014}

With a viable model of gauge-Higgs EW unification at hand, it is natural
and necessary to extend it to gauge-Higgs grand unification to
incorporate strong interaction. 
Since the idea of grand unification was proposed \cite{Georgi:1974sy}, 
a lot of grand unified theories based on grand unified gauge groups in
four and higher dimensional spaces have been discussed.
(See, e.g.,  Refs.\ \cite{Mambrini:2015vna}-\cite{Abe:2016eyh} for recent works 
and Refs.\ \cite{Slansky:1981yr, Yamatsu:2015npn} for review.)
The mere fact of the charge quantization in the quark-lepton spectrum
strongly indicates the grand unification.
Such attempt to construct gauge-Higgs grand unification has been made recently.  
$SO(11)$ gauge-Higgs grand unification in the RS space with fermions in the spinor and vector
representations of $SO(11)$ has been proposed.\cite{HY2015a, HY2015b, Yamatsu2016a}
The model carries over good features of
the $SO(5) \times U(1)_X$ EW unification.
In this paper we present detailed analysis of the $SO(11)$ gauge-Higgs grand unification.
Particularly we present how to obtain the observed quark-lepton mass spectrum 
in the combination of the Hosotani mechanism and brane interactions on the Planck brane.
It will be shown  that the proton decay can be forbidden by the new fermion number conservation.

There have been many  proposals of gauge-Higgs grand unification in the literature, but they
are not completely satisfactory in the points of the realistic spectrum and the symmetry breaking 
structure.\cite{Burdman2003}-\cite{Serra2011} 
In the current model $SO(11)$ symmetry is broken to $SO(4) \times SO(6)$ by orbifold
boundary conditions, which breaks down to $SU(2)_L \times U(1)_Y \times SU(3)_C$  by
a brane scalar on the Planck brane.  Finally $SU(2)_L \times U(1)_Y$ is dynamically broken to
$U(1)_\EM$ by the Hosotani mechanism.  The quark-lepton mass spectrum is reproduced.
However, unwanted exotic fermions appear.  Further elaboration of the scenario is necessary 
to achieve a completely realistic grand unification model.
We note that there have been many advances in the gauge-Higgs unification
both in electroweak theory and grand unification.\cite{DM2014}-\cite{Hasegawa2016}
Mechanism for dynamically selecting orbifold boundary conditions has been 
explored.\cite{Yamamoto2014}   The gauge symmetry breaking by the Hosotani mechanism 
has been examined not only in the continuum theory, but also
on the lattice by nonperturbative simulations.\cite{Cossu2014, Forcrand2015, Knechtli2016}

The paper is organized as follows. In Section 2 the $SO(11)$ model is introduced.
The symmetry breaking structure and fermion content are explained in detail.
The proton stability is also shown.  In Section 3 the mass spectrum of gauge fields
is determined.  In Section 4 the mass spectrum of fermion fields are determined.
With these results the effective potential $V_\eff (\theta_H)$ is evaluated in Section 5.
Conclusion and discussions are given in Section 6.

\section{$SO(11)$ Model}
The $SO(11)$ gauge theory is defined in the Randall-Sundrum (RS) space whose metric is
given by
\beeq
ds^2=  G_{MN} dx^M dx^N = e^{- 2 \sigma(y)} \eta_{\mu\nu}dx^{\mu} dx^{\nu} + dy^2 ,
\label{metric1}
\eneq
where $M,N=0,1,2,3,5$, $\mu,\nu=0,1,2,3$, $y=x^5$,
$\eta_{\mu\nu}=\mbox{diag}(-1,+1,+1,+1)$, $\sigma(y) = \sigma(y+ 2L)=\sigma(-y)$,
and  $\sigma(y)=ky$ for $0 \le y \le L$.  The topological structure of the  RS space is $S_1/ \mathbb{Z}_2$.
In terms of the conformal coordinate $z = e^{ky}$ ($1 \le z \le z_L=e^{kL}$) in the region
$0 \le y \le L$ 
\beeq
ds^2=  \frac{1}{z^2} \bigg(\eta_{\mu\nu}dx^{\mu} dx^{\nu} + \frac{dz^2}{k^2}\bigg) .
\label{metric2}
\eneq
The bulk region $0 < y < L$ ($1 < z < z_L$) is anti-de Sitter (AdS) spacetime 
with a cosmological constant $\Lambda= - 6k^2$, which is sandwiched by the Planck brane
at $y=0$ ($z=1$) and the TeV brane at $y=L$ ($z=z_L$).  
The KK mass scale is $m_\KK = \pi k/ (z_L -1) \sim \pi k z_L^{-1}$ for $z_L \gg 1$.

\subsection{Action and boundary conditions}

The model consists of $SO(11)$ gauge fields $A_M$,  fermion multiplets
in the spinor representation  $\Psi_{\bf 32}$ and  in the vector representation $\Psi_{\bf 11}$, 
and a brane scalar field $\Phi_{\bf 16}$.\cite{HY2015a}    
In each generation of quarks and leptons, 
one $\Psi_{\bf 32}$ and two $\Psi_{\bf 11}$'s are introduced.
$\Phi_{\bf 16} (x)$, in the spinor representation of $SO(10)$,  is defined on the Planck brane.

The bulk part of the action is given by
\beqn
&&\hskip -1.cm
S_{\text{bulk}}=\int d^5x\sqrt{-\mbox{det}G} \, \Bigl[-\tr 
\Big\{ \frac{1}{\;4\;}F^{MN} F_{MN}+\frac{1}{2\xi}
(f_{\text{gf}})^2+\mathcal{L}_{\text{gh}}\Big\}  \cr
\noalign{\kern 10pt}
&&\hskip 1.cm
+\sum_{a=1}^3 \Big\{ \overline{\Psi} {}_{\bf 32}^a 
\mathcal{D}(c_{\Psi_{\bf 32}^a} ) \Psi_{\bf 32}^a
+\overline{\Psi} {}_{\bf 11}^a 
\mathcal{D}(c_{\Psi_{\bf 11}^a} ) \Psi_{\bf 11}^a
+\overline{\Psi} {}_{\bf 11}^{\prime a} 
\mathcal{D}(c_{\Psi_{\bf 11}^{\prime a}} ) 
      \Psi_{\bf 11}^{\prime a} \Big\} ,\cr
\noalign{\kern 10pt}
&&\hskip -1.cm
\mathcal{D}(c)= \gamma^A {e_A}^M
\Big(\partial_M+\frac{1}{8}\omega_{MBC}[\gamma^B,\gamma^C] - ig A_M \Big) -c \, \sigma'(y) ,
\label{action1}
\eeqn
where $M,N,A,B,C=0,1,2,3,5$, $a=1,2,3$. $a$ stands for the generation
index. $c_{\Psi_{\bf 32}^a}$, $c_{\Psi_{\bf 11}^a}$,
$c_{\Psi_{\bf 32}^{\prime a}}$ represent bulk mass parameters.
We employ the background field method, separating ~$A_M$ 
into the classical part~$A^{\rm c}_M$  and the quantum part~$A^{\rm q}_M$ ;
$A_M = A^{\rm c}_M+A^{\rm q}_M$.
The gauge-fixing function and the associated ghost term are given, in the conformal coordinate,  by
\beqn
&&\hskip -1.cm
 f_{\rm gf} =
 z^2 \bigg\{ \eta^{\mu\nu} {\cal D}^{\rm c}_\mu A^{\rm q}_\nu
 +\xi k^2 z  {\cal D}^{\rm c}_z \Big( \frac{1}{z} A^{\rm q}_z \Big) \bigg\} ~~, \cr
\noalign{\kern 10pt}
&&\hskip -1.cm
{\cal L}_{\rm gh} 
= \bar c \Big\{ \eta^{\mu\nu} {\cal D}_\mu^{\rm c} {\cal D}_\nu
+ \xi k^2 z {\cal D}_z^{\rm c} \frac{1}{z} {\cal D}_z \Big\} c ~, 
\label{gaugefixing1}
\eeqn
where ${\cal D}^{\rm c}_M A^{\rm q}_N \equiv 
\dd_M A^{\rm q}_N-ig_A[A^{\rm c}_M,A^{\rm q}_N]$,
${\cal D}_M c \equiv  \dd_M c - ig  [ A_M , c ]$ etc.
We adopt the convention $\{ \gamma^A, \gamma^B \} = 2 \eta^{AB}$, 
$\eta^{AB} = \diag (-1, 1, 1, 1, 1)$, $G_{MN} = e_{A M} {e^A}_N$, and 
$\overline{\Psi} = i \Psi^\dagger \gamma^0$.

Generators $T_{jk}= - T_{kj}$ of $SO(11)$ are summarized in Appendix A
in the vectorial and spinorial representations.
We adopt the normalization 
$A_M = 2^{-1/2} \sum_{j<k} A_M^{(jk)} \, T_{jk}$ and
$F_{MN} = \dd_M A_N - \dd_N A_M - ig [ A_M,  A_N ] = 2^{-1/2}\sum_{j<k} F_{MN}^{(jk)} \, T_{jk}$,
With this normalization the $SU(2)_L$ weak gauge coupling constant is given by $g_w = g/\sqrt{L}$.
The orbifold boundary conditions for the gauge fields are given, in the $y$-coordinate, by
\beqn
&&\hskip -1.cm
\begin{pmatrix} A_\mu \cr A_y \end{pmatrix} (x, y_j - y) =
P_j \begin{pmatrix} A_\mu \cr - A_y \end{pmatrix} (x, y_j + y) P_j^{-1} ~,  \cr
\noalign{\kern 10pt}
&&\hskip -1.cm
A_M(x, y+ 2L) = U A_M (x, y) U^{-1} ~, \quad U = P_1 P_0 ~.
\label{BCgauge1}
\eeqn
where $(y_0, y_1) =(0, L)$.  The ghost fields, $c$ and $\bar c$, satisfy the same boundary
conditions as $A_\mu$.  
$P_j = P_j^\dagger = P_j^{-1}$ is given in the vectorial representation by
\beeq
P_0^{\rm vec}=\mbox{diag}(I_{10},-I_1) ~,~~ P_1^{\rm vec} =\mbox{diag}(I_4,-I_7)~, 
\label{BCP1}
\eneq
and in the spinorial representation by
\beqn
&&\hskip -1.cm
P_0^{\rm sp} = \sigma^0 \otimes \sigma^0 \otimes \sigma^0 \otimes \sigma^0 \otimes \sigma^3
= I_{16} \otimes \sigma^3 ~, \cr
\noalign{\kern 10pt}
&&\hskip -1.cm
P_1^{\rm sp} = \sigma^0 \otimes \sigma^3 \otimes \sigma^0 \otimes \sigma^0 \otimes \sigma^0
= I_2 \otimes \sigma^3 \otimes I_8 ~.
\label{BCP2}
\eeqn
The $SO(11)$ symmetry is broken down to $SO(10)$ by $P_0$ at $y=0$, and
to $SO(4) \times SO(7)$ by $P_1$ at $y=L$.    With these two combined, 
there remains $SO(4) \times SO(6) \simeq SU(2)_L \times SU(2)_R \times SU(4)$ symmetry,
which is further broken to $G_\SM = SU(2)_L \times SU(3)_C \times U(1)_Y$ by $\la \Phi_{\bf 16} \ra $
on the Planck brane as described below.
$SU(2)_L  \times U(1)_Y$ is dynamically broken to $U(1)_\EM$ by the Hosotani mechanism.

Fermion fields obey the following boundary conditions;
\beqn
&&\hskip -1.cm
\Psi_{\bf 32}^a (x,y_j - y)=  - \gamma^5 P_j^{\rm sp}  \Psi_{\bf 32}^a (x,y_j+y) ~, \cr
\noalign{\kern 10pt}
&&\hskip -1.cm
\Psi_{\bf 11}^a (x,y_j - y) = (-1)^j  \gamma^5 P_j^{\rm vec} \Psi_{\bf 11}^a (x,y_j+y) ~, \cr
\noalign{\kern 10pt}
&&\hskip -1.cm
\Psi_{\bf 11}^{\prime a}  (x,y_j - y) 
=  (-1)^{j+1}  \gamma^5 P_j^{\rm vec} \Psi_{\bf 11}^{\prime a} (x,y_j+y) ~.
\label{BCF1}
\eeqn
Eigenstates of $\gamma^5$ with $\gamma^5= \pm1$  correspond to 
right- and left-handed components in four dimensions.
For  $\Psi_{\bf 11}^a$ and $\Psi_{\bf 11}^{\prime a} $ one might impose  alternative
boundary conditions given by
\beqn
&&\hskip -1.cm
\Psi_{\bf 11}^a (x,y_j - y) =  \gamma^5 P_j^{\rm vec} \Psi_{\bf 11}^a (x,y_j+y) ~, \cr
\noalign{\kern 10pt}
&&\hskip -1.cm
\Psi_{\bf 11}^{\prime a}  (x,y_j - y) 
=  -  \gamma^5 P_j^{\rm vec} \Psi_{\bf 11}^{\prime a} (x,y_j+y) ~.
\label{BCF2}
\eeqn
It turns out that the model with \eqref{BCF1} is easier to analyze in reproducing
the mass spectrum of quarks and leptons.

On the Planck brane (at $y=0$) the brane scalar field $\Phi_{\bf 16}$ has 
an $SO(10)$ invariant action given by
\beqn
&&\hskip -1.cm
S_{\Phi_{\bf 16}}=
\int d^5x \sqrt{-\mbox{det}G} \, \delta(y)
\Big\{ -(D_\mu \Phi_{\bf 16})^\dag D^\mu \Phi_{\bf 16}
-\lambda_{\Phi_{\bf 16}} (\Phi_{\bf 16}^\dag \Phi_{\bf 16}-w^2)^2 \Big\}, \cr
\noalign{\kern 10pt}
&&\hskip -1.cm
D_\mu \Phi_{\bf 16}= ( \dd_\mu - ig A_\mu^{SO(10)} ) \Phi_{\bf 16} =
\Big\{ \dd_\mu- \frac{ig}{\sqrt{2}} \sum_{j < k}^{10} A_\mu^{(jk)} T_{jk}^{\rm sp} \Big\} \Phi_{\bf 16}~.
\label{braneaction1}
\eeqn
$\Phi_{\bf 16}$ develops VEV.  Without loss of generality one can take
\beeq
\la \Phi_{\bf 16} \ra =
\begin{pmatrix} 0_4 \cr  0_4  \cr v_4 \cr 0_4 \end{pmatrix} ~, ~~
v_4 = \begin{pmatrix} 0 \cr 0 \cr 0 \cr w \end{pmatrix} ~.
\label{vevPhi}
\eneq
On the Planck brane $SO(10)$ symmetry is spontaneously broken to $SU(5)$ by 
$\la \Phi_{\bf 16} \ra \not= 0$.  With the orbifold boundary condition $P_1$, 
$SO(11)$ symmetry is broken to the SM symmetry $G_\SM= SU(3)_C \times SU(2)_L \times U(1)_Y$.

To see it more explicitly, we note that mass terms for gauge fields are generated 
from \eqref{braneaction1} in the form 
$- g^2 \la \Phi_{\bf 16}^\dagger \ra \eta^{\mu\nu} A_\mu^{SO(10)} A_\nu^{SO(10)} \la \Phi_{\bf 16} \ra$.
Making use of \eqref{Clifford1} and \eqref{SpinorRep1} for $T_{jk}^{\rm sp}$, one obtains 
\beqn
&&\hskip -1.cm
{\cal L}^{\rm gauge}_{{\rm brane~mass}} = -  \delta(y) \, \frac{g^2 w^2}{8} \bigg\{ 
 (A_\mu^{15} - A_\mu^{26} )^2 + (A_\mu^{16} + A_\mu^{25} )^2 \cr
\noalign{\kern 10pt}
&&\hskip -1.0cm
+ (A_\mu^{17} - A_\mu^{28} )^2 + (A_\mu^{18} + A_\mu^{27} )^2  
+ (A_\mu^{19} - A_\mu^{2,10} )^2 + (A_\mu^{1,10} + A_\mu^{29} )^2 \cr
\noalign{\kern 10pt}
&&\hskip -1.cm
+ (A_\mu^{35} + A_\mu^{46} )^2 + (A_\mu^{36} - A_\mu^{45} )^2 
+ (A_\mu^{37} + A_\mu^{48} )^2 + (A_\mu^{38} - A_\mu^{47} )^2   \cr
\noalign{\kern 10pt}
&&\hskip -1.cm
+ (A_\mu^{39} + A_\mu^{4,10} )^2 + (A_\mu^{3,10} - A_\mu^{49} )^2 
+ (A_\mu^{57} - A_\mu^{68} )^2 + (A_\mu^{58} + A_\mu^{67} )^2  \cr
\noalign{\kern 10pt}
&&\hskip -1.cm
+ (A_\mu^{59} - A_\mu^{6,10} )^2  + (A_\mu^{5,10} + A_\mu^{69} )^2 
+ (A_\mu^{79} - A_\mu^{8,10} )^2 + (A_\mu^{7,10} + A_\mu^{89} )^2 \cr
\noalign{\kern 10pt}
&&\hskip -1.cm
+ (A_\mu^{23} - A_\mu^{14})^2 + (A_\mu^{31} - A_\mu^{24} )^2  
+ ( A_\mu^{12} -  A_\mu^{34}  + A_\mu^{56} + A_\mu^{78} + A_\mu^{9\,10} )^2 
~ \bigg\} ~.
\label{branemass1}
\eeqn
In all,   21 components in $SO(10)/SU(5)$ 
acquire large brane masses by $\la \Phi_{\bf 16} \ra$, which effectively
alters the Neumann boundary condition at $y=0$ to the Dirichlet boundary condition
for their low-lying modes ($m_n \ll gw/\sqrt{L}$) as will be seen in Section 3.
It follows that the $SU(5)$ generators are given, up to normalization,  by
\beqn
&&\hskip -1.cm
{\rm (i)}\quad   SU(2)_L: \cr
\noalign{\kern 10pt}
&&
T_L^1= \onehalf(T_{23} +  T_{14})  ~,~T_L^2=  \onehalf(T_{31} +  T_{24}) ~,~
T_L^3= \onehalf( T_{12} +  T_{34}) ~, \cr
\noalign{\kern 10pt}
&{\rm (ii)}&  SU(3)_C:  \cr
\noalign{\kern 10pt}
&&
 \begin{pmatrix}T_{57} +  T_{68} \cr T_{58} -  T_{67}\end{pmatrix} , ~
\begin{pmatrix} T_{59} +  T_{6\, 10} \cr T_{5\, 10} -  T_{69}\end{pmatrix}, ~
\begin{pmatrix} T_{79} +  T_{8\, 10} \cr T_{7\, 10} -  T_{89} \end{pmatrix},  \cr
\noalign{\kern 10pt}
&& T_{56} -  T_{78} ~, ~ T_{56} + T_{78}- 2T_{9\, 10} ~,  \cr
\noalign{\kern 10pt}
&&\hskip -1.cm
{\rm (iii)} \quad U(1)_Y : \cr
\noalign{\kern 10pt}
&&
Q_Y = \onehalf (T_{12} -  T_{34} ) - \onethird ( T_{56} + T_{78} + T_{9\, 10}) ~,  \cr
\noalign{\kern 10pt}
&{\rm (iv)}&  SU(5)/SU(3)_C \times SU(2)_L \times U(1)_Y:  \cr
\noalign{\kern 10pt}
&& \begin{pmatrix}T_{15} +  T_{26} \cr T_{16} -  T_{25}\end{pmatrix} , ~
\begin{pmatrix} T_{17} +  T_{28} \cr T_{18} -  T_{27}\end{pmatrix}, ~
\begin{pmatrix} T_{19} +  T_{2\, 10} \cr T_{1\, 10} -  T_{29} \end{pmatrix},  \cr
\noalign{\kern 5pt}
&& \begin{pmatrix}T_{35} -  T_{46} \cr T_{36} +  T_{45}\end{pmatrix} , ~
\begin{pmatrix} T_{37} -  T_{48} \cr T_{38} +  T_{47}\end{pmatrix}, ~
\begin{pmatrix} T_{39} -  T_{4\, 10} \cr T_{3\, 10} +  T_{49} \end{pmatrix} .
\label{SU5generators}
\eeqn
12 components of the gauge fields $A_\mu$  in the class (iv) have no zero modes 
by the boundary conditions.   This leaves $SU(2)_L \times SU(3)_C \times U(1)_Y$ symmetry.

It will be seen later that $SU(2)_L \times U(1)_Y$ symmetry is dynamically broken to $U(1)_\EM$
by the Hosotani mechanism.  The AB phase $\theta_H$ associated with $A_z^{4,11}$  becomes nontrivial 
so that $A_\mu^{34}$ picks up an additional mass term. 
Consequently the surviving massless gauge boson, the photon, is given by
\beqn
&&\hskip -1.cm
A_\mu^\EM = \frac{\sqrt{3}}{2} A_\mu^{12} - \frac{1}{2} A_\mu^{0_C} ~, ~~
A_\mu^{0_C} = \frac{1}{\sqrt{3}} (A_\mu^{56} + A_\mu^{78} + A_\mu^{9\,10}) ~, \cr
\noalign{\kern 10pt}
&&\hskip -1.cm
Q_\EM = T_{12} - \onethird ( T_{56} + T_{78} + T_{9\, 10}) 
= T^3_L + T^3_R - \onethird (T_{56} + T_{78} + T_{9,10} ) ~,
\label{photon1}
\eeqn
where $T_R^a$ are generators of $SU(2)_R$.
The orthogonal component 
$\tilde A_\mu = \frac{1}{2}A_\mu^{12} +  \frac{\sqrt{3}}{2} A_\mu^{0_C}$ and
$A_\mu^{34}$ mix with each other for $\theta_H \not= 0$.  
More rigorous and detailed reasoning is given in Section 3 in the twisted gauge.
The $U(1)_Y$ gauge boson, $B_\mu^Y$,  is given by 
\beeq
B_\mu^Y = \sqrt{\frac{3}{5}} A_\mu^{3_R} - \sqrt{\frac{2}{5}} A_\mu^{0_C} ~,~~
A_\mu^{3_L, 3_R} = \frac{1}{\sqrt{2}} ( A_\mu^{12} \pm A_\mu^{34} ) ~.
\label{BY1}
\eneq
The gauge couplings become
\beqn
&&\hskip -1.cm 
g A_\mu = g \Big\{ A_\mu^{3_L} T_{3_L} + A_\mu^{3_R} T_{3_R}
+ \frac{1}{\sqrt{2}} ( A_\mu^{56} T_{56} + A_\mu^{78} T_{78}  + A_\mu^{9\, 10} T_{9\, 10} )
+ \cdots \Big\} \cr
\noalign{\kern 10pt}
&&\hskip -0.cm
= g \bigg\{ \frac{\sqrt{3}}{2 \sqrt{2}} \,   A_\mu^\EM \, Q_\EM + \cdots  \bigg\}\cr
\noalign{\kern 10pt}
&&\hskip -0.cm
= g \Big\{ A_\mu^{3_L} T_{3_L} +   \sqrt{\frac{3}{5}} \, B_\mu^Y Q_Y + \cdots \Big\}.
\eeqn
In other words 4D gauge couplings and the Weinberg angle are given, 
at the grand unification scale,  by
\beeq
g_w = \frac{g}{\sqrt{L}} ~,~ e = \sqrt{\frac{3}{8}} \, g_w ~,~
g_Y = \sqrt{\frac{3}{5}}  \, g_w ~,~ \sin^2 \theta_W = \frac{g_Y{}^2}{g_w^2 + g_Y{}^2} = \frac{3}{8} ~.
\label{Wangle1}
\eneq

The content $\Psi_{\bf 32},  \Psi_{\bf 11}, \Psi_{\bf 11}'$ is determined 
with the EM charge given by \eqref{photon1}.
In the spinorial representation
\beqn
&&\hskip -1.cm
Q_\EM = 
\frac{1}{2} \sigma^3 \otimes \sigma^0 \otimes \sigma^0 \otimes \sigma^0 \otimes \sigma^0 \cr
\noalign{\kern 10pt}
&&\hskip -.5cm
- \frac{1}{6} \sigma^0 \otimes \Big\{ \sigma^3 \otimes \sigma^3 \otimes \sigma^0 \otimes \sigma^0
+ \sigma^0 \otimes \sigma^3 \otimes \sigma^3 \otimes \sigma^0
+ \sigma^0 \otimes \sigma^0 \otimes \sigma^3 \otimes \sigma^3 \Big\}~.
\label{EMcharge2}
\eeqn
We tabulate the content of $\Psi_{\bf 32}$ in Table \ref{32Content}.
We note that zero modes appear only for particles with the same quantum numbers 
as quarks and leptons in the SM, but not for anything else.

\ignore{
$\Psi_{\bf 11}$ decomposes into  an $SO(10)$ vector and a singlet.  The former  further
decomposes into an $SO(4)$ vector $\psi_j$ ($j=1 \sim 4$) and 
an $SO(6)$ vector $\psi_j$ ($j=5 \sim 10$).
An $SO(4)$ vector $\psi_j$ ($j=1 \sim 4$) transforms as $({\bf 2}, {\bf 2})$ of $SU(2)_L \times SU(2)_R$.
Under $\Omega_L \in SU(2)_L$ and $\Omega_R \in SU(2)_R$,
\beqn
&&\hskip -1.cm
\hat \psi \go \Omega_L \hat \psi  \, \Omega_R^\dagger ~, \cr
\noalign{\kern 10pt}
&&\hskip -1.cm
\hat \psi = \frac{1}{\sqrt{2}} (\psi_4 + i \vec \psi \vec\sigma) \cdot  i \sigma_2 \cr
\noalign{\kern 10pt}
&&\hskip -.5cm
= \frac{1}{\sqrt{2}} \begin{pmatrix} \psi_2 + i \psi_1 & -\psi_4 - i\psi_3 \cr
\psi_4 - i \psi_3 & \psi_2 - i \psi_1 \end{pmatrix}
\leftrightarrow
\begin{pmatrix} \hat E & N\cr \hat N & E \end{pmatrix} ~.
\label{vector1}
\eeqn
An $SO(6)$ vector $\psi_j$ ($j=5 \sim 10$) decomposes into two parts;
\beeq
\begin{pmatrix} D_j \cr \hat D_j \end{pmatrix}
=  \frac{1}{\sqrt{2}} ( \psi_{3 + 2j} \mp i \psi_{4 + 2j} ) \quad (j=1,2,3).
\label{vector2}
\eneq
With this notation the content of $\Psi_{\bf 11}$ and $\Psi_{\bf 11}'$ in \eqref{BCF1}
is summarized in Table \ref{11Content}.
}

\def\mynoalign1{\noalign{\kern 4pt}}
{\small
\begin{table}[p]
\begin{center}
\begin{tabular}{lclclclclclclc|c|}
\multicolumn{9}{c}   %
{\normalsize $\Psi_{\bf 32}$}\\
\noalign{\kern 3pt}
\hline
$SO(10)$ & $SU(5)_Z$ & $G_{227}$ & $SO(6)$ & $G_{SM}$ &$Q_\EM$ & name 
& $\begin{matrix} {\rm zero} \cr {\rm mode}\end{matrix}$
&$\begin{matrix}{\rm parity}\cr {\rm (left)}\end{matrix}$\\
\hline\hline
\mynoalign1
${\bf 16}$
 & $\begin{matrix}{\bf \overline{5}}_{-3} \cr {\bf \overline{5}}_{-3} \end{matrix}$ 
 & $({\bf 2,1,8})$ & ${\bf 4}$ & $({\bf 1,2})^{-1/2}_{-1/2}$ 
 & $\begin{matrix} 0 \cr -1 \end{matrix}$ 
 & $\begin{matrix} \nu \cr e \end{matrix}$ 
 & $\begin{matrix} \nu_L \cr e_L \end{matrix}$ & $(+,+)$ \\
\mynoalign1
 & $\begin{matrix}{\bf \overline{5}}_{-3} \cr{\bf 10}_{+1}\end{matrix} $ 
 & $({\bf 1,2,8})$ & ${\bf \overline{4}}$ & $({\bf \overline{3},1})^{+1/3}_{-2/3} $ 
&$\begin{matrix} \onethird  \cr -\twothird  \end{matrix} $ 
& $\begin{matrix} \hat d_1 \cr \hat u_1 \end{matrix} $ &  & $(+,-)$\\
\mynoalign1
 & $\begin{matrix}{\bf 10}_{+1} \cr{\bf 10}_{+1}\end{matrix}$ 
 & $({\bf 2,1,8})$ & ${\bf 4}$ & $({\bf 3,2})^{+1/6}_{+1/6}  $ 
& $\begin{matrix} \twothird  \cr -\onethird  \end{matrix}$
 & $\begin{matrix} u_3 \cr d_3 \end{matrix}$ 
 & $\begin{matrix} u_{3L} \cr d_{3L} \end{matrix}$ & $(+,+)$\\
\mynoalign1
 & $\begin{matrix}{\bf \overline{5}}_{-3} \cr{\bf 10}_{+1}\end{matrix} $ 
 & $({\bf 1,2,8})$ & ${\bf \overline{4}}$ & $({\bf \overline{3},1})^{+1/3}_{-2/3} $ 
 &$\begin{matrix} \onethird  \cr -\twothird  \end{matrix} $ 
 & $\begin{matrix} \hat d_2 \cr \hat u_2 \end{matrix} $ &  & $(+,-)$\\
\mynoalign1
& $\begin{matrix}{\bf 10}_{+1} \cr{\bf 10}_{+1}\end{matrix}$ 
 & $({\bf 2,1,8})$ & ${\bf 4}$ & $({\bf 3,2})^{+1/6}_{+1/6}  $ 
& $\begin{matrix} \twothird  \cr -\onethird  \end{matrix}$
 & $\begin{matrix} u_1 \cr d_1 \end{matrix}$ 
 & $\begin{matrix} u_{1L} \cr d_{1L} \end{matrix}$ & $(+,+)$\\
\mynoalign1
& $\begin{matrix}{\bf 10}_{+1} \cr{\bf 1}_{+5}\end{matrix}$ 
& $({\bf 1,2,8})$ & ${\bf \overline{4}}$ & $({\bf 1,1})^{+1}_{~0}$ 
& $\begin{matrix} 1 \cr 0 \end{matrix}$ 
& $\begin{matrix} \hat e \cr \hat \nu \end{matrix}$ &  & $(+,-)$\\
\mynoalign1
 & $\begin{matrix}{\bf 10}_{+1} \cr{\bf 10}_{+1}\end{matrix}$ 
 & $({\bf 2,1,8})$ & ${\bf 4}$ & $({\bf 3,2})^{+1/6}_{+1/6}  $ 
 & $\begin{matrix} \twothird  \cr -\onethird  \end{matrix}$
 & $\begin{matrix} u_2 \cr d_2 \end{matrix}$ 
 & $\begin{matrix} u_{2L} \cr d_{2L} \end{matrix}$ & $(+,+)$\\
\mynoalign1
 & $\begin{matrix}{\bf \overline{5}}_{-3} \cr{\bf 10}_{+1}\end{matrix} $ 
 & $({\bf 1,2,8})$ & ${\bf \overline{4}}$ & $({\bf \overline{3},1})^{+1/3}_{-2/3} $ 
 &$\begin{matrix} \onethird  \cr -\twothird  \end{matrix} $ 
 & $\begin{matrix} \hat d_3 \cr \hat u_3 \end{matrix} $ &  & $(+,-)$\\
\mynoalign1
\hline
\mynoalign1
$\overline{\bf 16}$ 
 & $\begin{matrix}{\bf \overline{10}}_{-1}\cr{\bf \overline{10}}_{-1}\end{matrix}$ 
 & $({\bf 2,1,8})$ & ${\bf \overline{4}}$ & $({\bf \overline{3},2})^{-1/6}_{-1/6} $ 
 &$\begin{matrix} \onethird  \cr -\twothird  \end{matrix} $ 
 & $\begin{matrix} \hat d'_3 \cr \hat u'_3 \end{matrix}$ & & $(-,+)$\\
\mynoalign1
 & $\begin{matrix}{\bf \overline{10}}_{-1}\cr {\bf 5}_{+3}\end{matrix}$ 
 & $({\bf 1,2,8})$ & ${\bf 4}$ & $({\bf 3,1})^{+2/3}_{-1/3} $ 
 & $\begin{matrix} \twothird  \cr -\onethird  \end{matrix}$
 & $\begin{matrix} u_2' \cr d_2' \end{matrix} $ 
 & $\begin{matrix} u_{2R} \cr d_{2R} \end{matrix}$ & $(-,-)$\\
\mynoalign1
 & $\begin{matrix}{\bf 5}_{+3}\cr {\bf 5}_{+3}\end{matrix}$ 
 & $({\bf 2,1,8})$ & ${\bf \overline{4}}$ & $({\bf 1,2})^{+1/2}_{+1/2}$ 
 & $\begin{matrix} 1 \cr 0 \end{matrix}$ 
 & $\begin{matrix} \hat e' \cr \hat \nu ' \end{matrix}$ & & $(-,+)$\\
\mynoalign1
 & $\begin{matrix}{\bf \overline{10}}_{-1}\cr {\bf 5}_{+3}\end{matrix}$ 
 & $({\bf 1,2,8})$ & ${\bf 4}$ & $({\bf 3,1})^{+2/3}_{-1/3} $ 
 & $\begin{matrix} \twothird  \cr -\onethird  \end{matrix}$
 & $\begin{matrix} u_1' \cr d_1' \end{matrix} $ 
 & $\begin{matrix} u_{1R} \cr d_{1R} \end{matrix}$ & $(-,-)$\\
\mynoalign1
 & $\begin{matrix}{\bf \overline{10}}_{-1}\cr{\bf \overline{10}}_{-1}\end{matrix}$ 
 & $({\bf 2,1,8})$ & ${\bf \overline{4}}$ & $({\bf \overline{3},2})^{-1/6}_{-1/6} $ 
 &$\begin{matrix} \onethird  \cr -\twothird  \end{matrix} $ 
 & $\begin{matrix} \hat d'_2 \cr \hat u'_2 \end{matrix}$ & & $(-,+)$ \\
\mynoalign1
 & $\begin{matrix}{\bf \overline{10}}_{-1}\cr {\bf 5}_{+3}\end{matrix}$ 
 & $({\bf 1,2,8})$ & ${\bf 4}$ & $({\bf 3,1})^{+2/3}_{-1/3} $ 
 & $\begin{matrix} \twothird  \cr -\onethird  \end{matrix}$
 & $\begin{matrix} u_3' \cr d_3' \end{matrix} $ 
 & $\begin{matrix} u_{3R} \cr d_{3R} \end{matrix}$ & $(-,-)$\\
\mynoalign1
 & $\begin{matrix}{\bf \overline{10}}_{-1}\cr{\bf \overline{10}}_{-1}\end{matrix}$ 
 & $({\bf 2,1,8})$ & ${\bf \overline{4}}$ & $({\bf \overline{3},2})^{-1/6}_{-1/6} $ 
 &$\begin{matrix} \onethird  \cr -\twothird  \end{matrix} $ 
 & $\begin{matrix} \hat d'_1 \cr \hat u'_1 \end{matrix}$  & & $(-,+)$ \\
\mynoalign1
 & $\begin{matrix}{\bf 1}_{-5}\cr{\bf \overline{10}}_{-1}\end{matrix}$ 
 & $({\bf 1,2,8})$ & ${\bf 4}$ & $({\bf 1,1})^{~0}_{-1}  $ 
 & $\begin{matrix} 0 \cr -1 \end{matrix}$ 
 & $\begin{matrix} \nu' \cr e' \end{matrix}$ 
 & $\begin{matrix} \nu_R \cr e_R \end{matrix}$ & $(-,-)$\\
\mynoalign1
\hline
\end{tabular}
\end{center}
\caption{The content of $\Psi_{\bf 32}$ is tabulated.
In the second and third columns $SU(5)_Z = SU(5) \times U(1)_Z$ and
$G_{227} =  SU(2)_L \times SU(2)_R \times SO(7)$ are shown.
In the fourth column $SO(6) $ in $G_{PS} = SU(2)_L \times SU(2)_R \times SO(6)$ is shown.
In the fifth column $G_{SM} = SU(3)_C \times SU(2)_L \times U(1)_Y$. Superscripts and subscripts
indicate $U(1)_Y$ charges. 
In the last column, parity at $y=0$ and $L$ is given for  left-handed components.  
The right-handed components have opposite parity.
\label{32Content}
}
\end{table}
}

$\Psi_{\bf 11}$ decomposes into  an $SO(10)$ vector and a singlet.  The former  further
decomposes into an $SO(4)$ vector $\psi_j$ ($j=1 \sim 4$) and 
an $SO(6)$ vector $\psi_j$ ($j=5 \sim 10$).
An $SO(4)$ vector $\psi_j$ ($j=1 \sim 4$) transforms as $({\bf 2}, {\bf 2})$ of $SU(2)_L \times SU(2)_R$.
Under $\Omega_L \in SU(2)_L$ and $\Omega_R \in SU(2)_R$,
\beqn
&&\hskip -1.cm
\hat \psi \go \Omega_L \hat \psi  \, \Omega_R^\dagger ~, \cr
\noalign{\kern 10pt}
&&\hskip -1.cm
\hat \psi = \frac{1}{\sqrt{2}} (\psi_4 + i \vec \psi \vec\sigma) \cdot  i \sigma_2 \cr
\noalign{\kern 10pt}
&&\hskip -.5cm
= \frac{1}{\sqrt{2}} \begin{pmatrix} \psi_2 + i \psi_1 & -\psi_4 - i\psi_3 \cr
\psi_4 - i \psi_3 & \psi_2 - i \psi_1 \end{pmatrix}
\leftrightarrow
\begin{pmatrix} \hat E & N\cr \hat N & E \end{pmatrix} ~.
\label{vector1}
\eeqn
An $SO(6)$ vector $\psi_j$ ($j=5 \sim 10$) decomposes into two parts;
\beeq
\begin{pmatrix} D_j \cr \hat D_j \end{pmatrix}
=  \frac{1}{\sqrt{2}} ( \psi_{3 + 2j} \mp i \psi_{4 + 2j} ) \quad (j=1,2,3).
\label{vector2}
\eneq
With this notation the contents of $\Psi_{\bf 11}$ and $\Psi_{\bf 11}'$ in \eqref{BCF1}
are summarized in Table \ref{11Content}. 
Notice that $\Psi_{\bf 11}$ and $ \Psi_{\bf 11}'$ have no components carrying the 
quantum numbers of $u$ quark.  
With the boundary conditions \eqref{BCF1}, only $(D_{jR}, \hat D_{jR})$ and 
$(D_{jL}', \hat D_{jL}' )$ have zero modes.
If the boundary conditions \eqref{BCF2}  were adopted, then 
$(N_R, E_R, \hat E_R, \hat N_R, S_L)$ and $(N_L', E_L', \hat E_L', \hat N_L', S_R')$
would have zero modes.

\def\mynoalign1{\noalign{\kern 4pt}}
{\small
\begin{table}[p]
\begin{center}
\begin{tabular}{lclclclclclc|c|}
\multicolumn{8}{c}   %
{\normalsize $\Psi_{\bf 11}$}\\
\noalign{\kern 3pt}
\hline
$SO(10)$ & $SU(5)_Z$ & $G_{PS}$  & $G_{SM}$ &$Q_\EM$ & name 
& $\begin{matrix} {\rm zero} \cr {\rm mode}\end{matrix}$
&$\begin{matrix}{\rm parity}\cr {\rm (left)}\end{matrix}$\\
\noalign{\kern 3pt}
\hline \hline
\mynoalign1
${\bf 10}$
 & ${\bf 5}_{+2}$ 
 & $({\bf 2,2,1})$  & $({\bf 1,2})_{+1/2}$ 
 & $\begin{matrix} +1 \cr 0 \end{matrix}$ 
 & $\begin{matrix} \hat E \cr \hat N \end{matrix}$ 
 &  & $(-,+)$ \\
\mynoalign1
 & ${\bf \overline{5}}_{-2}$ 
 & $({\bf 2,2,1})$  & $({\bf 1,2})_{-1/2}$ 
 & $\begin{matrix} 0 \cr -1 \end{matrix}$ 
 & $\begin{matrix} N \cr E \end{matrix}$ 
 &  & $(-,+)$ \\
\mynoalign1
 & ${\bf 5}_{+2}$ 
 & $({\bf 1,1,6})$  & $({\bf 3,1})_{-1/3}$ 
 & $- \onethird$ 
 & $D_j$ 
 & $D_{jR} $ & $(-,-)$ \\
\mynoalign1
 & ${\bf \overline{5}}_{-2}$ 
 & $({\bf 1,1,6})$  & $({\bf \overline{3},1})_{+1/3}$ 
 & $ + \onethird$ 
 & $\hat D_j$ 
 & $\hat D_{jR} $ & $(-,-)$ \\
\mynoalign1
\hline
\mynoalign1
$~{\bf 1}$ 
 & ${\bf 1}_{0}$ 
 & $({\bf 1,1,1})$ & $({\bf 1,1})_0$ &$~0$
 &$S$ 
 & & $(+,-)$\\
\mynoalign1
\hline 
\end{tabular}
\vskip 20pt
\begin{tabular}{lclclclclclc|c|}
\multicolumn{8}{c}   %
{\normalsize $\Psi_{\bf 11}'$}\\
\noalign{\kern 3pt}
\hline
$SO(10)$ & $SU(5)_Z$ & $G_{PS}$  & $G_{SM}$ &$Q_\EM$ & name 
& $\begin{matrix} {\rm zero} \cr {\rm mode}\end{matrix}$
&$\begin{matrix}{\rm parity}\cr {\rm (left)}\end{matrix}$\\
\noalign{\kern 3pt}
\hline \hline
\mynoalign1
${\bf 10}$
 & ${\bf 5}_{+2}$ 
 & $({\bf 2,2,1})$  & $({\bf 1,2})_{+1/2}$ 
 & $\begin{matrix} +1 \cr 0 \end{matrix}$ 
 & $\begin{matrix} \hat E' \cr \hat N' \end{matrix}$ 
 &  & $(+,-)$ \\
\mynoalign1
 & ${\bf \overline{5}}_{-2}$ 
 & $({\bf 2,2,1})$  & $({\bf 1,2})_{-1/2}$ 
 & $\begin{matrix} 0 \cr -1 \end{matrix}$ 
 & $\begin{matrix} N' \cr E' \end{matrix}$ 
 &  & $(+,-)$ \\
\mynoalign1
 & ${\bf 5}_{+2}$ 
 & $({\bf 1,1,6})$  & $({\bf 3,1})_{-1/3}$ 
 & $- \onethird$ 
 & $D_j'$ 
 & $D_{jL}' $ & $(+,+)$ \\
\mynoalign1
 & ${\bf \overline{5}}_{-2}$ 
 & $({\bf 1,1,6})$  & $({\bf \overline{3},1})_{+1/3}$ 
 & $ + \onethird$ 
 & $\hat D_j'$ 
 & $\hat D_{jL}' $ & $(+,+)$ \\
\mynoalign1
\hline
\mynoalign1
$~{\bf 1}$ 
 & ${\bf 1}_{0}$ 
 & $({\bf 1,1,1})$ & $({\bf 1,1})_0$ &$~0$
 &$S'$ 
 & & $(-,+)$\\
\mynoalign1
\hline
\end{tabular}
\end{center}
\caption{The contents of $\Psi_{\bf 11}$ and $\Psi_{\bf 11}'$
with the boundary conditions \eqref{BCF1}   are tabulated.
The same notation is adopted as in Table \ref{32Content}.
\label{11Content}
}
\end{table}
}

\subsection{Brane interactions}

In addition to \eqref{action1} and \eqref{braneaction1}, there appear brane mass-Yukawa
 interactions among $\Psi_{\bf 32}^a, \Psi_{\bf 11}^a, \Psi_{\bf 11}^{\prime a}$
and $\Phi_{\bf 16}$ on the Planck brane at $y=0$.
On the Planck brane the $SO(10)$ local gauge invariance must be manifestly preserved.
$\Psi_{\bf 32}$ ($\Psi_{\bf 11}$) field decomposes to 
$\Psi_{\bf 16}$ and $\Psi_{\overline{\bf 16}}$ ($\Psi_{\bf 10}$ and $\Psi_{\bf 1}$), 
as indicated in Table \ref{32Content} (Table \ref{11Content}), under $SO(10)$ transformations.
Only fields of even parity at $y=0$ participate in the brane mass-Yukawa interactions.
We need to write down $SO(10)$ invariant terms in terms of $\Psi_{{\bf 16}L}^a$, 
$\Psi_{\overline{\bf 16}R}^a$, $\Psi_{{\bf 10}R}^a$,  $\Psi_{{\bf 1}L}^a$,  
$\Psi_{{\bf 10}L}^{\prime a}$,  $\Psi_{{\bf 1}R}^{\prime a}$, and $\Phi_{\bf 16}$.
Further we impose the condition that the action be invariant under a global $U(1)$
$\Psi$ fermion number ($N_\Psi$) transformation
\beeq
\Psi_{\bf 32}^a \go e^{i\alpha} \Psi_{\bf 32}^a ~,~~
\Psi_{\bf 11}^a \go e^{i\alpha} \Psi_{\bf 11}^a ~,~~
\Psi_{\bf 11}^{\prime a} \go e^{i\alpha} \Psi_{\bf 11}^{\prime a} ~.
\label{FermionNo1}
\eneq

Six types of brane interactions are allowed.
\beqn
&&\hskip -1.cm
S_{\rm brane} = \int d^5 x \sqrt{- \det G} \, \delta(y) 
\Big( \cL_1 + \cL_2 + \cL_3 + \cL_4 + \cL_5 + \cL_6 \Big) ~, \cr
\noalign{\kern 10pt}
&&\hskip -1.cm
\cL_1 = -  \Big\{ \kappa_{[1,16]}^{ab}
\overline{\Psi}{}^{\prime a}_{{\bf 1}R}  \,\Phi_{\bf 16}^\dagger  \, \Psi_{{\bf 16}L}^b
+ \kappa_{[1,16]}^{ab}{}^* \,
\overline{\Psi}{}_{{\bf 16}L}^{b}  \, \Phi_{\bf 16}  \, \Psi^{\prime a}_{{\bf 1}R} \Big\} , \cr
\noalign{\kern 10pt}
&&\hskip -1.cm
\cL_2 = -  \Big\{ \kappa_{[1,\overline{16}]}^{ab}
\overline{\Psi}{}^a_{{\bf 1}L}  \,\tilde \Phi_{\overline{\bf 16}}^\dagger  \, \Psi_{\overline{\bf 16}R}^b
+ \kappa_{[1,\overline{16}]}^{ab}{}^* \, 
\overline{\Psi}{}_{\overline{\bf 16}R}^b  \, \tilde \Phi_{\overline{\bf 16}}  \, \Psi^a_{{\bf 1}L} \Big\} , \cr
\noalign{\kern 10pt}
&&\hskip -1.cm
\cL_3= -  \Big\{ \kappa_{[10,16]}^{ab}
 (\overline{\Psi}{}^a_{{\bf 10}R})_j   \, \hat{\tilde \Phi}_{\overline{\bf 16}}^\dagger  \,  
\Gamma^j \, \hat \Psi_{{\bf 16}L}^b
+  \kappa_{[10,16]}^{ab}{}^* \,  \overline{\hat \Psi}{}_{{\bf 16}L}^b  \, \Gamma^j \, 
\hat{\tilde \Phi}_{\overline{\bf 16}}  \, (\Psi^a_{{\bf 10}R})_j \Big\} , \cr
\noalign{\kern 10pt}
&&\hskip -1.cm
\cL_4 = -   \Big\{ \kappa_{[10,\overline{16}]}^{ab}
 (\overline{\Psi}{}^{\prime a}_{{\bf 10}L} )_j   \, \hat\Phi_{\bf 16}^\dagger  \,  
\Gamma^j \, \hat \Psi_{\overline{\bf 16}R}^b
+ \kappa_{[10,\overline{16}]}^{ab} {}^* \, 
\overline{\hat \Psi}{}^b_{\overline{\bf 16}R}  \, \Gamma^j \, 
\hat \Phi_{\bf 16}  \, (\Psi^{\prime a}_{{\bf 10}L})_j \Big\} , \cr
\noalign{\kern 10pt}
&&\hskip -1.cm
\cL_5 = -   \Big\{ \mu_{[1,1]}^{ab}
\overline{\Psi}{}^{\prime a}_{{\bf 1}R}   \, \Psi_{{\bf 1}L}^b
+ \mu_{[1,1]}^{ab}{}^* \,   \overline{\Psi}{}_{{\bf 1}L}^b  \,  \Psi^{\prime a}_{{\bf 1}R} \Big\} , \cr
\noalign{\kern 10pt}
&&\hskip -1.cm
\cL_6= -   \Big\{ \mu_{[10,10]}^{ab}
\overline{\Psi}{}^{\prime a}_{{\bf 10}L}   \, \Psi_{{\bf 10}R}^b
+ \mu_{[10,10]}^{ab}{}^* \,  \overline{\Psi}{}_{{\bf 10}R}^b  \,  \Psi^{\prime a}_{{\bf 10}L} \Big\} .
\label{BraneInt1}
\eeqn
Here $\tilde \Phi_{\overline{\bf 16}} =\hat R \, \Phi_{\bf 16}^*$ with  $\hat R$ defined in \eqref{Rtransform1}
transforms as $\overline{\bf 16}$, and we have employed 32-component notation given by
\beeq
\hat \Phi_{\bf 16} = \begin{pmatrix} \Phi_{\bf 16} \cr 0 \end{pmatrix},~~
\hat {\tilde \Phi}_{\overline{\bf 16}} =  \begin{pmatrix} 0 \cr \tilde \Phi_{\overline{\bf 16}} \end{pmatrix} , ~~
\hat \Psi_{\bf 16} = \begin{pmatrix} \Psi_{\bf 16}  \cr 0 \end{pmatrix} , ~~
\hat \Psi_{\overline{\bf 16}} = \begin{pmatrix}  0\cr \Psi_{\overline{\bf 16}} \end{pmatrix} .
\label{32fermion}
\eneq
In general all  coefficients $\kappa$ and $\mu$ in \eqref{BraneInt1} have matrix structure 
in the generation space, which induces flavor mixing.  
In the present paper we restrict ourselves to diagonal $\kappa$ and $\mu$.

The total action is given by
\beeq
S = S_{\rm bulk} + S_{\Phi_{\bf 16}} + S_{\rm brane} ~.
\label{totalAction}
\eneq

\subsection{EW Higgs boson}

The orbifold boundary condition \eqref{BCgauge1}  reduces the $SO(11)$ gauge symmetry to 
$SO(4) \times SO(6) \simeq SU(2)_L \times SU(2)_R \times SO(6)$.  It is easy to see that 
terms bilinear in fields in the gauge field part of the action $S_{\rm bulk}$, \eqref{action1}, become
\beqn
&&\hskip -1.cm
 \int d^4 x \frac{dz}{kz}  \, \sum_{j<k} \bigg[ \frac{1}{2} A_\mu^{(jk)} 
\Big\{ \eta^{\mu\nu}  \big( \Box + k^2 {\cal P}_4 \big)
  - \big( 1 - \frac{1}{\xi} \big) \dd^\mu \dd^\nu \Big\}  A_\nu^{(jk)} \cr
\noalign{\kern 10pt}
&&\hskip 1.3cm
+ \frac{1}{2} k^2 A_z^{(jk)} \big( \Box+\xi k^2 {\cal P}_z \big) A_z^{(jk)} 
 +\bar c^{(jk)}  \big( \Box + \xi k^2 {\cal P}_4 \big) c^{(jk)} \bigg] , \cr
\noalign{\kern 10pt}
&&\hskip -1.cm
\Box \equiv \eta^{\mu\nu} \dd_\mu \dd_\nu ~,~~
\dd^\mu = \eta^{\mu\nu} \dd_\nu ~,~~ 
 {\cal P}_4 \equiv z \frac{\dd}{\dd z} \frac{1}{z} \frac{\dd}{\dd z} ~,~~
 {\cal P}_z \equiv \frac{\dd}{\dd z} z\frac{\dd}{\dd z} \frac{1}{z} ~.  
\label{gauge-action2}
\eeqn
For four-dimensional components $A_\mu$ of $SO(10)$ gauge fields have additional 
bilinear terms coming from the brane scalar interaction $S_{\Phi_{\bf 16}}$, \eqref{braneaction1},
with $\la \Phi_{\bf 16} \ra \not= 0$.
Parity of $A_\mu$ and $A_z$ is summarized in Table \ref{gaugeContent1}.
For  $A_z$, only  four components  $({\bf 2,2,1})$ in 
$G_{PS} = SU(2)_L \times SU(2)_R \times SO(6)$
have parity $(+,+)$, and therefore zero modes corresponding to the Higgs doublet in the SM.
For  $A_\mu$,  components $({\bf 3,1,1})$, $({\bf 1,3,1})$, and $({\bf 1,1,15})$, 
corresponding to $SU(2)_L$, $SU(2)_R$, and $SO(6)$, respectively,  have parity $(+,+)$. 
$SU(2)_R \times SO(6)$ symmetry is spontaneously broken to $SU(3)_C \times U(1)_Y$
by $\la \Phi_{\bf 16} \ra \not= 0$, reducing to the SM gauge symmetry.

{
\begin{table}
\begin{center}
\renewcommand{\arraystretch}{1.2}
\begin{tabular}{|c|c|c|}
\hline
$G_{PS}$ & $A_\mu$ & $A_z$ \\
\hline 
$({\bf 3,1,1})$ &$(+,+)$&$(-,-)$\\
$({\bf 1,3,1})$ &$(+,+)$ &$(-,-)$\\
$({\bf 1,1,15})$ &$(+,+)$&$(-,-)$\\
$({\bf 2,2,6})$ &$(+,-)$&$(-,+)$\\
$({\bf 2,2,1})$ &$(-,-)$&$(+,+)$\\
$({\bf 1,1,6})$ &$(-,+)$&$(+,-)$\\
\hline
\end{tabular}
\end{center}
\caption{Parity of $A_\mu$ and $A_z$ is classified with the content 
in $G_{PS} = SU(2)_L \times SU(2)_R \times SO(6)$.}
\label{gaugeContent1}
\end{table}
}

Mode functions of $A_z$ in the fifth dimension are determined by 
\beeq
{\cal P}_z h_n(z) = - \lambda_n^2 h_n(z) ~,~~ 
\int_1^{z_L} \frac{kdz}{z} \, h_n(z) \, h_\ell (z) = \delta_{n \ell} ~, 
\label{zeromodeHiggs1}
\eneq
where boundary conditions at $z=1$ and $z_L$ are given by $(d/dz) (h_n/z) = 0$ or 
$h_n =0$  for parity even or odd fields, respectively.
In particular,  the zero mode ($\lambda_0=0$) function  is given by
\beeq
h_0^{(++)} (z) = u_H(z) =  \frac{1}{kz} \tilde u_H(y) = 
\sqrt{\myfrac{2}{k(z^2_L-1)}} ~ z  
\label{zeromodeHiggs2}
\eneq
for $1\le z \le z_L$ ($0 \le y \le L$).
The mode function in the $y$-coordinate satisfies $\tilde u_H(-y) = \tilde u_H(y) = \tilde u_H(y + 2L)$.
We note that
\beeq
\int_0^L dy \,  \tilde u_H(y)=  \int_1^{z_L} dz   \, u_H(z)  = \sqrt{ \frac{z^2_L-1}{2k}} ~.
\label{zeromodeHiggs3}
\eneq
The zero modes of $A_z$ are physical degrees of freedom, being unable to be gauged away.  
They are   in $A_z^{a \, 11} (x,z)$ ($a=1 \sim 4$).  In terms of mode functions $\{ h_n^{(++)} (z) \}$
for parity $(+,+)$ boundary condition, 
\beeq
A_z^{a \, 11} (x,z) = \phi_H^a (x) u_H(z) +  \sum_{n=1}^\infty \phi_H^{a (n)} (x) h_n^{(++)} (z) 
\quad (a= 1 \sim 4),
\label{KKexpansion1}
\eneq
where the four-component real field, $\phi_H^a (x)$,  plays the role of the EW Higgs doublet field in the SM.
It will be shown that $\phi_H \equiv \phi_H^4$ dynamically develops vev $\la \phi_H \ra \not= 0$, 
breaking the SM symmetry to $U(1)_\EM$.
$\phi_H^{1,2,3}$ are absorbed by $W$ and $Z$ bosons.

\subsection{AB phase $\theta_H$}

Under a general gauge transformation 
$A_M' = \Omega A_M \Omega^{-1} + (i/g)\Omega \dd_M \Omega^{-1}$
new gauge potentials satisfy new boundary conditions
\beqn
&&\hskip -1.cm
\begin{pmatrix} A_\mu' \cr A_y' \end{pmatrix} (x, y_j - y) =
P_j '\begin{pmatrix} A_\mu \cr - A_y \end{pmatrix} (x, y_j + y) P_j^{\prime -1} 
+ \frac{i}{g} P_j ' \begin{pmatrix} \dd_\mu \cr - \dd_y \end{pmatrix}  P_j^{\prime -1}  ~,  \cr
\noalign{\kern 10pt}
&&\hskip -1.cm
P_j ' = \Omega (x, y_j -y) P_j \Omega (x, y_j +y)^{-1} ~.
\label{BCgauge2}
\eeqn
The original boundary conditions \eqref{BCgauge1} are maintained if and only if $P_j' = P_j$.
Such a class of gauge transformations define the residual gauge invariance.\cite{YH2, HHKY2004}

There are a class of ``large'' gauge transformations which transform $A_y$
nontrivially.  Consider
\beeq
\Omega(y; \alpha)
= \exp \bigg\{- i \frac{g\alpha}{\sqrt{2}} \int_y^L dy \, \tilde u_H(y) \, T_{4,11}  \bigg\}  ~.
\label{LargeGT0}
\eneq
For $A_y = \frac{1}{\sqrt{2}} A^{(4,11)}_y (x,y) T_{4,11}$, 
\beqn
&&\hskip -1.cm
A_y' = A_y +\frac{\alpha}{\sqrt{2}} \, \tilde u_H(y) \, T_{4,11}  ~~, \cr
\noalign{\kern 10pt}
&&\hskip -1.cm
\phi_H(x) \go \phi_H' (x) = \phi_H (x) + \alpha ~.
\label{LargeGT1}
\eeqn
The new boundary condition matrices $P_j'$ are evaluated, with the aid of 
$\{ P_j , T_{4, 11} \} = 0$, to be $P_j' = \Omega (y_j -y ; \alpha) \Omega (y_j +y ; \alpha) P_j$,
that is, 
\beqn
&&\hskip -1.cm
P_0' = \Omega ( -y ; \alpha) \Omega (y ; \alpha) P_0 = \Omega(0 ; 2\alpha) P_0 ~, \cr
\noalign{\kern 10pt}
&&\hskip -1.cm
P_1' = \Omega ( L -y ; \alpha) \Omega (L + y ; \alpha) P_1 =  P_1 ~.
\label{LargeGT2}
\eeqn
The boundary conditions are preserved provided $\Omega(0 ; 2\alpha) = 1$.
As $( T_{4,11}^{\rm sp} )^2 =\frac{1}{4} \, I_{32}$ in the spinorial representation, 
the boundary conditions are preserved provided 
\beeq
\frac{g\alpha}{\sqrt{2}} \int_0^L dy \, \tilde u_H(y) 
= \frac{g\alpha}{\sqrt{2}}  \sqrt{ \frac{z^2_L-1}{2k}} = 2\pi n \quad (n = \hbox{an integer}) ~.
\label{LargeGT3}
\eneq

Aharonov-Bohm (AB)   phases along the fifth dimension are defined by  phases of  eigenvalues of
\beeq
\hat W = P \exp \bigg\{ ig \int_{-L}^L dy \, A_y \bigg\} \cdot P_1 P_0 ~.
\eneq
They are gauge-invariant.
$A_y^{4,11} (x,-y) = A_y^{4,11} (x,y) $ and the orthonormality relation \eqref{zeromodeHiggs1}
implies that $\int_1^{z_L} dz \, h_n^{(++)} (z) =0$ for $n \not= 0$.
Hence for $A_y = (1/\sqrt{2}) A_y^{4,11} T_{4,11}$
\beeq
\hat W = \exp \bigg\{ i \frac{g \phi_H}{\sqrt{2}}  \int_0^L dy \, \tilde u_H(y) \, 2 \,T_{4,11} \bigg\} 
\cdot \begin{pmatrix} I_4 &&\cr &- I_6 &\cr &&1 \end{pmatrix}
\eneq
in the vectorial representation so that the relevant phase $\hat \theta_H (x) $ is given by
\beqn
&&\hskip -1.cm
\hat \theta_H (x) = \frac{g \phi_H (x)}{\sqrt{2}}  \int_0^L dy \, \tilde u_H(y) 
= \frac{\phi_H(x)}{f_H} ~,  \cr
\noalign{\kern 10pt}
&&\hskip -1.cm
f_H = \frac{2}{g} \sqrt{\frac{k}{z_L^2 -1}}
=  \frac{2}{g_w} \sqrt{\frac{k}{L(z_L^2 -1)}}  ~.
\label{AB1}
\eeqn
Under a large gauge transformation satisfying \eqref{LargeGT2}, 
\beeq
\hat \theta_H (x) \go \hat \theta_H' (x) = \hat \theta_H (x) + 2 \pi n ~. 
\label{AB2}
\eneq
Denoting  $\theta_H = \la \hat \theta_H(x)\ra$, one has
\beeq
A^{(4,11)}_z (x,z)=\big\{ \theta_H f_H + H(x) \big\}  \, u_H(z) +\cdots ~.
\label{AB3}
\eneq
$H(x)$ corresponds to the neutral Higgs boson found at LHC.
The value of $\theta_H$ is undetermined at the tree level, but is determined
dynamically at the one loop level.  The effective potential $V_\eff (\theta_H)$ is
evaluated in Section 5.

\subsection{Twisted gauge}

Under a gauge transformation \eqref{LargeGT0} $\hat \theta_H (x)$ is transformed to
$\hat \theta_H' (x) = \hat \theta_H (x) + (\alpha/f_H)$.   
In the new gauge given with $\alpha = - \theta_H f_H$, 
\beqn
&&\hskip -1.cm
\Omega (y)  = \exp \bigg\{ i \frac{g \theta_H f_H}{\sqrt{2}} \int_y^L dy \, \tilde u_H(y)\cdot T_{4,11} \bigg\} \cr
\noalign{\kern 10pt}
&&\hskip -0.cm
= \exp \big\{ i \theta(z) T_{4,11} \big\} ~, \quad 
\theta (z) = \theta_H \frac{z_L^2 - z^2}{z_L^2 - 1}  \quad {\rm for~} 1 \le z \le z_L ~, 
\label{twisted1}
\eeqn
$\la \theta_H' (x) \ra =0$.  Hereafter this new gauge is called the twisted gauge, and 
quantities in the twisted gauge are denoted with tildes.\cite{HS2007, Falkowski2007}
According to \eqref{LargeGT2} the boundary conditions become
\beeq
\tilde P_0 = \Omega(0)^2 P_0 = e^{2i \theta_H T_{4,11}} P_0 ~,~~ \tilde P_1 = P_1  ~.
\label{BCtwisted1}
\eneq
In the vectorial representation
\beeq
\tilde P_0^{\rm vec} = 
\begin{cases} \begin{pmatrix} \cos 2 \theta_H & -\sin 2\theta_H \cr 
       - \sin 2\theta_H & - \cos 2 \theta_H  \end{pmatrix} &\hbox{in the 4-11 subspace,} \cr
       \noalign{\kern 10pt}
~~ I_9 &\hbox{otherwise.}  \end{cases}
\label{BCtwisted2}
\eneq
Components of  $\Psi_{\bf 32}$ split into two groups;
\beqn
&&\hskip -1.cm
\chi  
= \begin{pmatrix} \nu \cr \nu' \end{pmatrix}, \begin{pmatrix} e \cr e' \end{pmatrix},
\begin{pmatrix}  u_j \cr  u_j' \end{pmatrix}, \begin{pmatrix}  d_j \cr  d_j' \end{pmatrix}, \cr
\noalign{\kern 10pt}
&&\hskip -1.cm
\hat \chi = 
\begin{pmatrix} {\hat\nu} \cr {\hat\nu}' \end{pmatrix}, 
       \begin{pmatrix} {\hat e} \cr {\hat e}' \end{pmatrix},
       \begin{pmatrix} {\hat u}_j \cr {\hat u}_j' \end{pmatrix}, 
       \begin{pmatrix} {\hat d}_j \cr {\hat d}_j' \end{pmatrix}. 
\label{spinor2}
\eeqn
The boundary condition matrix becomes
\beeq
\tilde P_0^{\rm sp} = \begin{pmatrix} \cos  \theta_H &  \mp i \sin \theta_H \cr 
      \pm  i \sin  \theta_H & - \cos  \theta_H  \end{pmatrix} 
\quad {\rm for~} \bigg\{ \,
\begin{matrix} \chi \cr \hat \chi \end{matrix} ~.
\label{BCtwisted3}
\eneq
It turns out very convenient to evaluate $V_\eff (\theta_H)$ in the twisted gauge.

\subsection{Proton stability}

In the present model of gauge-Higgs grand unification, the proton decay is forbidden.
Fields of up quark quantum number, $u$ and $u'$,  are contained only in $\Psi_{\bf 32}$.
Fields of down quark quantum number  are in $\Psi_{\bf 32}$ ($d$ and $d'$) and 
in $\Psi_{\bf 11}$ and  $\Psi_{\bf 11}'$ ($D$ and $D'$), 
fields of electron quantum number are in $\Psi_{\bf 32}$ ($e$ and $e'$) and 
in $\Psi_{\bf 11}$ and  $\Psi_{\bf 11}'$ ($E$ and $E'$),
and fields of neutrino quantum number are in $\Psi_{\bf 32}$ ($\nu$, $\hat \nu$,  $\nu '$ and $\hat \nu '$) and 
in $\Psi_{\bf 11}$ and  $\Psi_{\bf 11}'$ ($N$, $\hat N$, $S$, $N'$, $\hat N '$, and $S'$).
The down quark at low energies, for instance, is a linear combination of $d$, $d'$, $D$ and $D'$.
All quarks and leptons have $N_\Psi =1$ fermion number.
The total  action preserves  the $N_\Psi =1$ fermion number. 
As the proton has $N_\Psi =3$ and the positron has $N_\Psi =-1$, the proton decay $p \go \pi^0 e^+$,
for instance, cannot occur.
This is contrasted to $SU(5)$ or $SO(10)$ GUT in four dimensions in which proton decay 
inevitably takes place.   The $SO(11)$ gauge-Higgs grand unification provides a natural framework
of grand unification in which proton decay is forbidden.

One comment is in order.  $S$ and $S'$ in $\Psi_{\bf 11}$ and  $\Psi_{\bf 11}'$ are $SO(10)$ singlets.
One could introduce Majorana masses such as $\overline{S} S^c \delta (y)$  on the Planck brane, which
break the $N_\Psi$ fermion number.  They would give rise to Majorana masses for neutrinos, and 
at the same time could induce proton decay at higher loops.

\section{Spectrum of gauge fields}

$V_\eff (\theta_H)$ at the one loop level is determined from the mass spectrum of all fields
when $\la \hat \theta_H (x) \ra = \theta_H$.  It is convenient to determine the spectrum in 
the twisted gauge in which $\la \tilde{\hat \theta}_H (x) \ra = 0$.  
The nontrivial $\theta_H$ dependence is transferred to the boundary conditions.
In this section we determine the spectrum of gauge fields.  

In the absence of the brane interactions with $\Phi_{\bf 16}$ in $S_{\Phi_{\bf 16}}$, 
\eqref{braneaction1}, the boundary conditions for gauge fields are given by
\beqn
&&\hskip -1.cm
\begin{cases}
N: ~~ \myfrac{\dd }{\dd z} \, A_\mu = 0 &\hbox{for parity} = + \cr
\noalign{\kern 5pt}
D: ~~  A_\mu = 0 &\hbox{for parity} = -
\end{cases}  \cr
\noalign{\kern 10pt}
&&\hskip -1.cm
\begin{cases}
N: ~~ \myfrac{\dd }{\dd z} \Big(\myfrac{1}{ z} \, A_z \Big)= 0 &\hbox{for parity} = + \cr
\noalign{\kern 5pt}
D: ~~ A_z = 0 &\hbox{for parity} = -
\end{cases}  
\label{BCgauge3}
\eeqn
at $z=1$ ($y=0$) and $z=z_L$ ($y=L$).
In the presence of $S_{\Phi_{\bf 16}}$, the brane mass terms \eqref{branemass1} for $A_\mu$
are induced, and the Neumann boundary condition $N$ is modified to an effective Dirichlet
condition $D_\eff$  for low-lying KK modes of the twenty-one components of $A_\mu$ as described below.
The boundary conditions for the gauge fields are summarized in Table \ref{gaugeContent2}.

{
\begin{table}
\begin{center}
\ovalbox{
\begin{tabular}{l}
\\[-1.em]
$SO(11)$\\[0.75em]
\ (5),(6)\\[0.5em]
\end{tabular}
\ovalbox{
\begin{tabular}{l}
\\[-1.0em]
$SO(10)$\\[0.75em]
\ \ (4)\\[0.5em]
\end{tabular}
\ovalbox{
\begin{tabular}{l}
\\[-1em]
$SU(5)$\ \ \ \ \ \ \ \  $G_{SM}$\\[0.5em]
\ \ (2)\ \ \ \ \ \ \ \ \ \ (1) \\[0.5em]
\end{tabular}
}
\hspace{-6em}
\ovalbox{
\begin{tabular}{l}
\\[-1.25em]
\hspace{5em}
$G_{PS}$\\[0.5em]
\hspace{5.3em}(3) \\[0.4em]
\end{tabular}
}}}
\end{center}
\begin{center}
\renewcommand{\arraystretch}{1.1}
\begin{tabular}{|c|c|c|c|c|}\hline
&&$\begin{matrix}{\rm No.~of}\cr{\rm generators} \end{matrix}$
&$\begin{matrix} A_\mu \cr z: (1, z_L)\end{matrix}$
&$\begin{matrix} A_z \cr z: (1, z_L)\end{matrix}$ \\
\hline 
(1)&$G_{SM}$ &12 &$(N,N)$ &$(D, D)$ \\
(2)&$SU(5) / G_{SM}$ &12  &$(N,D)$&$(D, N)$\\
(3)&$G_{PS} /  G_{SM}$ &9 &$(D_{\rm eff},N)$ &$(D,D)$ \\
(4)&$SO(10) / (SU(5)\cup G_{PS}) $ &12 &$(D_{\rm eff},D)$ &$(D, N)$ \\
(5)&$SO(5)/SO(4)$ &4 &$(D,D)$ &$(N,N)$ \\
(6)&$SO(7)/SO(6)$ &6 &$(D,N)$ &$(N,D)$ \\\hline
\end{tabular}
\end{center}
\caption{The Venn diagram of gauge group structure is displayed in the top and
boundary conditions in each category are summarized in the bottom table.
Here $G_{SM}=SU(3)_C\times SU(2)_L\times U(1)_Y$, 
$G_{PS}=SU(2)_L\times SU(2)_R\times SO(6)$, and  $SU(5)\cap G_{PS}=G_{SM}$.
$N$ and $D$ stand for Neumann and Dirichlet conditions, respectively. 
$D_{\rm eff}$ represents the effective Dirichlet condition explained 
in Section 3.1, (i), (ii), (v), (vii) and (viii).
}
\label{gaugeContent2}
\end{table}
}

In the twisted gauge 
$\tilde A_M = \Omega(z) A_M \Omega(z)^{-1} + \frac{i}{g} \Omega (z) \dd_M \Omega (z)^{-1}$
where $\Omega(z)$ is given by \eqref{twisted1}.  One finds
\beqn
&&\hskip -1.cm
A_M^{k4} = \cos \theta(z) \tilde A_M^{k4} - \sin  \theta(z) \tilde A_M^{k,11} ~, \quad (k \not= 4) , \cr
\noalign{\kern 10pt}
&&\hskip -1.cm
A_M^{k,11} = \sin \theta(z) \tilde A_M^{k4} + \cos  \theta(z) \tilde A_M^{k,11} ~,  \cr
\noalign{\kern 10pt}
&&\hskip -1.cm
A_z^{4,11} = \tilde A_z^{4,11} - \frac{\sqrt{2}}{g} \, \theta' (z) 
=  \tilde A_z^{4,11} +  \frac{2\sqrt{2}}{g} \, \theta_H \, \frac{z}{z_L^2 - 1} ~.
\label{twisted2}
\eeqn
All other components are unchanged.
At $z = z_L$, $\theta(z_L) = 0$, and $A_M^{k4}$ and $A_M^{k,11}$ always have opposite parity.
It follows that $\tilde A_M$ satisfies the same boundary condition as $A_M$ at $z=z_L$.
In the bulk ($1< z <  z_L$) the bilinear part of the action is the same as in the free theory
in the twisted gauge.  Hence, depending on the BC at $z=z_L$,  
wave functions for $\tilde A_\mu$  and $\tilde A_z$ are given by
\beqn
\tilde A_\mu&&
N: ~ C(z; \lambda) ~, ~~ D:~ S(z; \lambda) ~, \cr
\noalign{\kern 5pt}
\tilde A_z&& 
N: ~ S'(z; \lambda) ~, ~~ D:~ C'(z; \lambda) ~,
\label{twistedMode1}
\eeqn
where $C(z;\lambda)$ and $S(z; \lambda)$ are defined in Appendix B.

\subsection{$A_\mu$ components}

\noindent
(i) $(\tilde A_\mu^{a_L}, \tilde A_\mu^{a_R}, \tilde A_\mu^{a, 11})$
 $(a=1,2)$: $W$ and $W_R$ towers

Original $ A_\mu^{a_L}$, $A_\mu^{a_R}$, and $A_\mu^{a, 11}$ have parity
$(+,+)$, $(+,+)$, and $(-,-)$, respectively.  $A_\mu^{a_R}$ picks up a brane mass.
To find its boundary condition at $z=1$, we need recall the equation of motion.
$\cL^{\rm gauge}_{\rm brane ~mass}$ in \eqref{branemass1} yields
$- (g^2 w^2/4) \delta(y) (A_\mu^{a_R} )^2$.
The equation of motion for $A_\mu^{a_R}$ in the $y$-coordinate is 
\beeq
\bigg\{ \eta^{\mu\nu} \Big( \Box + \frac{\dd}{\dd y} e^{-2\sigma(y)} \frac{\dd}{\dd y}  - 
\frac{g^2 w^2}{2} \delta (y) \Big) - \Big(1- \frac{1}{\xi} \Big) \dd^\mu \dd^\nu \bigg\}
A_\nu^{a_R} = \cdots ,
\label{AmuReq1}
\eneq
where interaction terms on the right side involve neither  total $y$ derivative nor $\delta(y)$.
By integrating the equation $\int_{-\ep}^\ep dy \cdots$  and taking the limit $\ep \go 0$, 
one finds that $\dd A_\mu^{a_R}/\dd y |_{y=\ep} = (g^2 w^2/4) A_\mu^{a_R}|_{y=0}$.
The brane mass gives rise to cusp behavior at $y=0$.
The boundary conditions for $(A_\mu^{a_L}, A_\mu^{a_R}, A_\mu^{a, 11})$ ($a=1,2$) at $z=1$ are
\beqn
&&\hskip -1.cm
\frac{\dd}{\dd z} A_\mu^{a_L} = 0 ~, \cr
\noalign{\kern 10pt}
&&\hskip -1.cm
\Big( \frac{\dd}{\dd z}  - \omega\Big) A_\mu^{a_R} = 0 ~, 
~~ \omega = \frac{g^2 w^2}{4k} ~, \cr
\noalign{\kern 10pt}
&&\hskip -1.cm
A_\mu^{a, 11} = 0 ~.
\label{BCgaugeW1}
\eeqn
Here it is understood that $\dd A_\mu^{a_R}/\dd z$ is evaluated at $z=1^+$.

Expressed in terms of fields in the twisted gauge, \eqref{BCgaugeW1} becomes
\beqn
&&\hskip -1.cm
\dd_z \tilde A_\mu^{23} + \dd_z \tilde A_\mu^{14}  \, \cos \theta_H 
- \dd_z \tilde A_\mu^{1,11} \sin \theta_H  = 0 ~, \cr
\noalign{\kern 10pt}
&&\hskip -1.cm
(\dd_z - \omega) \tilde A_\mu^{23} - (\dd_z - \omega) \tilde A_\mu^{14} \cos \theta_H 
+ (\dd_z - \omega) \tilde A_\mu^{1,11} \sin \theta_H = 0 ~, \cr
\noalign{\kern 10pt}
&&\hskip -1.cm
\tilde A_\mu^{14} \sin \theta_H +  \tilde A_\mu^{1,11} \cos \theta_H = 0 ~.
\label{BCgaugeW2}
\eeqn
From the boundary conditions at $z_L$, one can set 
\[
[ \tilde A_\mu^{23}, \tilde A_\mu^{14}, \tilde A_\mu^{1,11}] = 
[ \alpha_{23} C(z; \lambda) , \alpha_{14} C(z; \lambda), \alpha_{1,11} S(z; \lambda) ] a_\mu(x)
\]
for each mode.  The boundary conditions \eqref{BCgaugeW2} are transformed to
\beqn
&&\hskip -1.cm
K \begin{pmatrix} \alpha_{23} \cr \alpha_{14} \cr \alpha_{1,11} \end{pmatrix} = 0 ~,\cr
\noalign{\kern 10pt}
&&\hskip -1.cm
K = \begin{pmatrix} C' & \cos \theta_H C' &- \sin \theta_H S' \cr
C' - \omega C & - \cos\theta_H ( C' - \omega C )& \sin\theta_H (S' - \omega S) \cr
0 & \sin\theta_H C & \cos\theta_H S 
\end{pmatrix} 
\label{BCgaugeW3}
\eeqn
where $C' = C'(1; \lambda)$ etc..  The spectrum $\{ \lambda_n \}$ is determined by 
$\det K = 0$, or by
\beeq
2 C' (SC' + \lambda \sin^2 \theta_H) - \omega C (2 SC' + \lambda \sin^2 \theta_H ) = 0 .
\label{GaugeSpectrumW1}
\eneq
For sufficiently large $w$, the spectrum $\{ m_n = k \lambda_n \}$ of low-lying KK modes 
is approximately determined by the second term in \eqref{GaugeSpectrumW1}.
This approximation for $z_L = 10^5$, for instance, is justified for $\omega > 10^{-3}$.
\beqn
&W ~\hbox{tower:} & 2 S(1;\lambda) C'(1; \lambda) + \lambda \sin^2 \theta_H = 0 ~, \cr
&W_R ~\hbox{tower:} & C(1; \lambda) = 0 ~.
\label{GaugeSpectrumW2}
\eeqn
Asymptotically the equations determining the spectra of $W$ and $W_R$ towers 
become $SC' + \lambda \sin^2 \theta_H =0$ and $C' = 0$, respectively.
The mass of $W$ boson, $m_W = m_{W^{(0)}}$,  is given by
\beeq
m_W \sim \sqrt{\frac{k}{L}} \, z_L^{-1} \sin \theta_H 
= \frac{\sin \theta_H}{\pi \sqrt{kL}} ~ m_\KK ~. 
\label{Wmass1}
\eneq


\noindent 
(ii)  $(\tilde A_\mu^{3_L}, \tilde C_\mu ,  \tilde B_\mu^Y ,\tilde A_\mu^{3, 11})$: 
$\gamma$, $Z$ and $Z_R$ towers

Here $C_\mu = \sqrt{1/5} (A_\mu^{12} - A_\mu^{34} + A_\mu^{56} + A_\mu^{78} 
+ A_\mu^{9\, 10}) = \sqrt{2/5} A_\mu^{3_R} + \sqrt{3/5} A_\mu^{0_C}$.
Original $ A_\mu^{3_L}$, $C_\mu$, $B_\mu^Y$, and $A_\mu^{3, 11}$ have parity
$(+,+)$, $(+,+)$, $(+,+)$, and $(-,-)$, respectively.  
$C_\mu$ picks up a brane mass $- \delta(y) (5 g^2 w^2/8) C_\mu^2$ so that 
boundary conditions at $z=1$ are 
\beqn
&&\hskip -1.cm
\frac{\dd}{\dd z} A_\mu^{3_L} = 0 ~, \cr
\noalign{\kern 10pt}
&&\hskip -1.cm
\Big( \frac{\dd}{\dd z}  - \frac{5}{2} \omega\Big) C_\mu = 0 ~,  \cr
\noalign{\kern 10pt}
&&\hskip -1.cm
A_\mu^{3, 11} = 0 ~, \cr
\noalign{\kern 10pt}
&&\hskip -1.cm
\frac{\dd}{\dd z} B_\mu^{Y} = 0 ~.
\label{BCgaugeZ1}
\eeqn
In the twisted gauge they become
\beqn
&&\hskip -1.cm
\dd_z \tilde A_\mu^{12} + \dd_z \tilde A_\mu^{34}  \, \cos \theta_H 
- \dd_z \tilde A_\mu^{3,11} \sin \theta_H = 0 ~, \cr
\noalign{\kern 10pt}
&&\hskip -1.cm
(\dd_z - \fivehalves \omega) \Big\{ \tilde A_\mu^{12} -  \tilde A_\mu^{34} \cos \theta_H 
+  \tilde A_\mu^{3,11} \sin \theta_H 
+ \sqrt{3}  \tilde A_\mu^{0_C} \Big\} = 0 ~, \cr
\noalign{\kern 10pt}
&&\hskip -1.cm
\tilde A_\mu^{34} \sin \theta_H +  \tilde A_\mu^{3,11} \cos \theta_H = 0 ~, \cr
\noalign{\kern 10pt}
&&\hskip -1.cm
\dd_z \tilde A_\mu^{12} - \dd_z \tilde A_\mu^{34}  \, \cos \theta_H 
+ \dd_z \tilde A_\mu^{3,11} \sin \theta_H  - \frac{2}{\sqrt{3}} \dd_z \tilde A_\mu^{0_C}= 0 ~ .
\label{BCgaugeZ2}
\eeqn
Adding the first and second equations, one gets
\[
(2 \dd_z - \fivehalves \omega) \tilde A_\mu^{12} 
+ \fivehalves \omega ( \tilde A_\mu^{34} \cos \theta_H - \tilde A_\mu^{3,11} \sin \theta_H )
+ \sqrt{3} ( \dd_z - \fivehalves \omega) \tilde A_\mu^{0_C}= 0 ~ .
\]
Adding the first and fourth equations, one gets
\[
\dd_z \tilde A_\mu^{12}  - \frac{1}{\sqrt{3}} \dd_z \tilde A_\mu^{0_C}= 0 ~.
\]
Writing $[ \tilde A_\mu^{12}, \tilde A_\mu^{34}, \tilde A_\mu^{3,11}, \tilde A_\mu^{0_C}] = 
[ \alpha_{12} C(z) , \alpha_{34} C(z), \alpha_{3,11} S(z),  \alpha_{0_C} C(z) ] a_\mu(x)$,  one finds that
\beqn
&&\hskip -1.cm
K \begin{pmatrix} \alpha_{12} \cr \alpha_{34} \cr \alpha_{3,11} \cr \alpha_{0_C}\end{pmatrix} = 0 ~. \cr
\noalign{\kern 10pt}
&&\hskip -1.cm
K = \begin{pmatrix} C' & \cos \theta_H C' &- \sin \theta_H S'  & 0 \cr
\frac{4}{5} C' - \omega C & \omega \cos\theta_H C & - \omega\sin\theta_H S 
& \sqrt{3} \,(\frac{2}{5} C' - \omega  C) \cr
0 & \sin\theta_H C & \cos\theta_H S & 0 \cr
C' & 0 & 0 & - \frac{1}{\sqrt{3}} C'
\end{pmatrix} .
\label{BCgaugeZ3}
\eeqn

The spectrum is determined by $\det K =0$;
\beeq
C' \big\{ 2 C' (S C' + \lambda \sin^2 \theta_H ) 
- \omega C (5 S C' + 4\lambda \sin^2 \theta_H ) \big\} = 0 ~.
\label{GaugeSpectrumZ1}
\eneq
For sufficiently large $w$, the spectrum  of low-lying KK modes 
is approximately determined by the second term in \eqref{GaugeSpectrumZ1}.  One finds that
\beqn
&\gamma ~\hbox{tower:} &C' (1; \lambda) = 0 ~, \cr
&Z ~\hbox{tower:} & 5 S(1;\lambda) C'(1; \lambda) + 4 \lambda \sin^2 \theta_H = 0 ~, \cr
&Z_R ~\hbox{tower:} & C(1; \lambda) = 0 ~.
\label{GaugeSpectrumZ2}
\eeqn
The mass of $Z$ boson, $m_Z = m_{Z^{(0)}}$,  is given by
\beeq
m_Z \sim \sqrt{\frac{8k}{5L}} \, z_L^{-1} \sin \theta_H = \frac{m_W}{\cos \theta_W} ~, \quad
\sin^2 \theta_W = \frac{3}{8} ~.
\label{GaugeSpectrumZ3}
\eneq

\noindent 
(iii) $\tilde A_\mu^{4, 11}$: $\hat A^{4}$ tower

$A_\mu^{4, 11}$ obeys $(D,D)$ and there is no zero mode.  Its spectrum is determined by
\beeq
\hat A^{4} ~\hbox{tower:} \quad S(1; \lambda) = 0 ~.
\label{GaugeSpectrumHatA}
\eneq

\noindent 
(iv) $SU(3)_C$ gluons 

The boundary condition is $(N,N)$ so that
\beeq
\hbox{gluon tower:} \quad C'(1; \lambda) = 0 ~.
\label{GaugeSpectrumGluon}
\eneq

\noindent 
(v) $X$-gluons

These are six components given by
\beeq
\frac{1}{\sqrt{2}} \begin{pmatrix} A_\mu^{57} - A_\mu^{68} \cr 
\noalign{\kern 5pt} A_\mu^{58} + A_\mu^{67} \end{pmatrix}, ~~
\frac{1}{\sqrt{2}} \begin{pmatrix} A_\mu^{59} - A_\mu^{6\, 10} \cr 
\noalign{\kern 5pt} A_\mu^{5\, 10} + A_\mu^{69} \end{pmatrix}, ~~
\frac{1}{\sqrt{2}} \begin{pmatrix} A_\mu^{79} - A_\mu^{8\, 10} \cr 
\noalign{\kern 5pt} A_\mu^{7\, 10} + A_\mu^{89} \end{pmatrix},
\label{Xgluon1}
\eneq
which originally obey $(N,N)$ boundary conditions.
They have brane masses of the form $- \delta(y) (g^2 w^2/4) A_\mu^2$ in \eqref{branemass1}
so that  boundary conditions at $z=1$ become $(\dd_z - \omega ) A_\mu =0$. 
Consequently the spectrum is determined by $C' - \omega C =0$.  For the low-lying KK modes
\beeq
\hbox{X-gluon tower:} \quad C (1; \lambda) = 0 ~.
\label{GaugeSpectrumXGluon}
\eneq

\noindent 
(vi) $X$-bosons

These are six components given by
\beeq
\frac{1}{\sqrt{2}} \begin{pmatrix} A_\mu^{15} + A_\mu^{26} \cr 
\noalign{\kern 5pt} A_\mu^{16} - A_\mu^{25} \end{pmatrix}, ~~
\frac{1}{\sqrt{2}} \begin{pmatrix} A_\mu^{17} + A_\mu^{28} \cr 
\noalign{\kern 5pt} A_\mu^{18} - A_\mu^{27} \end{pmatrix}, ~~
\frac{1}{\sqrt{2}} \begin{pmatrix} A_\mu^{19} + A_\mu^{2\, 10} \cr 
\noalign{\kern 5pt} A_\mu^{1\, 10} - A_\mu^{29} \end{pmatrix},
\label{Xboson1}
\eneq
which originally obey $(N,D)$ boundary conditions.  There is no brane mass, and
the spectrum is determined by
\beeq
X\hbox{-boson tower:} \quad S' (1; \lambda) = 0 ~.
\label{GaugeSpectrumXboson}
\eneq

\noindent 
(vii) $X'$-bosons

These are six components given by
\beeq
\frac{1}{\sqrt{2}} \begin{pmatrix} A_\mu^{15} - A_\mu^{26} \cr 
\noalign{\kern 5pt} A_\mu^{16} + A_\mu^{25} \end{pmatrix}, ~~
\frac{1}{\sqrt{2}} \begin{pmatrix} A_\mu^{17} - A_\mu^{28} \cr 
\noalign{\kern 5pt} A_\mu^{18} + A_\mu^{27} \end{pmatrix}, ~~
\frac{1}{\sqrt{2}} \begin{pmatrix} A_\mu^{19} - A_\mu^{2\, 10} \cr 
\noalign{\kern 5pt} A_\mu^{1\, 10} + A_\mu^{29} \end{pmatrix},
\label{XPboson1}
\eneq
which originally obey $(N,D)$ boundary conditions.  
They have brane masses of the form $- \delta(y) (g^2 w^2/4) A_\mu^2$ in \eqref{branemass1}
so that  boundary conditions at $z=1$ become $(\dd_z - \omega ) A_\mu =0$. 
Consequently the spectrum is determined by $S' - \omega S =0$.  For the low-lying KK modes
\beeq
X'\hbox{-boson tower:} \quad S (1; \lambda) = 0 ~.
\label{GaugeSpectrumXPboson}
\eneq

\noindent 
(viii) $Y$, $Y'$-bosons

There are three classes;
\beqn
Y: &&
\frac{1}{\sqrt{2}} \begin{pmatrix} A_\mu^{35} - A_\mu^{46} \cr 
\noalign{\kern 5pt} A_\mu^{36} + A_\mu^{45} \end{pmatrix}, ~~
\frac{1}{\sqrt{2}} \begin{pmatrix} A_\mu^{37} - A_\mu^{48} \cr 
\noalign{\kern 5pt} A_\mu^{38} + A_\mu^{47} \end{pmatrix}, ~~
\frac{1}{\sqrt{2}} \begin{pmatrix} A_\mu^{39} - A_\mu^{4\, 10} \cr 
\noalign{\kern 5pt} A_\mu^{3\, 10} + A_\mu^{49} \end{pmatrix}, \cr
\noalign{\kern 10pt}
Y': &&
\frac{1}{\sqrt{2}} \begin{pmatrix} A_\mu^{35} + A_\mu^{46} \cr 
\noalign{\kern 5pt} A_\mu^{36} - A_\mu^{45} \end{pmatrix}, ~~
\frac{1}{\sqrt{2}} \begin{pmatrix} A_\mu^{37} + A_\mu^{48} \cr 
\noalign{\kern 5pt} A_\mu^{38} - A_\mu^{47} \end{pmatrix}, ~~
\frac{1}{\sqrt{2}} \begin{pmatrix} A_\mu^{39} + A_\mu^{4\, 10} \cr 
\noalign{\kern 5pt} A_\mu^{3\, 10} - A_\mu^{49} \end{pmatrix}, \cr
\noalign{\kern 10pt}
\hat Y: &&
A_\mu^{b \,11} \quad (b=5 \sim 10)~.
\label{Yboson1}
\eeqn
Original fields in $Y$, $Y'$, and $\hat Y$ satisfy $(N,D)$, $(N,D)$, and $(D,N)$
boundary conditions.  $A_\mu^{4 b}$ and $A_\mu^{b \, 11}$ ($b=5 \sim 10$)
mix with each other by $\theta_H \not= 0$.    Fields in $Y'$ have brane masses 
of the form $- \delta(y) (g^2 w^2/4) A_\mu^2$ in \eqref{branemass1}.
Boundary conditions at $z=1$ for $(A_\mu^{35}, A_\mu^{46}, A_\mu^{6 \, 11})$,
for instance,  are
\beeq
\dd_z (A_\mu^{35} - A_\mu^{46}) = 0 ~,~~
(\dd_z - \omega ) ( A_\mu^{35} + A_\mu^{46} ) = 0 ~,~~
A_\mu^{6 \, 11} = 0 ~.
\label{BCgaugeY1}
\eneq
Wring $[ \tilde A_\mu^{35}, \tilde A_\mu^{46}, \tilde A_\mu^{11,6}] = 
[ \alpha_{35} S(z;\lambda) , \alpha_{46} S(z;\lambda), \alpha_{11,6} C(z;\lambda) ] a_\mu(x) $,  
one finds that
\beeq
\begin{pmatrix} S' & - \cos \theta_H S' &  \sin \theta_H C' \cr
S' - \omega S &  \cos\theta_H (S' - \omega S) & -\sin\theta_H (C' - \omega C) \cr
0 & \sin\theta_H S & \cos\theta_H C 
\end{pmatrix}
\begin{pmatrix} \alpha_{35} \cr \alpha_{46} \cr \alpha_{11,6} \end{pmatrix} = 0 ~.
\eneq
The spectrum is determined by
\beqn
&&\hskip -1.cm
2 S' ( C S' - \lambda \sin^2 \theta_H ) - \omega S (2C S' - \lambda \sin^2 \theta_H) \cr
\noalign{\kern 5pt}
&&\hskip -1.cm
=2 S' ( S C' + \lambda \cos^2 \theta_H ) 
- \omega S \big\{ 2S C' + \lambda (1 + \cos^2 \theta_H) \big\}= 0 ~.
\label{GaugeSpectrumYboson1}
\eeqn
For the low-lying modes one finds
\beqn
&Y ~\hbox{boson tower:} & 2 C(1;\lambda) S'(1; \lambda) - \lambda \sin^2 \theta_H = 0 ~, \cr
&Y' ~\hbox{boson tower:} & S(1; \lambda) = 0 ~.
\label{GaugeSpectrumYboson2}
\eeqn

\subsection{$A_z$ components}

Similarly one can find the spectrum for $A_z$.  
The evaluation is simpler as $A_z$ does not couple to the brane scalar field $\Phi_{\bf 16}$.

\noindent 
(i) $A_z^{ab}~ (1 \le a < b \le 3)$ and  $ A_z^{jk} ~ (5 \le j < k \le 10)$:  

These components satisfy boundary conditions are  $(D,D)$ so that 
\beeq
C' (1; \lambda) = 0 ~.
\label{GaugeZSpectrum1}
\eneq

\noindent 
(ii) $ A_z^{a4}, A_z^{a\, 11}~ (a= 1\sim 3)$:  

Boundary conditions of $(A_z^{a4}, A_z^{a\, 11})$ are  $(D,D)$ and $(N,N)$, respectively. 
$A_z^{a4}$ and $A_z^{a\, 11}$ mix with each other by $\theta_H$.
Writing $[ \tilde A_z^{a4}, \tilde A_z^{a11}] = 
[ \beta_{a4} C' (z; \lambda) , \beta_{a11} S'(z; \lambda)] a_z(x)$,  one finds that
at $z=1$ 
\beeq
\begin{pmatrix}  \cos \theta_H C' &- \sin \theta_H S' \cr
 \sin\theta_H C &  \cos\theta_H S 
\end{pmatrix}
\begin{pmatrix} \beta_{a4} \cr \beta_{a11} \end{pmatrix} = 0 ~.
\label{GaugeZSpectrum2}
\eneq
The spectrum is determined by
\beeq
 S(1;\lambda) C'(1; \lambda) + \lambda  \sin^2 \theta_H  = 0 ~.
\label{GaugeZSpectrum3}
\eneq

\noindent 
(iii) $A_z^{4,11}$:  Higgs tower

It obeys $(N,N)$ boundary conditions so that 
\beeq
\hbox{Higgs tower:} \quad  S (1; \lambda) = 0 ~.
\label{HiggsSpectrum1}
\eneq
There always is a zero mode, which will acquire a mass at the 1-loop level.

\noindent 
(iv) $A_z^{ak}$$(a = 1\sim 3 ,  k = 5\sim 10)$:  

These components obey $(D,N)$ boundary conditions so that 
\beeq
S' (1; \lambda) = 0 ~.
\label{GaugeZSpectrum4}
\eneq

\noindent 
(v)  $A_z^{k4}$, $A_z^{k11}$ $(k= 5\sim 10)$:  

$A_z^{k4}$ and $A_z^{k11}$ satisfy $(D,N)$ and $(N,D)$ boundary conditions, respectively.  
Writing $[ \tilde A_z^{k4}, \tilde A_z^{k11}] = 
[ \beta_{a4} S' (z; \lambda) , \beta_{a\,11} C'(z; \lambda)] a_z(x) $,  one finds that
\beeq
\begin{pmatrix}  \cos \theta_H S' &- \sin \theta_H C' \cr
\sin\theta_H S &  \cos\theta_H C 
\end{pmatrix}
\begin{pmatrix} \beta_{k4} \cr \beta_{k11} \end{pmatrix} = 0 ~.
\eneq
The spectrum is determined by
\beeq
 S(1;\lambda) C'(1; \lambda) + \lambda  \cos^2 \theta_H  = 0 ~.
\label{GaugeZSpectrum5}
\eneq

\section{Spectrum of fermion fields}

We take Dirac matrices $\gamma^A$ in the spinor representation in \eqref{action1}.
\beeq
\gamma^\mu = \begin{pmatrix} & \sigma^\mu \cr \bar \sigma^\mu & \end{pmatrix} , ~~
\gamma^5 = \begin{pmatrix} I_2 & \cr & - I_2\end{pmatrix}, ~~
\sigma^\mu = ( I_2, \vec \sigma ) , ~~ \bar \sigma^\mu = (- I_2, \vec \sigma ).
\eneq
The fermion  action becomes 
\beqn
&&\hskip -1.cm
\int d^5x \sqrt{-\det G} ~ \bar{\Psi} {\cal D}(c) \Psi   \cr
\noalign{\kern 10pt}
&&\hskip -1.cm
= \int d^4 x \int_1^{z_L} \frac{dz}{k} ~ \overline{\check \Psi} \,  \Bigg[
\begin{matrix} -k ( D_- (c) + ig A_z)  &\sigma^\mu (\dd_\mu - ig A_\mu) \cr 
\noalign{\kern 5pt}
\bar \sigma^\mu (\dd_\mu - ig A_\mu) & - k (D_+ (c) - ig A_z)  \end{matrix}  \Bigg] \, \check \Psi ~, 
\label{Faction1}
\eeqn
where $\check \Psi = z^{-2}  \, \Psi $ and $D_\pm (c) $ is defined in \eqref{BesselF5}.

\subsection{Brane mass terms}

In addition to \eqref{action1}, the fermion fields have brane interactions given by
$S_{\rm brane}$ in \eqref{BraneInt1}.  With $\la \Phi_{\bf 16} \ra \not=0$ in \eqref{vevPhi},
$S_{\rm brane}$ generates fermion mass terms on the Planck brane.
As indicated in \eqref{BraneInt1}, the mass terms have matrix structure in the three generations.
In the present paper we restrict ourselves to the case of diagonal mass matrices, and consider
each generation of quarks and leptons separately.  We shall drop the generation index henceforth.
Each interaction Lagrangian $\cL_j$ ($j= 1 \sim 6$) in \eqref{BraneInt1} generates a brane
mass $\cL_j^m$.
\beqn
&&\hskip -1.cm
S_{\rm brane}^m = \int d^5 x \sqrt{- \det G} \, \delta(y) 
\Big\{ \cL_1^m + \cL_2^m + \cL_3^m + \cL_4^m + \cL_5^m + \cL_6^m  \Big\} ~, \cr
\noalign{\kern 10pt}
&&\hskip -1.cm
\cL_1^m = - 2 \mu_1 \, ( \overline{S}{}_R' \hat \nu_L + \overline{\hat \nu}_L S_R' ) ~, \cr
\noalign{\kern 10pt}
&&\hskip -1.cm
\cL_2^m =  -  2 \mu_2 \, 
( \overline{S}{}_L   \nu_R' + \overline{\nu'}_R   S_L ) ~, \cr
\noalign{\kern 10pt}
&&\hskip -1.cm
\cL_3^m =  - 2 \mu_3 \,
\Big\{ i (\overline{ E}{}_R  e_L - \overline{ e}_L  E_R )
+ i (\overline{ N}{}_R   \nu_L - \overline{\nu}{}_L   N_R )
\cr
\noalign{\kern 10pt}
&&\hskip 0.cm
+ (\overline{\hat D}{}_{1R} \hat d_{1L} + \overline{\hat d}{}_{1L} \hat D_{1R} )
+ (\overline{\hat D}{}_{2R} \hat d_{2L} + \overline{\hat d}{}_{2L} \hat D_{2R} )
+ (\overline{\hat D}{}_{3R} \hat d_{3L} + \overline{\hat d}{}_{3L} \hat D_{3R} ) \Big\} , \cr
\noalign{\kern 10pt}
&&\hskip -1.cm
\cL_4^m =  -2 \mu_4 \,
\Big\{ i (\overline{\hat E}{}_L' \hat e_R' - \overline{\hat e '}{}_R  \hat E_L' )
+ i (\overline{\hat N}{}_L' \hat \nu_R' - \overline{\hat \nu '}{}_R  \hat N_L' )
\cr
\noalign{\kern 10pt}
&&\hskip 0.cm
- (\overline{D}{}_{1L}' d_{1R}' + \overline{d}{}_{1R}' D_{1L}' )
+ (\overline{D}{}_{2L}' d_{2R}' + \overline{d}{}_{2R}' D_{2L}' )
- (\overline{D}{}_{3L}' d_{3R}' + \overline{d}{}_{3R}' D_{3L}' ) \Big\} , \cr
\noalign{\kern 10pt}
&&\hskip -1.cm
\cL_5^m = -  2\mu_5 \,    ( \overline{S}{}_R'   S_L + \overline{S}{}_L S_R' ) ~, \cr
\noalign{\kern 10pt}
&&\hskip -1.cm
\cL_6^m =  -  2\mu_6 \,    \Big\{  \overline{E}{}_L'   E_R + \overline{E}{}_R E_L'  
+ \overline{\hat E}{}_L'   \hat E_R + \overline{\hat E}{}_R \hat E_L' \cr
\noalign{\kern 10pt}
&&\hskip 1.0cm
+ \overline{N}{}_L'   N_R + \overline{N}{}_R N_L'  
+ \overline{\hat N}{}_L'   \hat N_R + \overline{\hat N}{}_R \hat N_L'  \cr
\noalign{\kern 10pt}
&&\hskip 1.0cm
+ \sum_{j=1}^3 \big( \overline{D}{}_{jL}'   D_{jR} + \overline{D}{}_{jR} D_{jL}' 
+ \overline{\hat D}{}_{jL}'   \hat D_{jR} + \overline{\hat D}{}_{jR} \hat D_{jL}' \big)  \Big\} ,
\label{BraneInt2}
\eeqn
where 
\beqn
&&\hskip -1.cm
2\mu_1 = \kappa_{[1,16]} \, w ~, \cr
&&\hskip -1.cm
2\mu_2 = \kappa_{[1,\overline{16}]}  \, w ~, \cr
&&\hskip -1.cm
2\mu_3 = \sqrt{2} \,  \kappa_{[10,16]}  \, w ~, \cr
&&\hskip -1.cm
2\mu_4 = \sqrt{2} \,  \kappa_{[10,\overline{16}]}  \, w ~, \cr
&&\hskip -1.cm
2\mu_5 =   \mu_{[1,1]} ~, \cr
&&\hskip -1.cm
2\mu_6 =   \mu_{[10,10]}  ~.
\label{BraneInt3}
\eeqn
All $\mu_k$'s are taken to be real without loss of generality.  Fermions with $Q_\EM = \pm \twothird$,
$u_j, u_j', \hat u_j, \hat u_j'$, do not appear in the brane masses in \eqref{BraneInt2}. 

\subsection{Quarks and leptons}

To derive the mass spectrum for fermions, we note that the components of $\Psi_{\bf 32}$
in the original and twisted gauges are related by
\beqn
&&\hskip -1.cm
\chi = \begin{pmatrix} \cos \onehalf \theta(z) & -i \sin \onehalf \theta(z) \cr
-i \sin \onehalf \theta(z) & \cos \onehalf \theta(z) \end{pmatrix} \tilde \chi ~, \cr
\noalign{\kern 10pt}
&&\hskip -1.cm
\hat \chi =  \begin{pmatrix} \cos \onehalf \theta(z) & i \sin \onehalf \theta(z) \cr
 i \sin \onehalf \theta(z) & \cos \onehalf \theta(z) \end{pmatrix} \tilde{\hat\chi} ~,
\label{spinor3}
\eeqn
where $\chi$ and $\hat \chi$ are defined in \eqref{spinor2}.  $\theta (z)$ is given in \eqref{twisted1}.
In the original gauge with $\theta_H$, one has
\beqn
&&\hskip -1.cm
g A_z^{cl} = \frac{g}{\sqrt{2}} \, A_z^{(4, 11)} \, T^{4 , 11} = - \theta' (z)  \, T^{4 , 11} ~, \cr
\noalign{\kern 10pt}
&&\hskip -1.cm 
T^{4 , 11} = \begin{cases} \onehalf \tau_1 &{\rm for ~} \chi , \cr
\noalign{\kern 5pt}
- \onehalf \tau_1 &{\rm for ~} \hat \chi . \end{cases}
\label{backgroundAz2}
\eeqn
We denote
\beqn
&&\hskip -1.cm
\hat{\cal D}^{cl} (c) = \begin{pmatrix} - k \hat D_- (c) & \sigma^\mu \dd_\mu \cr
\bar \sigma^\mu \dd_\mu & - k \hat D_+ (c) \end{pmatrix} , \cr
\noalign{\kern 10pt}
&&\hskip -1.cm
\hat D_\pm (c)  
= \pm \Big(\frac{d}{dz} + i\theta '(z) \, T^{4 , 11} \Big) + \frac{c}{z}  ~, \cr
\noalign{\kern 10pt}
&&\hskip -1.cm
{\cal D}_0 (c) = \begin{pmatrix} - k D_- (c) & \sigma^\mu \dd_\mu \cr
\bar \sigma^\mu \dd_\mu & - k D_+ (c) \end{pmatrix} .
\label{spinor4}
\eeqn
To simplify the notation the bulk mass parameters are denoted as 
\beeq
c_0 = c_{\Psi_{\bf 32}} ~, ~~
c_1 = c_{\Psi_{\bf 11}'} ~, ~~
c_2 = c_{\Psi_{\bf 11}} ~.
\label{bulkmass2}
\eneq

\noindent
(i) $Q_\EM = + \frac{2}{3}$ : $u_j, u_j'$

\def\spm{\noalign{\kern 3pt}}
\def\spl{\noalign{\kern 5pt}}
There are no brane mass terms.  The boundary conditions are
$D_+ \check u_{jL} = 0$, $\check u_{jR}=0$, $\check u_{jL}' = 0$, and $D_- \check u_{jR}' = 0$ at $z=1, z_L$.
The equations of motion in the twisted gauge are 
\beqn
&&\hskip -1.cm
-k D_- (c_0) \begin{pmatrix} \tilde{\check u}{}_{jR} \cr \noalign{\kern 5pt} 
\tilde{\check u}{}_{jR}' \end{pmatrix}
 + \sigma^\mu \dd_\mu \begin{pmatrix} \tilde{\check u}{}_{jL} \cr \noalign{\kern 5pt} 
\tilde{\check u}{}_{jL}' \end{pmatrix} = 0~, \cr
\noalign{\kern 10pt}
&&\hskip -1.cm
-k D_+ (c_0) \begin{pmatrix} \tilde{\check u}{}_{jL} \cr \noalign{\kern 5pt} 
\tilde{\check u}{}_{jL}' \end{pmatrix}
 + \bar \sigma^\mu \dd_\mu \begin{pmatrix} \tilde{\check u}{}_{jR} \cr \noalign{\kern 5pt} 
\tilde{\check u}{}_{jR}' \end{pmatrix} = 0~.
\label{eq-up1}
\eeqn
$(\tilde{\check u}_j, \tilde{\check u}_j')$ satisfy the same boundary conditions  at $z=z_L$ as 
$({\check u}_j, {\check u}_j')$ so that one can write, for each mode, as
\beeq
\begin{pmatrix} \tilde{\check{u}}_{jR} \cr \tilde{\check{u}}_{jR}'   \end{pmatrix} =
\begin{pmatrix}
\alpha_u S_R(z; \lambda, c_0) \cr \spm \alpha_{u'} C_R(z; \lambda, c_0)   \end{pmatrix} f_R(x)
~,~~
\begin{pmatrix} \tilde{\check{u}}_{jL} \cr \tilde{\check{u}}_{jL}'  \end{pmatrix} =
\begin{pmatrix}
\alpha_u C_L(z; \lambda, c_0) \cr \spm \alpha_{u'} S_L(z; \lambda, c_0)  \end{pmatrix} f_L(x) ~,
\label{wavef-up1}
\eneq
where $\bar \sigma \dd f_R(x) = k \lambda f_L (x)$ and $\sigma \dd f_L(x) = k \lambda f_R (x)$.
Both right- and left-handed modes have the same coefficients $\alpha_{u}$ and $ \alpha_{u'}$ as 
a result of the equations of motion.  

The boundary conditions at $z=1$ for the right-handed components, $\check u_{jR}=0$
and $D_- \check u_{jR}' = 0$,  become
\beeq
\begin{pmatrix} \cos \onehalf \theta_H S_R^{c_0} & -i \sin \onehalf \theta_H C_R^{c_0} \cr
\noalign{\kern 5pt}
-i \sin \onehalf \theta_H  C_L^{c_0} & \cos \onehalf \theta_H  S_L^{c_0} \end{pmatrix}
\begin{pmatrix} \alpha_u \cr \noalign{\kern 5pt} \alpha_{u'} \end{pmatrix} = 0 
\label{BCup1}
\eneq
so that the spectrum is determined by
\beeq
 S_L^{c_0}  S_R^{c_0}  + \sin^2 \onehalf \theta_H  = 0 ~,
\label{mass-u1}
\eneq
where  $S_L^c=  S_L (1; \lambda , c) $ etc..
The mass of the lowest mode, $m= k\lambda$,  is given by
\beeq
m_u = \begin{cases} \pi^{-1} \sqrt{1 - 4c_0^2} \, \sin \onehalf \theta_H  \, m_\KK &{\rm for~} c_0 < \onehalf , \cr
\noalign{\kern 10pt}
 \pi^{-1} \sqrt{4c_0^2 - 1} \, z_L^{-c_0+0.5} \sin \onehalf \theta_H  \, m_\KK &{\rm for~} c_0 > \onehalf .
 \end{cases}
 \label{mass-u2}
\eneq
$c_0=c_{\Psi_{\bf 32}}$ is determined from the up-type quark mass.
For the top quark $c_0 < \onehalf$, whereas for the charm and up quarks $c_0 > \onehalf$.
Note that
\beeq
\frac{m_t}{m_W} \sim \frac{\sqrt{kL (1 - 4 c_0^2)}}{2 \cos \onehalf \theta_H} ~.
\eneq

\noindent
(ii) $Q_\EM = - \frac{2}{3}$ : $\hat u_j, \hat u_j'$

There are no brane mass terms.  The boundary conditions are
$D_+ \check{\hat u}_{jL} = 0$, $\check{\hat u}_{jR}=0$, $\check{\hat u}_{jL}' = 0$, and $D_- \check{\hat u}_{jR}' = 0$ at $z=1$, and $\check{\hat u}_{jL} = 0$, $D_- \check{\hat u}_{jR}=0$, $D_+ \check{\hat u}_{jL}' = 0$, 
and $\check{\hat u}_{jR}' = 0$ at $z=z_L$.  Wave functions of each mode are given by
\beeq
\begin{pmatrix} \tilde{\check{\hat u}}_{jR} \cr \tilde{\check{\hat u}}_{jR}'   \end{pmatrix} =
\begin{pmatrix}
\alpha_{\hat u} C_R(z; \lambda, c_0) \cr \spm \alpha_{\hat u'} S_R(z; \lambda, c_0)   \end{pmatrix} f_R(x)
~,~~
\begin{pmatrix} \tilde{\check{\hat u}}_{jL} \cr \tilde{\check{\hat u}}_{jL}'  \end{pmatrix} =
\begin{pmatrix}
\alpha_{\hat u} S_L(z; \lambda, c_0) \cr \spm \alpha_{\hat u'} C_L(z; \lambda, c_0)  \end{pmatrix} f_L(x) ~.
\label{wavef-hatup1}
\eneq
Boundary conditions at $z=1$ lead to
\beeq
\begin{pmatrix} \cos \onehalf \theta_H C_R^{c_0} &  i \sin \onehalf \theta_H S_R^{c_0} \cr
\noalign{\kern 5pt}
 i \sin \onehalf \theta_H  S_L^{c_0} & \cos \onehalf \theta_H  C_L^{c_0} \end{pmatrix}
\begin{pmatrix} \alpha_{\hat u} \cr \noalign{\kern 5pt} \alpha_{\hat u'} \end{pmatrix} = 0 
\label{BChatup1}
\eneq
so that the spectrum is determined by
\beeq
 S_L^{c_0}  S_R^{c_0}  + \cos^2 \onehalf \theta_H  = 0 ~.
\label{mass-hatu1}
\eneq
The mass of the lowest mode  is given by
\beeq
m_{\hat u} = \begin{cases} \pi^{-1} \sqrt{1 - 4c_0^2} \, \cos \onehalf \theta_H  \, m_\KK &{\rm for~} c_0 < \onehalf , \cr
\noalign{\kern 10pt}
 \pi^{-1} \sqrt{4c_0^2 - 1} \, z_L^{-c_0+0.5} \cos \onehalf \theta_H  \, m_\KK &{\rm for~} c_0 > \onehalf .
 \end{cases}
\label{mass-hatu2}
\eneq
Note that
\beeq
\frac{m_{\hat u}}{m_u} = \cot \onehalf \theta_H ~.
\label{mass-hatu3}
\eneq

\noindent
(iii) $Q_\EM = - \frac{1}{3}$ : $d_j, d_j', D_j, D_j'$

\ignore{
The relevant part of the action is
\beqn
&&\hskip -1.cm
z^4 {\cal L} \sim 
i ( - {\check{d}}_{jL}^\dagger , {\check{d}}_{jR}^\dagger, 
   - {\check{d}}_{jL}^{\prime \dagger} , {\check{d}}_{jR}^{\prime \dagger} ) \hat{\cal D}^{cl} (c_0)
\begin{pmatrix}  {\check{d}}_{jR} \cr  {\check{d}}_{jL} \cr
     {\check{d}}_{jR}' \cr  {\check{d}}_{jL}' \end{pmatrix}   \cr
     \noalign{\kern 10pt}
&&\hskip -1.cm
+ i ( - \check{D}_{jL}^{\prime\dagger} , \check{D}_{jR}^{\prime\dagger} ) {\cal D}_0 (c_1)
\begin{pmatrix}  \check{D}_{jR}' \cr  \check{D}_{jL}' \end{pmatrix}
+  i ( - \check{D}_{jL}^{\dagger} , \check{D}_{jR}^{\dagger} ) {\cal D}_0 (c_2)
\begin{pmatrix}  \check{D}_{jR} \cr  \check{D}_{jL} \end{pmatrix}\cr
\noalign{\kern 10pt}
&&\hskip -1.cm
- 2 \mu_4 \delta(y) \big\{   -i \check{D}_{jL}^{\prime\dagger} {\check d}_{jR}'  
+ i {\check d}_{jR}^{\prime\dagger}  \check{D}_{jL}' \big\} 
-2 \mu_6 \delta(y) \big\{  -i \check{D}_{jL}^{\prime\dagger}  \check{D}_{jR}
+ i \check{D}_{jR}^{\dagger}  \check{D}_{jL}' \big\} ~.
\label{action-down1}
\eeqn
}

Equations of motion are
\beqn
\begin{matrix} (a) \cr \noalign{\kern 2pt} (b)  \end{matrix}
&&
-k \hat D_-  \begin{pmatrix} \check{d}_{jR} \cr \check{d}_{jR}' \end{pmatrix}
+ \sigma^\mu \dd_\mu \begin{pmatrix} \check{d}_{jL} \cr \check{d}_{jL}' \end{pmatrix} = 0 ~, \cr
\noalign{\kern 10pt}
\begin{matrix} (c) \cr \noalign{\kern 2pt} (d)  \end{matrix}
&&
-k \hat D_+  \begin{pmatrix} \check{d}_{jL} \cr \check{d}_{jL}' \end{pmatrix}
+ \bar \sigma^\mu \dd_\mu \begin{pmatrix} \check{d}_{jR} \cr \check{d}_{jR}' \end{pmatrix} 
= 2 \mu_4 \delta(y) \begin{pmatrix} 0 \cr \check{D}_{jL}'   \end{pmatrix}~, \cr
\noalign{\kern 10pt}
(e) &&
-k  D_-   \check{D}_{jR}'  + \sigma^\mu \dd_\mu  \check{D}_{jL}' 
=  2\mu_4 \delta(y)  \check{d}_{jR}'  +2 \mu_6 \delta(y) \check{D}_{jR} ~, \cr
\noalign{\kern 10pt}
(f) &&
-k  D_+   \check{D}_{jL}'  + \bar \sigma^\mu \dd_\mu  \check{D}_{jR}'  = 0 ~, \cr
\noalign{\kern 10pt}
(g)&&
-k  D_-   \check{D}_{jR}  + \sigma^\mu \dd_\mu  \check{D}_{jL}  =0 ~, \cr
\noalign{\kern 10pt}
(h)&&
-k  D_+   \check{D}_{jL}  + \bar \sigma^\mu \dd_\mu  \check{D}_{jR} 
= 2\mu_6 \delta(y) \check{D}_{jL}' ~.
\label{eq-down1}
\eeqn
Here $D_+$ acting on $\check d_{jL}, \check D_{jL},  \check D_{jL}' $ means
$D_+(c_0)$, $D_+(c_2)$, $D_+(c_1)$, respectively.
Brane interactions affect boundary conditions at $y=0$.
$\check{d}_{jR}, \check{d}_{jL}' , \check{D}_{jR}',  \check{D}_{jL}$ are parity-odd at $y=0$, 
whereas $\check{d}_{jL}, \check{d}_{jR}' , \check{D}_{jL}',  \check{D}_{jR}$ are parity-even.
Recall that $D_\pm (c)$ is parity-odd at $y=0$;
\beeq
D_\pm (c) =  \frac{ e^{-\sigma (y)} }{k} \Big( \pm \frac{d}{dy} + c \sigma '(y) \Big)
= e^{-\sigma(y)}  \Big( \pm \frac{1}{k} \frac{d}{dy} + c \ep(y) \Big) ~.
\eneq
Noting that $ A_z^{(4, 11)}$ is parity-even and  integrating over $y$ from $-\ep$ to $\ep$
in \eqref{eq-down1}, one finds
\beqn
 (a) &\Rightarrow& 
2 \check{d}_{jR} (x, \ep) = 0 ~, \cr
\noalign{\kern 5pt}
 (d) &\Rightarrow& 
-2 \check{d}_{jL}' (x,\ep) = 2 \mu_4 \check{D}_{jL}' (x, 0) ~, \cr
\noalign{\kern 5pt}
(e) &\Rightarrow& 
2 \check{D}_{jR}' (x,\ep) = 2\mu_4 \check{d}_{jR}' (x,0) + 2\mu_6 \check{D}_{jR} (x,0) ~, \cr
\noalign{\kern 5pt}
 (h) &\Rightarrow& 
-2 \check{D}_{jL} (x,\ep) = 2 \mu_6 \check{D}_{jL}' (x,0) ~.
\label{BC-down1}
\eeqn
For parity-even fields we evaluate the equations \eqref{eq-down1} at $y=\ep >0$,
with the help of \eqref{BC-down1}, to find
\beqn
 (c) &\Rightarrow& 
 \hat D_+  \check{d}_{jL} =0 ~, \cr
 \noalign{\kern 5pt}
 (b) &\Rightarrow& 
\hat D_-  \check{d}_{jR}'  +  \mu_4 D_- (c_1) \check{D}_{jR}' = 0 ~, \cr
 \noalign{\kern 5pt}
 (f) &\Rightarrow& 
 D_+  \check{D}_{jL}' -  \mu_4 \hat D_+  \check{d}_{jL}'
 -   \mu_6 D_+  \check{D}_{jL} = 0 ~, \cr
 \noalign{\kern 5pt}
 (g) &\Rightarrow& 
D_-  \check{D}_{jR} + \mu_6 D_-  \check{D}_{jR}' = 0~.
\label{BC-down2}
\eeqn
To summarize, the boundary conditions at $z=1^+$ ($y=\ep$)  are given by
\beeq
\begin{array}{lll}
\text{(right-handed)}  &\hskip .2cm &\text{(left-handed)} \cr
\noalign{\kern 7pt}
\check{d}_{jR}  = 0 ~, & 
& \hat D_+  \check{d}_{jL}  = 0 ~, \cr
\noalign{\kern 7pt}
\hat D_-  \check{d}_{jR}'  +  \mu_4 D_-  \check{D}_{jR}' = 0  ~,  & 
&\check{d}_{jL}' +  \mu_4 \check{D}_{jL}' = 0 ~, \cr
\noalign{\kern 7pt}
\check{D}_{jR}'   -  \mu_4 \check{d}_{jR}'  -  \mu_6 \check{D}_{jR} =0~, & 
& D_+   \check{D}_{jL}'   -  \mu_4 \hat D_+    \check{d}_{jL}' 
-  \mu_6 D_+   \check{D}_{jL} =0 ~, \cr
\noalign{\kern 7pt}
D_-  \check{D}_{jR} + \mu_6 D_-   \check{D}_{jR}' = 0 ~, & 
& \check{D}_{jL} +  \mu_6 \check{D}_{jL}' = 0 ~.
\end{array}
\label{BC-down3}
\eneq

\def\spm{\noalign{\kern 3pt}}
\def\spl{\noalign{\kern 5pt}}
In the twisted gauge all fields obey free equations in the bulk so that
eigenmodes are expressed, with the boundary conditions at the TeV brane
taken into account, as
\beeq
\begin{pmatrix} \tilde{\check{d}}_{jR} \cr \tilde{\check{d}}_{jR}' \cr 
\check{D}_{jR}' \cr \check{D}_{jR} \end{pmatrix} =
\begin{pmatrix}
\alpha_d S_R(z; \lambda, c_0) \cr \spm \alpha_{d'} C_R(z; \lambda, c_0) \cr \spm
\alpha_{D'} S_R(z; \lambda, c_1) \cr \spm \alpha_{D} C_R(z; \lambda, c_2)  \end{pmatrix}
f_R(x)
~,~~
\begin{pmatrix} \tilde{\check{d}}_{jL} \cr \tilde{\check{d}}_{jL}' \cr 
\check{D}_{jL}' \cr \check{D}_{jL} \end{pmatrix} =
\begin{pmatrix}
\alpha_d C_L(z; \lambda, c_0) \cr \spm \alpha_{d'} S_L(z; \lambda, c_0) \cr \spm
\alpha_{D'} C_L(z; \lambda, c_1) \cr \spm \alpha_{D} S_L(z; \lambda, c_2)  \end{pmatrix} 
f_L(x)~.
\label{wavef-down1}
\eneq
The boundary conditions \eqref{BC-down3} for the right-handed components 
are converted to
\beqn
&&\hskip -1.cm
K \begin{pmatrix}
\alpha_d \cr  \alpha_{d'}  \cr \alpha_{D'}  \cr  \alpha_{D}   \end{pmatrix} = 0 ~, \cr
\noalign{\kern 10pt}
&&\hskip -1.cm
K = \begin{pmatrix} 
\cos \frac{\theta_H}{2} S_R^{c_0} & -i \sin \frac{\theta_H}{2} C_R^{c_0} & 0 & 0 \cr \spl
 -i \sin \frac{\theta_H}{2} C_L^{c_0} & \cos \frac{\theta_H}{2} S_L^{c_0} 
 & \mu_4 C_L^{c_1} & 0 \cr \spl
i \mu_4 \sin \frac{\theta_H}{2} S_R^{c_0} & - \mu_4 \cos \frac{\theta_H}{2} C_R^{c_0}
& S_R^{c_1} & -  \mu_6 C_R^{c_2} \cr \spl
0 & 0 & \mu_6 C_L^{c_1} & S_L^{c_2}
\end{pmatrix} ~,
\label{spectrum-down1}
\eeqn
where $S_R^c = S_R(1; \lambda, c)$ etc..
The spectrum $\{ \lambda_n \}$ is determined from $\det K = 0$. 
\beeq
\Big\{ \cos^2  \onehalf \theta_H  S_L^{c_0} S_R^{c_0} 
+ \sin^2 \onehalf \theta_H C_L^{c_0} C_R^{c_0}\Big\} 
\Big\{ S_R^{c_1} S_L^{c_2} +  \mu_6^2 C_L^{c_1} C_R^{c_2} \Big\}
+  \mu_4^2 S_R^{c_0} C_R^{c_0} C_L^{c_1} S_L^{c_2} = 0 ~,
\label{spectrum-down2}
\eneq
or, by making use of $C_L C_R - S_L S_R = 1$ one finds that
\beeq
\Big\{ S_L^{c_0} S_R^{c_0} + \sin^2 \onehalf \theta_H \Big\} 
\Big\{  S_R^{c_1} S_L^{c_2} + \mu_6^2  C_L^{c_1} C_R^{c_2} \Big\}
+  \mu_4^2 S_R^{c_0} C_R^{c_0} C_L^{c_1} S_L^{c_2} = 0 ~.
\label{spectrum-down3}
\eneq
The same result is obtained from the boundary conditions for the left-handed 
components in \eqref{BC-down3}.

For the mode with the lowest mass, the down-type quark, one can suppose that 
$\lambda z_L \ll 1$, $\sin^2 \onehalf \theta_H \gg |S_L^{c_0} S_R^{c_0}|$ and
$\mu_6^2  C_L^{c_1} C_R^{c_2} \gg | S_R^{c_1} S_L^{c_2}|$ so that
\beeq
- S_R^{c_0} S_L^{c_2} \simeq 
\frac{\mu_6^2 C_R^{c_2} }{\mu_4^2 C_R^{c_0}} \,  \sin^2 \onehalf \theta_H ~. 
\label{mass-d1}
\eneq
In the first and second generations $c_j > \onehalf$, whereas
in the third generation $c_j < \onehalf$.   The mass is given by
\beeq
m_d = \begin{cases} \myfrac{1}{\pi} \myfrac{\mu_6}{\mu_4}  \sqrt{(1 - 2c_0)(1+2c_2)} \, 
z_L^{c_0 - c_2} \sin \onehalf \theta_H  \, m_\KK &{\rm for~} c_0, c_2 < \onehalf , \cr
\noalign{\kern 10pt}
\myfrac{1}{\pi} \myfrac{\mu_6}{\mu_4}  \sqrt{(2c_0 -1)(1+2c_2)} \, 
 \, z_L^{-c_2+0.5} \sin \onehalf \theta_H  \, m_\KK &{\rm for~} c_0, c_2  > \onehalf .
 \end{cases}
 \label{mass-d2}
\eneq
In either case one finds that
\beeq
\frac{m_d}{m_u} = \frac{\mu_6}{\mu_4} \sqrt{\frac{1+2c_2}{1+2c_0}} \, z_L^{c_0 - c_2} ~.
\label{mass-d3}
\eneq

\noindent
(iv) $Q_\EM = + \frac{1}{3}$ : $\hat d_j, \hat d_j', \hat D_j, \hat D_j'$

$\hat D_{j R}$ and $\hat D_{j L}'$ have zero modes. 
Equations of motion are
\beqn
&&\hskip -1.cm
-k \hat D_-  \begin{pmatrix} \check{\hat d}_{jR} \cr \check{\hat d}_{jR}' \end{pmatrix}
+ \sigma^\mu \dd_\mu \begin{pmatrix} \check{\hat d}_{jL} \cr \check{\hat d}_{jL}' \end{pmatrix} =
2 \mu_3 \delta(y) \begin{pmatrix}  \check{\hat D}_{jR} \cr 0  \end{pmatrix}~,   \cr
\noalign{\kern 10pt}
&&\hskip -1.cm
-k \hat D_+  \begin{pmatrix} \check{\hat d}_{jL} \cr \check{\hat d}_{jL}' \end{pmatrix}
+ \bar \sigma^\mu \dd_\mu \begin{pmatrix} \check{\hat d}_{jR} \cr \check{\hat d}_{jR}' \end{pmatrix} 
= 0~, \cr
\noalign{\kern 10pt}
&&\hskip -1.cm
-k  D_-  \check{\hat D}_{jR}' + \sigma^\mu \dd_\mu  \check{\hat D}_{jL}'
=  2 \mu_6 \delta(y) \check{\hat D}_{jR}  ~, \cr
\noalign{\kern 10pt}
&&\hskip -1.cm
-k  D_+   \check{\hat D}_{jL}' + \bar \sigma^\mu \dd_\mu  \check{\hat D}_{jR}'  = 0 ~, \cr
\noalign{\kern 10pt}
&&\hskip -1.cm
-k  D_-   \check{\hat D}_{jR}  + \sigma^\mu \dd_\mu  \check{\hat D}_{jL} =0 ~, \cr
\noalign{\kern 10pt}
&&\hskip -1.cm
-k  D_+   \check{\hat D}_{jL}  + \bar \sigma^\mu \dd_\mu  \check{\hat D}_{jR} 
= 2\mu_3 \delta(y)  \check{\hat d}_{jL}  + 2\mu_6 \delta(y) \check{\hat D}_{jL}' ~.
\label{eq-hatdown1}
\eeqn
At the Planck brane ($y=0$), $\hat d_R, \hat d_L', \hat D_R', \hat D_L$ are parity-odd,
whereas $\hat d_L, \hat d_R', \hat D_L', \hat D_R$ are parity-even.
The boundary conditions at $y=\ep$ ($z=1^+$) become
\beeq
\begin{array}{lll}
\text{(right-handed)}  &\hskip .2cm &\text{(left-handed)} \cr
\noalign{\kern 7pt}
\hat D_-  \check{\hat d}_{jR}'  = 0 ~, & 
& \check{\hat d}_{jL}'   = 0 ~, \cr
\noalign{\kern 7pt}
\check{\hat d}_{jR}  -  \mu_3  \check{\hat D}_{jR} = 0  ~,  & 
&\hat D_+ \check{\hat d}_{jL} - \mu_3 D_+ \check{\hat D}_{jL} = 0 ~, \cr
\noalign{\kern 7pt}
D_-  \check{\hat D}_{jR}   +  \mu_3 \hat D_-  \check{\hat d}_{jR}
  +  \mu_6 D_- \check{\hat D}_{jR}' =0~, & 
& \check{\hat D}_{jL}   + \mu_3   \check{\hat d}_{jL}
+  \mu_6  \check{\hat D}_{jL}'  =0 ~, \cr
\noalign{\kern 7pt}
\check{\hat D}_{jR}' - \mu_6 \check{\hat D}_{jR} = 0 ~, & 
& D_+ \check{\hat D}_{jL}' -  \mu_6 D_+ \check{\hat D}_{jL}  = 0 ~.
\end{array}
\label{BC-hatdown1}
\eneq

Eigenmodes are given by
\beeq
\begin{pmatrix} \tilde{\check{\hat d}}_{jR}' \cr  \tilde{\check{\hat d}}_{jR} \cr 
\check{\hat D}_{jR} \cr \check{\hat D}_{jR}'  \end{pmatrix} =
\begin{pmatrix}
 \alpha_{\hat d'} S_R(z; \lambda, c_0) \cr \spm \alpha_{\hat d} C_R(z; \lambda, c_0) \cr \spm
 \alpha_{\hat D} C_R(z; \lambda, c_2) \cr \spm \alpha_{\hat D'} S_R(z; \lambda, c_1)  \end{pmatrix}
f_R(x) ~,~~
\begin{pmatrix} \tilde{\check{\hat d}}_{jL}' \cr  \tilde{\check{\hat d}}_{jL} \cr 
\check{\hat D}_{jL} \cr \check{\hat D}_{jL}'    \end{pmatrix} =
\begin{pmatrix}
\alpha_{\hat d'} C_L(z; \lambda, c_0) \cr \spm \alpha_{\hat d} S_L(z; \lambda, c_0) \cr \spm 
\alpha_{\hat D} S_L(z; \lambda, c_2)  \cr \spm \alpha_{\hat D'} C_L(z; \lambda, c_1)  \end{pmatrix} 
f_L(x)~.
\label{wavef-hatdown1}
\eneq
The boundary conditions \eqref{BC-hatdown1} lead to
\beqn
&&\hskip -1.cm
K \begin{pmatrix}
\alpha_{\hat d'} \cr  \alpha_{\hat d}  \cr \alpha_{\hat D}  \cr  \alpha_{\hat D'}   \end{pmatrix} = 0 ~, \cr
\noalign{\kern 10pt}
&&\hskip -1.cm
K = \begin{pmatrix} 
\cos \frac{\theta_H}{2} C_L^{c_0} &  i \sin \frac{\theta_H}{2} S_L^{c_0} & 0 & 0 \cr \spl
 i \sin \frac{\theta_H}{2} S_R^{c_0} & \cos \frac{\theta_H}{2} C_R^{c_0} 
 & - \mu_3 C_R^{c_2} & 0 \cr \spl
i \mu_3 \sin \frac{\theta_H}{2} C_L^{c_0} &   \mu_3 \cos \frac{\theta_H}{2} S_L^{c_0}
& S_L^{c_2} &    \mu_6 C_L^{c_1} \cr \spl
0 & 0 & - \mu_6 C_R^{c_2} & S_R^{c_1}
\end{pmatrix} ~.
\label{spectrum-hatdown1}
\eeqn
The spectrum is determined by
\beeq
\Big\{ \cos^2  \onehalf \theta_H  C_L^{c_0} C_R^{c_0} 
+ \sin^2 \onehalf \theta_H S_L^{c_0} S_R^{c_0}\Big\} 
\Big\{ S_R^{c_1} S_L^{c_2} +  \mu_6^2 C_L^{c_1} C_R^{c_2} \Big\}
+  \mu_3^2 S_L^{c_0} C_L^{c_0} S_R^{c_1} C_R^{c_2} = 0 ~,
\label{spectrum-hatdown2}
\eneq
or
\beeq
\Big\{   S_L^{c_0} S_R^{c_0} +\cos^2  \onehalf \theta_H  \Big\} 
\Big\{ S_R^{c_1} S_L^{c_2} +  \mu_6^2 C_L^{c_1} C_R^{c_2} \Big\}
+  \mu_3^2 S_L^{c_0} C_L^{c_0} S_R^{c_1} C_R^{c_2} = 0 ~.
\label{spectrum-hatdown3}
\eneq
For the mode with the lowest mass, 
\beeq
- S_L^{c_0} S_R^{c_1} \simeq 
\frac{\mu_6^2 C_L^{c_1} }{\mu_3^2 C_L^{c_0}} \,  \cos^2 \onehalf \theta_H ~. 
\label{mass-hatdown1}
\eneq
The mass is given by
\beeq
m_{\hat d} = \begin{cases} \myfrac{1}{\pi} \myfrac{\mu_6}{\mu_3}  \sqrt{(1 - 2c_1)(1+2c_0)} \, 
z_L^{c_1 - c_0} \cos \onehalf \theta_H  \, m_\KK &{\rm for~} c_0, c_1 < \onehalf , \cr
\noalign{\kern 10pt}
\myfrac{1}{\pi} \myfrac{\mu_6}{\mu_3}  \sqrt{(2c_0 +1)(2c_1-1)} \, 
 \, z_L^{-c_0+0.5} \cos \onehalf \theta_H  \, m_\KK &{\rm for~} c_0, c_1  > \onehalf .
 \end{cases}
 \label{mass-hatdown2}
\eneq
One finds that
\beeq
\frac{m_{\hat d}}{m_u} = \frac{\mu_6}{\mu_3} \sqrt{ \frac{1-2c_1}{1-2c_0} }
\, \cot \onehalf \theta_H 
\times \begin{cases} z_L^{c_1 - c_0} &{\rm for~} c_0, c_1 < \onehalf , \cr
\noalign{\kern 5pt}
1 &{\rm for~} c_0, c_1  > \onehalf .
\end{cases}
\label{mass-hatdown3}
\eneq

\noindent
(v) $Q_\EM = - 1$ : $e, e', E, E'$

\def\sps{\noalign{\kern 2pt}}
\def\spm{\noalign{\kern 3pt}}
\def\spl{\noalign{\kern 5pt}}

The spectrum in the $Q_\EM = - 1$ sector is found in a similar manner. 
Boundary conditions at $y=\ep$ ($z=1^+$)  are given by
\beeq
\begin{array}{lll}
\text{(right-handed)}  &\hskip .2cm &\text{(left-handed)} \cr
\noalign{\kern 7pt}
\hat D_-  \check{e}_{R}' = 0  ~,  & 
&\check{e}_{L}'  = 0 ~, \cr
\noalign{\kern 7pt}
\check{e}_{R} + i \mu_3   \check{E}_{R}  = 0 ~, & 
&\hat D_+  \check{e}_{L}  + i \mu_3  D_+  \check{E}_{L}   = 0 ~, \cr
\noalign{\kern 7pt}
D_-  \check{E}_{R} + i \mu_3 \hat D_-  \check{e}_{R}
+ \mu_6 D_-  \check{E}_{R}' = 0 ~, & 
& \check{E}_{L} + i \mu_3 \check{e}_{L} +  \mu_6 \check{E}_{L}' = 0 ~, \cr
\noalign{\kern 7pt}
\check{E}_{R}'    -  \mu_6 \check{E}_{R} =0~, & 
& D_+  \check{E}_{L}'  -  \mu_6 D_+  \check{E}_{L}  =0 ~,
\end{array}
\label{BC-electron1}
\eneq
and mode functions in the twisted gauge are given by
\beeq
\begin{pmatrix} \tilde{\check{e}}_{R}' \cr \sps \tilde{\check{e}}_{R} \cr \sps
 \check{E}_{R} \cr \sps \check{E}_{R}'  \end{pmatrix} =
\begin{pmatrix}
\alpha_{e'} C_R(z; \lambda, c_0) \cr \spm \alpha_{e} S_R(z; \lambda, c_0) \cr \spm
\alpha_{E} S_R(z; \lambda, c_2) \cr \spm \alpha_{E'} C_R(z; \lambda, c_1)  \end{pmatrix}
f_R(x) ~,~~
\begin{pmatrix} \tilde{\check{e}}_{L}' \cr \sps \tilde{\check{e}}_{L} \cr \sps
 \check{E}_{L} \cr \sps \check{E}_{L}'  \end{pmatrix} =
\begin{pmatrix}
\alpha_{e'} S_L(z; \lambda, c_0) \cr \spm \alpha_{e} C_L(z; \lambda, c_0) \cr \spm
\alpha_{E} C_L(z; \lambda, c_2) \cr \spm \alpha_{E'} S_L(z; \lambda, c_1)  \end{pmatrix} 
f_L(x) ~.
\label{wavef-electron1}
\eneq
The boundary conditions in \eqref{BC-electron1} lead to
\beeq
\begin{pmatrix} 
\cos \frac{\theta_H}{2} S_L^{c_0} & -i \sin \frac{\theta_H}{2} C_L^{c_0} & 0 & 0 \cr \spl
 -i \sin \frac{\theta_H}{2} C_R^{c_0} & \cos \frac{\theta_H}{2} S_R^{c_0} 
 & i \mu_3 S_R^{c_2} & 0 \cr \spl
 \mu_3 \sin \frac{\theta_H}{2} S_L^{c_0} &  i \mu_3 \cos \frac{\theta_H}{2} C_L^{c_0}
& C_L^{c_2} &   \mu_6 S_L^{c_1} \cr \spl
0 & 0 & - \mu_6 S_R^{c_2} & C_R^{c_1}
\end{pmatrix}
\begin{pmatrix}
\alpha_{e'} \cr \spl  \alpha_{e}  \cr \spl  \alpha_{E}  \cr \spl
\alpha_{E'}   \end{pmatrix} = 0 ~.
\eneq
Consequently the spectrum is determined by
\beeq
\Big\{ S_L^{c_0} S_R^{c_0} + \sin^2 \onehalf \theta_H \Big\} 
\Big\{ C_R^{c_1} C_L^{c_2} +  \mu_6^2 S_L^{c_1} S_R^{c_2} \Big\}
+  \mu_3^2 S_L^{c_0} C_L^{c_0} C_R^{c_1} S_R^{c_2} = 0 ~.
\label{spectrum-electron1}
\eneq
For the lowest mode, the electron, 
\beeq
- S_L^{c_0} S_R^{c_2} \simeq \frac{C_L^{c_2}}{\mu_3^2 C_L^{c_0}}  \sin^2 \onehalf \theta_H ~,
\label{mass-electron1}
\eneq
so that
\beeq
m_e = \begin{cases} \myfrac{1}{\pi} \myfrac{1}{\mu_3}  \sqrt{(1 - 2c_2)(1+2c_0)} \, 
z_L^{c_2 - c_0} \sin \onehalf \theta_H  \, m_\KK &{\rm for~} c_0, c_2 < \onehalf , \cr
\noalign{\kern 10pt}
\myfrac{1}{\pi} \myfrac{1}{\mu_3}  \sqrt{(2c_2 -1)(1+2c_0)} \, 
 \, z_L^{-c_0+0.5} \sin \onehalf \theta_H  \, m_\KK &{\rm for~} c_0, c_2  > \onehalf .
 \end{cases}
 \label{mass-electron2}
\eneq
One finds that
\beeq
\frac{m_{e}}{m_u} = \frac{1}{\mu_3} \sqrt{ \frac{1-2c_2}{1-2c_0} }
\times \begin{cases} z_L^{c_2 - c_0} &{\rm for~} c_0, c_2 < \onehalf , \cr
\noalign{\kern 5pt}
1 &{\rm for~} c_0, c_2  > \onehalf .
\end{cases}
\label{mass-electron3}
\eneq

\noindent
(vi) $Q_\EM = + 1$ : $\hat e, \hat e', \hat E, \hat E'$

There are no zero modes.
Boundary conditions at $y=\ep$ ($z=1^+$) for right-handed components are given by
\beqn
&&\hskip -1.cm
\check{\hat e}_{R}  = 0 ~, \cr
\noalign{\kern 5pt}
&&\hskip -1.cm
\hat D_- \check{\hat e}_{R}' - i \mu_4 D_- \check{\hat E}_{R}' = 0 ~, \cr
\noalign{\kern 5pt}
&&\hskip -1.cm
\check{\hat E}_{R}'   - i \mu_4   \check{\hat e}_{R}'
-  \mu_6  \check{\hat E}_{R} =0 ~, \cr
\noalign{\kern 5pt}
&&\hskip -1.cm
D_- \check{\hat E}_{R} +  \mu_6 D_- \check{\hat E}_{R}' = 0~,
\label{BC-hatelectron1}
\eeqn
and wave functions in the twisted gauge are given by
\beeq
\begin{pmatrix} \tilde{\check{\hat e}}_{R} \cr \tilde{\check{\hat e}}_{R}' \cr 
\check{\hat E}_{R}' \cr \check{\hat E}_{R} \end{pmatrix} =
\begin{pmatrix}
\alpha_{\hat e} C_R(z; \lambda, c_0) \cr \spm \alpha_{\hat e'} S_R(z; \lambda, c_0) \cr \spm
\alpha_{\hat E'} C_R(z; \lambda, c_1) \cr \spm \alpha_{\hat E} S_R(z; \lambda, c_2)  \end{pmatrix}
f_R(x)~.
\label{wavef-hatelectron1}
\eneq
Expressions for the left-handed components are obtained by simple replacement which
would be obvious from the cases for $Q_\EM = \pm  \onethird$ etc.
The boundary conditions in \eqref{BC-hatelectron1} are converted to 
\beeq
\begin{pmatrix} 
\cos \frac{\theta_H}{2} C_R^{c_0} &  i \sin \frac{\theta_H}{2} S_R^{c_0} & 0 & 0 \cr \spl
 i \sin \frac{\theta_H}{2} S_L^{c_0} & \cos \frac{\theta_H}{2} C_L^{c_0} 
 & - i \mu_4 S_L^{c_1} & 0 \cr \spl
\mu_4 \sin \frac{\theta_H}{2} C_R^{c_0} &   -i \mu_4 \cos \frac{\theta_H}{2} S_R^{c_0}
& C_R^{c_1} &    -\mu_6 S_R^{c_2} \cr \spl
0 & 0 &   \mu_6 S_L^{c_1} & C_L^{c_2}
\end{pmatrix} 
\begin{pmatrix}   \alpha_{\hat e} \cr  \spm \alpha_{\hat e'}  \cr  \spm 
\alpha_{\hat E'}  \cr \spm  \alpha_{\hat E}   \end{pmatrix} = 0 ~.
\label{spectrum-hatelectron1}
\eneq
The spectrum is determined by
\beeq
\Big\{   S_L^{c_0} S_R^{c_0} +\cos^2  \onehalf \theta_H  \Big\} 
\Big\{ C_R^{c_1} C_L^{c_2} +  \mu_6^2 S_L^{c_1} S_R^{c_2} \Big\}
+  \mu_4^2 S_R^{c_0} C_R^{c_0} S_L^{c_1} C_L^{c_2} = 0 ~.
\label{spectrum-hatelectron2}
\eneq
For the lowest mode 
\beeq
- S_R^{c_0} S_L^{c_1} \simeq \frac{C_R^{c_1}}{\mu_4^2 C_R^{c_0}}  \cos^2 \onehalf \theta_H ~,
\label{mass-hatelectron1}
\eneq
so that
\beeq
m_{\hat e} = \begin{cases} \myfrac{1}{\pi} \myfrac{1}{\mu_4}  \sqrt{(1 - 2c_0)(1+2c_1)} \, 
z_L^{c_0 - c_1} \cos \onehalf \theta_H  \, m_\KK &{\rm for~} c_0, c_1 < \onehalf , \cr
\noalign{\kern 10pt}
\myfrac{1}{\pi} \myfrac{1}{\mu_4}  \sqrt{(2c_0 -1)(1+2c_1)} \, 
 \, z_L^{-c_1+0.5} \cos \onehalf \theta_H  \, m_\KK &{\rm for~} c_0, c_1  > \onehalf .
 \end{cases}
 \label{mass-hatelectron2}
\eneq
One finds that
\beeq
\frac{m_{\hat e}}{m_u} = \frac{1}{\mu_4} \sqrt{ \frac{1+2c_1}{1+2c_0} }
 \, z_L^{c_0 - c_1} \cot \onehalf \theta_H 
\label{mass-hatelectron3}
\eneq
both for $c_0, c_1 < \onehalf$ and for $c_0, c_1 > \onehalf$.

\noindent
(vii) $Q_\EM = 0$ : $\nu, \nu', N, N',  S, S', \hat \nu, \hat \nu', \hat N, \hat N'$

Only $\nu_L, \nu_R'$ have zero modes. 
In general all these ten components mix with each other.
It is convenient to split them into two sets;
\beqn
&&\hskip -1.cm
{\rm Set ~1}: ~ \nu, \nu', N, N',  S, \cr
&&\hskip -1.cm
{\rm Set ~2}: ~ \hat \nu, \hat \nu', \hat N, \hat N', S' .
\nonumber
\eeqn
Boundary conditions at $y=\ep$ ($z=1^+$) become
\beqn
&&\hskip -1.cm
\check{\nu}_{R} + i \mu_3  \check{N}_{R} = 0 ~,   \cr
\noalign{\kern 7pt}
&&\hskip -1.cm
\hat D_- \check{\nu}_{R}' + \mu_2 D_- \check{S}_{R} = 0 ~,  \cr
\noalign{\kern 7pt}
&&\hskip -1.cm
\check{N}_{R}' - \mu_6  \check{N}_{R} = 0 ~,  \cr
\noalign{\kern 7pt}
&&\hskip -1.cm
\check{S}_{R}  - \mu_5 \check{S}_{R}' - \mu_2 \check{\nu}_{R}' = 0 ~,  \cr
\noalign{\kern 7pt}
&&\hskip -1.cm
D_- \check{N}_{R} + i \mu_3 \hat D_- \check{\nu}_{R} + \mu_6 D_- \check{N}_{R}' = 0~,
\label{BC-neutral1}
\eeqn
for Set 1, and 
\beqn
&&\hskip -1.cm
\hat D_- \check{\hat \nu}_{R}' - i  \mu_4 D_- \check{\hat N}_{R}' = 0 ~,  \cr
\noalign{\kern 7pt}
&&\hskip -1.cm
\check{\hat \nu}_{R} - \mu_1  \check{S}_{R}' = 0 ~,   \cr
\noalign{\kern 7pt}
&&\hskip -1.cm
D_- \check{\hat N}_{R} + \mu_6  D_- \check{\hat N}_{R}' = 0 ~,  \cr
\noalign{\kern 7pt}
&&\hskip -1.cm
D_- \check{S}_{R}' +   \mu_5  D_- \check{S}_{R} + \mu_1 \hat D_- \check{\nu}_{R} = 0~,  \cr
\noalign{\kern 7pt}
&&\hskip -1.cm
\check{\hat N}_{R}'  - i \mu_4 \check{\hat \nu}_{R}' - \mu_6 \check{\hat N}_{R} = 0 ~, 
\label{BC-neutral2}
\eeqn
for Set 2.   When $\mu_5=0$, the two sets of boundary conditions decouple from
each other.    We set $\mu_5=0$ in the following analysis.

Wave functions in the twisted gauge are given by
\beeq
\begin{pmatrix} \tilde{\check{\nu}}_{R} \cr \sps \tilde{\check{\nu}}_{R}' \cr \sps
 \check{N}_{R} \cr \sps \check{N}_{R}'   \cr \sps \check{S}_{R} \end{pmatrix} =
\begin{pmatrix}
\alpha_{\nu} S_R(z; \lambda, c_0) \cr \spm \alpha_{\nu'} C_R(z; \lambda, c_0) \cr \spm
\alpha_{N} S_R(z; \lambda, c_2) \cr \spm \alpha_{N'} C_R(z; \lambda, c_1) \cr \spm 
\alpha_{S} C_R(z; \lambda, c_2)  \end{pmatrix}
~,~~
\begin{pmatrix} \tilde{\check{\hat \nu}}_{R}' \cr \sps \tilde{\check{\hat \nu}}_{R} \cr \sps
\check{\hat N}_{R}'  \cr \sps \check{\hat N}_{R} \cr  \sps  \check{S}_{R}' \end{pmatrix} 
 =
\begin{pmatrix}
\alpha_{\hat \nu'} S_R(z; \lambda, c_0) \cr \spm \alpha_{\hat \nu} C_R(z; \lambda, c_0) \cr \spm 
\alpha_{\hat N'} C_R(z; \lambda, c_1) \cr \spm \alpha_{\hat N} S_R(z; \lambda, c_2) \cr \spm 
\alpha_{S'} S_R(z; \lambda, c_1)  \end{pmatrix}
~.
\label{wavef-neutral1}
\eneq
The boundary conditions \eqref{BC-neutral1} and \eqref{BC-neutral2} for $\mu_5=0$  lead to
\beqn
&&\hskip -1.cm
\begin{pmatrix} 
\cos \frac{\theta_H}{2} S_R^{c_0} & -i \sin \frac{\theta_H}{2} C_R^{c_0} & i \mu_3 S_R^{c_2} & 0 & 0 \cr \spl
 -i \sin \frac{\theta_H}{2} C_L^{c_0} & \cos \frac{\theta_H}{2} S_L^{c_0} &0 & 0 &\mu_2 S_L^{c_2} \cr \spl
i \mu_2 \sin \frac{\theta_H}{2} S_R^{c_0} &  - \mu_2 \cos \frac{\theta_H}{2} C_R^{c_0}
& 0 &  0 & C_R^{c_2} \cr \spl
0 & 0 & - \mu_6 S_R^{c_2} & C_R^{c_1} & 0 \cr \spl
i \mu_3 \cos \frac{\theta_H}{2} C_L^{c_0} &  \mu_3 \sin \frac{\theta_H}{2} S_L^{c_0}
& C_L^{c_2} &   \mu_6 S_L^{c_1} & 0 
\end{pmatrix}
 \begin{pmatrix}
\alpha_{\nu} \cr  \alpha_{\nu'}  \cr \alpha_{N}  \cr  \alpha_{N'} \cr \alpha_{S} \end{pmatrix} = 0 , \cr
\noalign{\kern 20pt}
&&\hskip -1.cm
\begin{pmatrix} 
\cos \frac{\theta_H}{2} C_L^{c_0} & i \sin \frac{\theta_H}{2} S_L^{c_0} & -i \mu_4 S_L^{c_1} & 0 & 0 \cr \spl
 i \sin \frac{\theta_H}{2} S_R^{c_0} & \cos \frac{\theta_H}{2} C_R^{c_0} &0 & 0 &-\mu_1 S_R^{c_1} \cr \spl
i \mu_1 \sin \frac{\theta_H}{2} C_L^{c_0} &   \mu_1 \cos \frac{\theta_H}{2} S_L^{c_0}
& 0 &  0 & C_L^{c_1} \cr \spl
0 & 0 &  \mu_6 S_L^{c_1} & C_L^{c_2} & 0 \cr \spl
- i \mu_4 \cos \frac{\theta_H}{2} S_R^{c_0} &  \mu_4 \sin \frac{\theta_H}{2} C_R^{c_0}
& C_R^{c_1} &   -\mu_6 S_R^{c_2} & 0 
\end{pmatrix} 
 \begin{pmatrix}
\alpha_{\hat \nu'} \cr  \alpha_{\hat \nu}  \cr 
\alpha_{\hat N'}  \cr  \alpha_{\hat N} \cr \alpha_{S'} \end{pmatrix} = 0.
\label{spectrum-neutral1}
\eeqn
The spectrum for Set 1 is determined by
\beqn
&&\hskip -1.5cm
\sin^2 \frac{\theta_H}{2} \big\{ C_L^{c_0} C_R^{c_2} + \mu_2^2 S_R^{c_0} S_L^{c_2} \big\} 
\big\{ C_R^{c_0} C_R^{c_1} C_L^{c_2} + \mu_3^2 S_L^{c_0} C_R^{c_1} S_R^{c_2}
+ \mu_6^2 C_R^{c_0} S_L^{c_1} S_R^{c_2} \big\} \cr
\noalign{\kern 5pt}
&&\hskip -1.5cm
+ \cos^2 \frac{\theta_H}{2} \big\{ S_L^{c_0} C_R^{c_2} + \mu_2^2 C_R^{c_0} S_L^{c_2} \big\} 
\big\{ S_R^{c_0} C_R^{c_1} C_L^{c_2} + \mu_3^2 C_L^{c_0} C_R^{c_1} S_R^{c_2}
+ \mu_6^2 S_R^{c_0} S_L^{c_1} S_R^{c_2} \big\} = 0 .
\label{spectrum-neutral2}
\eeqn

As will be seen shortly, $\mu_2$ needs to be very large to have small neutrino masses.
Careful evaluation of each term in \eqref{spectrum-neutral2} is necessary to find
approximate formulas for neutrinos. 
For the lowest mode with $\lambda  z_L \ll 1$, 
\beqn
&&\hskip -1.cm
C_L^{c_0} C_R^{c_2} + \mu_2^2 S_R^{c_0} S_L^{c_2} \simeq z_L^{c_0 - c_2} 
\bigg\{ 1 - \frac{\mu_2^2 (\lambda z_L)^2 z_L^{2(c_2 - c_0)}}{(1-2c_0)(1+2c_2)} \bigg\}, \cr
\noalign{\kern 10pt}
&&\hskip -1.cm
S_L^{c_0} C_R^{c_2} + \mu_2^2 C_R^{c_0} S_L^{c_2} 
\simeq - \lambda z_L^{1 + c_0 - c_2} 
\bigg\{ \frac{1}{1 + 2 c_0} + \frac{\mu_2^2 z_L^{2(c_2 - c_0)}}{1+2c_2} \bigg\}, \cr
\noalign{\kern 10pt}
&&\hskip -1.cm
 S_R^{c_0}  C_L^{c_2} + \mu_3^2 C_L^{c_0}  S_R^{c_2}
\simeq  \lambda z_L^{1 + c_2 - c_0} \bigg\{ \frac{1}{1-2c_0}
+ \frac{\mu_3^2 z_L^{2(c_0 - c_2)}}{1-2c_2} \bigg\} 
\label{spectrum-neu1}
\eeqn
for $c_0, c_2 < \onehalf$,  and 
\beqn
&&\hskip -1.cm
C_L^{c_0} C_R^{c_2} + \mu_2^2 S_R^{c_0} S_L^{c_2} \simeq z_L^{c_0 - c_2} 
\bigg\{ 1 - \frac{\mu_2^2 (\lambda z_L)^2 z_L^{2c_2 - 1}}{(2c_0-1)(1+2c_2)} \bigg\}, \cr
\noalign{\kern 10pt}
&&\hskip -1.cm
S_L^{c_0} C_R^{c_2} + \mu_2^2 C_R^{c_0} S_L^{c_2} 
\simeq - \lambda z_L^{1 + c_0 - c_2} 
\bigg\{ \frac{1}{1 + 2 c_0} + \frac{\mu_2^2 z_L^{2(c_2 - c_0)}}{1+2c_2} \bigg\}, \cr
\noalign{\kern 10pt}
&&\hskip -1.cm
 S_R^{c_0}  C_L^{c_2} + \mu_3^2 C_L^{c_0}  S_R^{c_2}
\simeq  \lambda z_L^{c_0 + c_2 } \bigg\{ \frac{1}{2c_0 - 1}
+ \frac{\mu_3^2 }{2c_2-1} \bigg\} 
\label{spectrum-neu2}
\eeqn
for $c_0, c_2 > \onehalf$.
For neutrinos $\lambda z_L = \pi m_\nu/m_\KK$.  In the third generation,
for which we choose $c_0,  c_2 < \onehalf$, it is found, a posteriori, that
$\mu_2 \lambda z_L^{1+c_2-c_0} \sim (m_\tau /m_t) \sin \onehalf \theta_H \ll 1$,
$\mu_2 z_L^{c_2-c_0} \sim m_\tau /m_{\nu_\tau} \gg 1$, and 
$\mu_3 z_L^{c_0 - c_2} \sim m_t/m_\tau \gg 1$.
In the first and second generations we have $c_0, c_2 > \onehalf$.
It is found, a posteriori, that
$\mu_2 \lambda z_L^{0.5+c_2} \sim (m_e /m_u) \sin \onehalf \theta_H \ll 1$,
$\mu_2 z_L^{c_2-c_0} \sim m_e /m_{\nu_e} \gg 1$, and 
$\mu_3^2  \sim (m_u/m_e)^2 \gg 1$.
Hence, in both cases the mass of the lowest mode, the neutrino, is determined approximately by
\beeq
- S_L^{c_2} S_R^{c_2} \simeq \frac{1}{\mu_2^2 \mu_3^2} \, \sin^2 \onehalf \theta_H ~,
\label{spectrum-neutral3}
\eneq
so that
\beeq
m_{\nu} = \begin{cases} \myfrac{1}{\pi} \myfrac{1}{\mu_2 \mu_3}  \sqrt{1 - 4c_2^2} \, 
\sin \onehalf \theta_H  \, m_\KK &{\rm for~} c_2 < \onehalf , \cr
\noalign{\kern 10pt}
\myfrac{1}{\pi} \myfrac{1}{\mu_2 \mu_3}  \sqrt{4c_2^2 -1} \, 
 \, z_L^{-c_2+0.5} \sin \onehalf \theta_H  \, m_\KK &{\rm for~} c_2  > \onehalf .
 \end{cases}
 \label{mass-neutral1}
\eneq
One finds that
\beeq
\frac{m_{\nu}}{m_u} = \frac{1}{\mu_2 \mu_3} \sqrt{ \frac{1-4c_2^2}{1-4c_0^2} }
\times \begin{cases} 1 &{\rm for~} c_0, c_2 < \onehalf , \cr
\noalign{\kern 5pt}
z_L^{c_0 - c_2} &{\rm for~} c_0, c_2  > \onehalf .
\end{cases}
\label{mass-neutral2}
\eneq
We note that
\beeq
\frac{m_\nu}{m_e} = \frac{1}{\mu_2} \sqrt{\frac{1+ 2c_2}{1+ 2 c_0}} \, z_L^{c_0 - c_2}
\label{mass-neutral3}
\eneq
for $c_0, c_2 < \onehalf $ and for $c_0, c_2 > \onehalf $.

The spectrum for Set 2 is determined by
\beqn
&&\hskip -1.5cm
\sin^2 \frac{\theta_H}{2} \big\{ S_R^{c_0} C_L^{c_1} + \mu_1^2 C_L^{c_0} S_R^{c_1} \big\} 
\big\{ S_L^{c_0} C_L^{c_2} C_R^{c_1} + \mu_4^2 C_R^{c_0} C_L^{c_2} S_L^{c_1}
+ \mu_6^2 S_L^{c_0} S_R^{c_2} S_L^{c_1} \big\} \cr
\noalign{\kern 5pt}
&&\hskip -1.5cm
+ \cos^2 \frac{\theta_H}{2} \big\{ C_R^{c_0} C_L^{c_1} + \mu_1^2 S_L^{c_0} S_R^{c_1} \big\} 
\big\{ C_L^{c_0} C_L^{c_2} C_R^{c_1} + \mu_4^2 S_R^{c_0} C_L^{c_2} S_L^{c_1}
+ \mu_6^2 C_L^{c_0} S_R^{c_2} S_L^{c_1} \big\} = 0 .
\label{spectrum-neutral4}
\eeqn
The mass of the lowest mode is approximately given  by
\beeq
m_{\hat \nu} = \begin{cases} 
\myfrac{\pi^{-1}  \, m_\KK   \cot \onehalf \theta_H}
{ \sqrt{  \Big( \mfrac{1}{1-2c_0} + \mfrac{\mu_1^2 z_L^{2(c_0 - c_1)}}{1-2c_1} \Big)
\Big( \mfrac{1}{1+2c_0} + \mfrac{\mu_4^2 z_L^{2(c_1 - c_0)}}{1+2c_1} \Big) } } 
 &{\rm for~} c_0, c_1 < \onehalf , \cr
\noalign{\kern 10pt}
\myfrac{\pi^{-1}  \, m_\KK   \cot \onehalf \theta_H}
{ \sqrt{  \Big( \mfrac{1}{2c_0 -1} + \mfrac{\mu_1^2 }{2c_1-1} \Big)
\Big( \mfrac{z_L^{2 c_0}}{1+2c_0} + \mfrac{\mu_4^2 z_L^{2c_1}}{1+2c_1} \Big) } } 
&{\rm for~} c_0, c_1  > \onehalf .
 \end{cases}
 \label{mass-neutral4}
\eneq

\subsection{Exotic particles}

In each generation one can reproduce the mass spectrum of quarks and leptons
at the unification scale by adjusting the parameters $c_0, c_1, c_2, \mu_2, \mu_3, \mu_4/\mu_6$
in \eqref{mass-u2}, \eqref{mass-d2}, \eqref{mass-electron2}, and \eqref{mass-neutral1}.
There are more than enough number of parameters.  

However, there also appear
new particles below the KK scale as shown 
in \eqref{mass-hatu2} in the $Q_\EM= - \twothird$ sector, 
in \eqref{mass-hatdown2} in the $Q_\EM= + \onethird$ sector, 
in \eqref{mass-hatelectron2} in the $Q_\EM= + 1$ sector, and
in \eqref{mass-neutral4} in the $Q_\EM= 0$ sector.
In particular, the exotic particle in the $Q_\EM= - \twothird$ sector causes
a severe problem.  As shown in \eqref{mass-hatu3}, the ratio of $m_{\hat u}$ to
$m_u$ is solely determined by $\theta_H$.  Phenomenologically $\theta_H < 0.1$.
It will be seen in the next section that with reasonable parameters it is not possible
to get a minimum of the effective potential $V_\eff (\theta_H)$ at very small $\theta_H$.
It is unavoidable to have unwanted light $\hat u$ particles
in the first and second generation.


\section{Effective potential}

In this section, 
we evaluate the Higgs effective potential $V_{\rm eff}(\theta_H)$ 
by using the mass spectrum formulas of $SO(11)$   gauge bosons and   fermions.
The contributions to the effective potential from the quark-lepton multiplets in the
first and second generations are negligibly small in the RS space, and can be ignored.
In numerical evaluation, we use the  mass parameters and
gauge coupling constants listed in PDG \cite{Agashe:2014kda}.

One-loop effective potential from each KK tower is given by\cite{HOOS2008, Falkowski2007, Hatanaka2012}
\begin{align}
V_{\rm eff}(\theta_H)&=
\pm\frac{1}{2}\int\frac{d^4p}{(2\pi)^4}
\sum_n \mbox{ln}\left(p^2+m_n(\theta_H)^2\right)
=\pm I\left[Q(q);f(\theta_H)\right],
\label{effV1}
\end{align}
where $\{ m_n(\theta_H) \}$  is the mass spectrum of the KK tower and  we take $+$ sign for bosons 
and $-$ sign for fermions.  $ I  [Q;f ]$ is given by
\begin{align}
I [Q(q); f(\theta_H) ]
:=\frac{(kz_L^{-1})^4}{(4\pi)^2}
\int_0^\infty dq \, q^3  \ln  [1+Q(q)f(\theta_H) ],
\label{Eq:I-function-def}
\end{align}
where $Q(q)=\tilde{Q}(iqz_L^{-1})$ when the mass spectrum ($m_n = k \lambda_n$) 
is determined by $\tilde{Q} (\lambda_n) =0$.
For example, when a mass spectrum is determined by the equation
$A(\lambda_n)+ B(\lambda_n) f(\theta_H)=0$, 
we   rewrite the equation as  $1+\tilde{Q}(\lambda_n)f(\theta_H)=0$  where
$\tilde{Q}(\lambda)={A(\lambda)}/{B(\lambda)}$.
The first and second derivatives of
$I\left[Q(q);f(\theta_H)\right]$ with respect to $\theta_H$ are given by
\begin{align}
\frac{\partial  I [Q(q);f(\theta_H) ] }{\partial\theta_H}
&=\frac{(kz_L^{-1})^4}{(4\pi)^2}
\int_0^\infty dq \, q^3
\frac{Q(q)   f^{(1)} (\theta_H)}{1+Q(q)f(\theta_H)} ~, \cr
\noalign{\kern 10pt}
\frac{\partial^2 I [Q(q);f(\theta_H) ]}{\partial\theta_H^2}
& =\frac{(kz_L^{-1})^4}{(4\pi)^2}
\int_0^\infty dq \, q^3
\bigg[
\frac{Q(q) f^{(2)} (\theta_H)}{1+Q(q)f(\theta_H)}
-\Big( \frac{Q(q)  f^{(1)} (\theta_H)}{1+Q(q)f(\theta_H)}\Big)^2
\bigg] ,
\label{I-func2}
\end{align}
where $f^{(n)}(\theta_H):=\partial^nf(\theta_H)/\partial\theta_H^n$.

The evaluation of the total effective potential 
$V_{\rm eff}(\theta_H) = V_{\rm eff}^{\rm gauge}(\theta_H)
+V_{\rm eff}^{\rm fermion}(\theta_H)$ is straightforward.
We are interested in the $\theta_H$-dependent part of $V_{\rm eff}$,
to which only KK towers with $\theta_H$-dependent spectra contribute.
$V_{\rm eff}^{\rm gauge} (\theta_H)$ in the $\xi=1$ gauge is decomposed as
\beeq
V_{\rm eff}^{\rm gauge}(\theta_H)
=V_{\rm eff}^{W^{\pm}} +V_{\rm eff}^{Z} +V_{\rm eff}^{Y}
+V_{\rm eff}^{A_z^{a4},A_z^{a\, 11}} +V_{\rm eff}^{A_z^{k4},A_z^{k\, 11}} ~.
\label{effVgauge1}
\eneq
The equations determining the spectra are given by
\eqref{GaugeSpectrumW2} for the $W^\pm$ tower, 
\eqref{GaugeSpectrumZ2} for the $Z$ tower,
\eqref{GaugeSpectrumYboson2} for the $Y$ boson tower,
\eqref{GaugeZSpectrum5} for the $A_z^{a4},A_z^{a\, 11}$ $(a=1,2,3)$ towers, and 
\eqref{GaugeZSpectrum3} for the $A_z^{k4},A_z^{k11}$ $(k=5,...,10)$ towers.
It has been confirmed that the use of the approximate formula \eqref{GaugeSpectrumW2}
in place of the exact formula \eqref{GaugeSpectrumW1}, for instance, is numerically
justified.  One finds that 
\begin{align}
V_{\rm eff}^{W^{\pm}}(\theta_H)
&=4
I \left[\onehalf Q_{0}\left(q, \onehalf \right);\sin^2\theta_H\right],  \cr
\noalign{\kern 5pt}
V_{\rm eff}^{Z}(\theta_H)
&=2
I\left[\fourfifths Q_{0}\left(q,\onehalf \right);\sin^2\theta_H\right],\cr
\noalign{\kern 5pt}
V_{\rm eff}^{Y}(\theta_H)
&=12
I\left[\onehalf Q_{0}\left(q,\onehalf \right);1+\cos^2\theta_H\right], \cr
\noalign{\kern 5pt}
V_{\rm eff}^{A_z^{a4},A_z^{a\, 11}}(\theta_H)
&=3
I\left[Q_{0}\left(q,\onehalf \right);\sin^2\theta_H\right], \cr
\noalign{\kern 5pt}
V_{\rm eff}^{A_z^{k4},A_z^{k\, 11}}(\theta_H)
&=6
I\left[Q_{0}\left(q,\onehalf \right); \cos^2\theta_H\right] .
\label{effVgauge2}
\end{align}
Here
\begin{align}
&Q_0\left(q,c\right) =\frac{z_L}{q^2}
\frac{1}{\hat{F}_{c}^{++}(q)\hat{F}_{c}^{--}(q)} ~,  \cr
\noalign{\kern 5pt}
&\hat{F}_{c}^{\pm\pm}(q) =
\hat{F}_{c\pm\frac{1}{2},c\pm\frac{1}{2}}(qz_L^{-1},q) ~,  \cr
\noalign{\kern 5pt}
&
\hat{F}_{\alpha,\beta}(u,v) =
I_\alpha(u)K_\beta(v)
-e^{-i(\alpha-\beta)\pi}K_\alpha(u)I_\beta(v),
\label{Fhat-function1}
\end{align}
where $I_\alpha(u)$ and $K_\alpha(u)$ are modified Bessel functions.

The fermion part $V_{\rm eff}^{\rm fermion}(\theta_H)$ is evaluated in a similar manner.
Following the classification based on $Q_\EM$  in the previous section, we decompose 
$V_{\rm eff}^{\rm fermion}$ into eight parts;
\begin{align}
&V_{\rm eff}^{\rm fermion}(\theta_H;c_0,c_1,c_2; \mu_k) \cr
\noalign{\kern 5pt}
&= V_{\rm eff}^{\rm (i)} +  V_{\rm eff}^{\rm (ii)} + V_{\rm eff}^{\rm (iii)} + 
V_{\rm eff}^{\rm (iv)} + V_{\rm eff}^{\rm (v)} +  V_{\rm eff}^{\rm (vi)} + 
V_{\rm eff}^{\rm (vii-1)} + V_{\rm eff}^{\rm (vii-2)} ~.
\label{effVf1}
\end{align}
$V_{\rm eff}^{\rm fermion}$ depends on the three bulk mass parameters ($c_j$)
and brane interaction mass parameters ($\mu_k$) in the third generation.
We set $\mu_5=0$ as before.
The equations determining the mass spectra are
(i) \eqref{mass-u1} for the $Q_\EM = + \twothird$ ($u$-type) quarks, 
(ii) \eqref{mass-hatu1} for the $Q_\EM = - \twothird$ ($\hat u$-type) quarks, 
(iii) \eqref{spectrum-down3} for the $Q_\EM = - \onethird$ ($d$-type) quarks,
(iv) \eqref{spectrum-hatdown3} for the $Q_\EM =+ \onethird$ ($\hat d$-type) quarks,
(v) \eqref{spectrum-electron1} for the $Q_\EM = - 1$ ($e$-type) leptons, 
(vi) \eqref{spectrum-hatelectron2} for the $Q_\EM = + 1$ ($\hat e$-type) leptons, 
(vii-1) \eqref{spectrum-neutral2} for the $Q_\EM = 0$ ($\nu$-type) leptons, and
(vii-2) \eqref{spectrum-neutral4} for the $Q_\EM =  0$ ($\hat \nu$-type) leptons.
$V_{\rm eff}^{\rm F(i)}$, $V_{\rm eff}^{\rm F(ii)}$, etc. are given by
\begin{align}
V_{\rm eff}^{\rm (i)}(\theta_H)
&=-4I \left[Q_0\left(q,c_0\right);\sin^2  \onehalf  \theta_H \right],\cr
\noalign{\kern 5pt}
V_{\rm eff}^{\rm (ii)}(\theta_H)
&=-4I \left[Q_0\left(q,c_0\right);\cos^2\onehalf \theta_H \right], \cr
\noalign{\kern 5pt}
V_{\rm eff}^{\rm (iii)}(\theta_H)
&=-4I\left[Q_{\rm (iii)}(q,c_0,c_1,c_2, \mu_4, \mu_6); \sin^2 \onehalf  \theta_H \right],\cr
\noalign{\kern 5pt}
V_{\rm eff}^{\rm (iv)}(\theta_H)
&=-4I\left[Q_{\rm (iv)}(q,c_0,c_1,c_2, \mu_3, \mu_6 );\cos^2  \onehalf \theta_H\right], \cr
\noalign{\kern 5pt}
V_{\rm eff}^{\rm (v)}(\theta_H)
&=-4I\left[Q_{\rm (v)}(q,c_0,c_1,c_2, \mu_3, \mu_6);\sin^2\onehalf \theta_H\right],\cr
\noalign{\kern 5pt}
V_{\rm eff}^{\rm (vi)}(\theta_H)
&=-4I\left[Q_{\rm (v)}(q,c_0,c_1,c_2, \mu_4, \mu_6);\cos^2\onehalf \theta_H \right], \cr
\noalign{\kern 5pt}
V_{\rm eff}^{\rm (vii-1)}(\theta_H)
&=-4I\left[Q_{\rm (vii-1)}(q,c_0,c_1,c_2, \mu_2, \mu_3, \mu_6);\cos^2\onehalf \theta_H\right],\cr
\noalign{\kern 5pt}
V_{\rm eff}^{\rm (vii-2)}(\theta_H)
&=-4I\left[Q_{\rm (vii-2)}(q,c_0,c_1,c_2, \mu_1, \mu_4, \mu_6);\sin^2 \onehalf \theta_H\right],
\label{effVf-2}
\end{align}
where 
\begin{align}
Q_{\rm (iii)}(q) =&
\frac{z_L}{q^2}  \bigg\{ \hat{F}_{c_0}^{++}\hat{F}_{c_0}^{--}
+\frac{\mu_4^2 \hat{F}_{c_2}^{++} \hat{F}_{c_0}^{--} \hat{F}_{c_1}^{+-} \hat{F}_{c_0}^{-+}}
{ \hat{F}_{c_2}^{++} \hat{F}_{c_1}^{--} +\mu_6^2 \hat{F}_{c_1}^{+-} \hat{F}_{c_2}^{-+}}
\bigg\}^{-1},  \cr
\noalign{\kern 10pt}
Q_{\rm (iv)}(q)=&
\frac{z_L}{q^2}
\bigg\{ \hat{F}_{c_0}^{++} \hat{F}_{c_0}^{--} 
+\frac{\mu_3^2 \hat{F}_{c_0}^{++} \hat{F}_{c_1}^{--} \hat{F}_{c_0}^{+-} \hat{F}_{c_2}^{-+}}
{\hat{F}_{c_2}^{++} \hat{F}_{c_1}^{--} +\mu_6^2 \hat{F}_{c_1}^{+-} \hat{F}_{c_2}^{-+}}
\bigg\}^{-1},  \cr
\noalign{\kern 10pt}
Q_{\rm (v)}(q)=&
\frac{z_L}{q^2}
\bigg\{
\hat{F}_{c_0}^{++} \hat{F}_{c_0}^{--}
+\frac{\mu_3^2 \hat{F}_{c_0}^{++} \hat{F}_{c_2}^{--} \hat{F}_{c_0}^{+-} \hat{F}_{c_1}^{-+}}
{\mu_6^2 \hat{F}_{c_1}^{++} \hat{F}_{c_2}^{--} + \hat{F}_{c_2}^{+-} \hat{F}_{c_1}^{-+}}
\bigg\}^{-1}, \cr
\noalign{\kern 10pt}
Q_{\rm (vi)}(q) =&
\frac{z_L}{q^2}
\bigg\{
\hat{F}_{c_0}^{++}
\hat{F}_{c_0}^{--}
+\frac{\mu_4^2
\hat{F}_{c_1}^{++}
\hat{F}_{c_0}^{--}
\hat{F}_{c_2}^{+-}
\hat{F}_{c_0}^{-+}}
{\mu_6^2
\hat{F}_{c_1}^{++}
\hat{F}_{c_2}^{--}
+
\hat{F}_{c_2}^{+-}
\hat{F}_{c_1}^{-+}}
\bigg\}^{-1}, \cr
\noalign{\kern 10pt}
Q_{\rm (vii-1)}(q) =&
-\frac{z_L}{q^2}
\frac{1}
{\mu_2^2
\hat{F}_{c_2}^{++}
\hat{F}_{c_0}^{--}
+
\hat{F}_{c_0}^{+-}
\hat{F}_{c_2}^{-+}} \cr
\noalign{\kern 5pt}
&
\times
\frac{
\left\{ (z_L/q^2) +(1-\mu_2^2\mu_3^2)
\hat{F}_{c_2}^{++}
\hat{F}_{c_2}^{--}
\right\}
\hat{F}_{c_1}^{-+}
+\mu_6^2
\hat{F}_{c_1}^{++}
\hat{F}_{c_2}^{--}
\hat{F}_{c_2}^{-+}}
{
\hat{F}_{c_2}^{+-}
\hat{F}_{c_0}^{-+}
\hat{F}_{c_1}^{-+}
+\mu_3^2
\hat{F}_{c_0}^{++}
\hat{F}_{c_2}^{--}
\hat{F}_{c_1}^{-+}
+\mu_6^2
\hat{F}_{c_1}^{++}
\hat{F}_{c_2}^{--}
\hat{F}_{c_0}^{-+} },  \cr
\noalign{\kern 10pt}
Q_{\rm (vii-2)}(q) =&
-\frac{z_L}{q^2}
\frac{1}
{\mu_1^2
\hat{F}_{c_0}^{++}
\hat{F}_{c_1}^{--}
+
\hat{F}_{c_1}^{+-}
\hat{F}_{c_0}^{-+}} \cr
\noalign{\kern 5pt}
&
\times
\frac{
\left\{(z_L/q^2)+(1-\mu_1^2\mu_4^2)
\hat{F}_{c_1}^{++}
\hat{F}_{c_1}^{--}
\right\}
\hat{F}_{c_2}^{+-}
+\mu_6^2
\hat{F}_{c_1}^{++}
\hat{F}_{c_2}^{--}
\hat{F}_{c_1}^{+-}}
{
\hat{F}_{c_0}^{+-}
\hat{F}_{c_2}^{+-}
\hat{F}_{c_1}^{-+}
+\mu_4^2
\hat{F}_{c_1}^{++}
\hat{F}_{c_0}^{--}
\hat{F}_{c_2}^{+-}
+\mu_6^2
\hat{F}_{c_1}^{++}
\hat{F}_{c_2}^{--}
\hat{F}_{c_0}^{+-}},
\label{effVf3}
\end{align}
and   $\hat{F}_{c}^{\pm\pm} = \hat{F}_{c}^{\pm\pm}(q)$.

The Higgs mass $m_H(\theta_H=\theta_H^{\rm min})$ is determined by 
\begin{align}
m_H^2(\theta_H)=\left.\frac{1}{f_H^2}
\frac{d^2V_{\rm eff}(\theta_H)}{d\theta_H^2}\right|_{\theta_H=\theta_1},
\label{Eq:Higgs_mass}
\end{align}
where $f_H$ is given by \eqref{AB1}, or by
\begin{align}
f_H=  \sqrt{\frac{\sin^2\theta_W}{\pi\alpha_{em}}
\frac{k^2}{(z_L^2-1)\mbox{log}(z_L)}}.
\end{align}
Here $\alpha_{em}$ is the fine-structure constant, and 
$\theta_W$ is the Weinberg angle.

In the following we give
example calculations for the effective potential and show
a result for the Higgs mass. 
As we remarked before, the current $SO(11)$ model necessarily contains 
light exotic particles, and therefore is not completely realistic.
With this in mind,  we do not insist on reproducing all of the observed values of the masses 
of the SM gauge bosons,   Higgs boson, quarks and leptons. 
Further  the GUT relation leads to  $\sin^2\theta_W = \frac{3}{8}$. 
The RGE effect must be taken into account to compare it with  the observed value at low energies.

From the mass relations  \eqref{mass-d3}, \eqref{mass-electron3}, and \eqref{mass-neutral3}
applied to the third generation, one finds the following constraints for the brane mass parameters
$(\mu_2,\mu_3,\mu_4,\mu_6)$;
\begin{align}
&\mu_2\simeq
\sqrt{\frac{1+ 2c_2}{1+ 2 c_0}} \, z_L^{c_0-c_2} \, \frac{m_\tau}{m_{\nu_\tau}}~, \cr
\noalign{\kern 10pt}
&\mu_3\simeq
\sqrt{\frac{1- 2c_2}{1- 2 c_0}} \, z_L^{c_2 - c_0} \, \frac{m_t}{m_\tau} ~, \cr
\noalign{\kern 10pt}
&\frac{\mu_6}{\mu_4}\simeq 
\sqrt{\frac{1+2c_0}{1+2c_2}} \, z_L^{c_2 - c_0} \, \frac{m_b}{m_t} ~.
\label{Brane-bulk-parameter-relation}
\end{align}
For $c_0=c_2$, these constraints lead typically to 
$\mu_2>O(10^{10})$, $\mu_3\sim 100$, $\mu_4\simeq 40\mu_6$ and $\mu_6=O(1)$. 
No constraint appears for $\mu_1$.
It should be noted that the brane parameters $\mu_k$ 
sensitively depend on the balk mass parameters $c_0$ and $c_2$, as demonstrated below.

\ignore{
\begin{align}
\mu_1=O(1),\ \ 
\mu_2>O(10^{10}),\ \
\mu_3\sim 100,\ \
\mu_4\simeq 40\mu_6,\ \
\mu_6=O(1),
\label{Eq:brane-mass-parameter-set}
\end{align}
}

To find a consistent set of the parameters we take  the following procedure.
\begin{enumerate}
\item[0.] We fix $z_L=e^{kL}$ and pick  $\theta_H=\theta_H^{\rm min}$.
\item We suppose that the minimum of $V_{\rm eff}(\theta_H)$ is
      located at $\theta_H=\theta_H^{\rm min}$.
      The equation (\ref{GaugeSpectrumZ2}) determine the  spectrum 
      $\{ \lambda_{Z^{(n)}}  \}$ of the $Z$ tower.  
      By using the zero mode mass $m_{Z^{(0)}}$ and the observed $Z$
      boson mass $m_Z^{\rm obs}$,   
      $k=m_{Z^{(0)}}/\lambda_{Z^{(0)}} = m_Z^{\rm obs}/\lambda_{Z^{(0)}}$ 
      is determined.
      \ignore{Its approximate $Z$ boson mass (\ref{GaugeSpectrumZ3})
      and $\sin^2\theta_W$
      determine $\lambda_{Z^{(0)}}$, which fixes $k$ by the $Z$ boson
      mass $m_Z=\lambda_{Z^{(0)}}k$.}
      The KK mass scale is also determined by 
      $m_{KK} = \pi k /(z_L - 1) \simeq \pi k z_L^{-1}$.
\item The observed top quark mass $m_t^{\rm obs}$ determines
      the bulk mass parameter of the third generation $SO(11)$ spinor
      fermion $c_0$ through \eqref{mass-u2}.
\item We choose a sample value of the bulk mass parameter of $SO(11)$
      vector fermion $c_2$.   Then the brane mass parameters $\mu_2$, $\mu_3$
      and $\mu_6/\mu_4$ are determined by \eqref{Brane-bulk-parameter-relation}.
      For $\mu_1$ and $\mu_4$ we have taken sample values $\mu_1=1$ and $\mu_4=0.5$.
\item At this stage $V_\eff (\theta_H)$ is determined, once 
the bulk mass parameters of $SO(11)$ vector fermions $c_1$ is given.
$c_1$ is fixed by demanding that $V_\eff (\theta_H)$ has the global minimum
at $\theta_H=\theta_H^{\rm min}$.
\begin{align}
0=\left.
\frac{d}{d\theta_H} V_{\rm eff}(\theta_H)
\right|_{\theta_H=\theta_H^{\rm min}}.
\end{align}
\item By using the above values of the bulk and brane parameters, we
      obtain the Higgs boson mass $m_H$ by using (\ref{Eq:Higgs_mass}).
\end{enumerate}
Depending on the initial values of $c_2$, $\mu_1$ and $\mu_4$, one may not
find a consistent solution at the step 4.  
In particular, we could not find consistent solutions with $\theta_H^{\rm min} \sim 0.1$
for $c_0=c_2$.  
Judicious choice of appropriate values for $c_2$, $\mu_1$ and $\mu_4$ is necessary.

We give a sample calculation for the effective potential 
$V_{\rm eff}(\theta_H)$ and $m_H(\theta_H)$.
Let us take, as a set of the input parameters,    $z_L=10^{10}$, $\theta_H=0.10$,
$\alpha_{em}^{-1}(m_Z)=127.916\pm 0.015$,
$\sin^2\theta_W(m_Z)=0.23116\pm 0.00013$,
$m_t=173.1\pm 1.22\,$GeV, $m_b=4.18\,$GeV, $m_\tau=1.776\,$GeV, and 
$m_{\nu_\tau}=0.1\,$eV.
For $m_b$ and $m_\tau$, we use the central values.
The value of $m_{\nu_\tau}$ is a reference value for our calculation.
$m_{KK}$, $k$, and $f_H$ are determined to be
\begin{align}
m_{KK}=1.088\times 10^4\ \mbox{GeV},\ \ 
k=3.464\times 10^{13}\ \mbox{GeV},\ \
f_H=2216\ \mbox{GeV}.
\label{mKK2}
\end{align}
For $m_t=165.0, 170.0, 175.0\,$GeV, consistent sets of 
the bulk and brane parameters are tabulated in Table \ref{parameter-result1}.
The Higgs boson mass $m_H$ is the output.  In the current model 
it comes out in the range $50\,{\rm GeV} < m_H < 55\,{\rm GeV}$, smaller than 
the observed value $m_H \simeq 125\,$GeV.
Even if one takes slightly different values for $c_2$, $\mu_1$ and $\mu_4$,
the value of $m_H$ does not change very much.

\begin{table}
\begin{center}
\renewcommand{\arraystretch}{1.2}
\begin{tabular}{|c|ccc|ccc|c|}
\hline
Top quark&\multicolumn{3}{c|}{Bulk parameters}&\multicolumn{3}{c|}{Brane parameters}&Higgs\\
$m_t$[GeV]&$c_0$&$c_1$&$c_2$ &$\mu_2$&$\mu_3$ &$\mu_6$&$m_H$[GeV]\\
\hline
165.0&0.3696&0.4286&0.2970 &9.05$\times 10^{10}$&21.8 &0.00249&50.96\\
170.0&0.3559&0.4293&0.3120 &5.20$\times 10^{10}$&36.8 &0.00420&51.77\\
175.0&0.3496&0.4286&0.3270 &2.95$\times 10^{10}$&62.8 &0.00719&53.52\\
\hline
\end{tabular}
\end{center}
\caption{Consistent parameter sets for $\theta_H=0.10$, $z_L=10^{10}$, 
$(\mu_1, \mu_4)=(1.0, 0.50)$ for various values of $m_t$.
The Higgs boson mass $m_H$ is the output.}
\label{parameter-result1}
\end{table}

\begin{figure}[tbh]
\begin{center}
\includegraphics[bb=0 0 427 305, height=6.2cm]{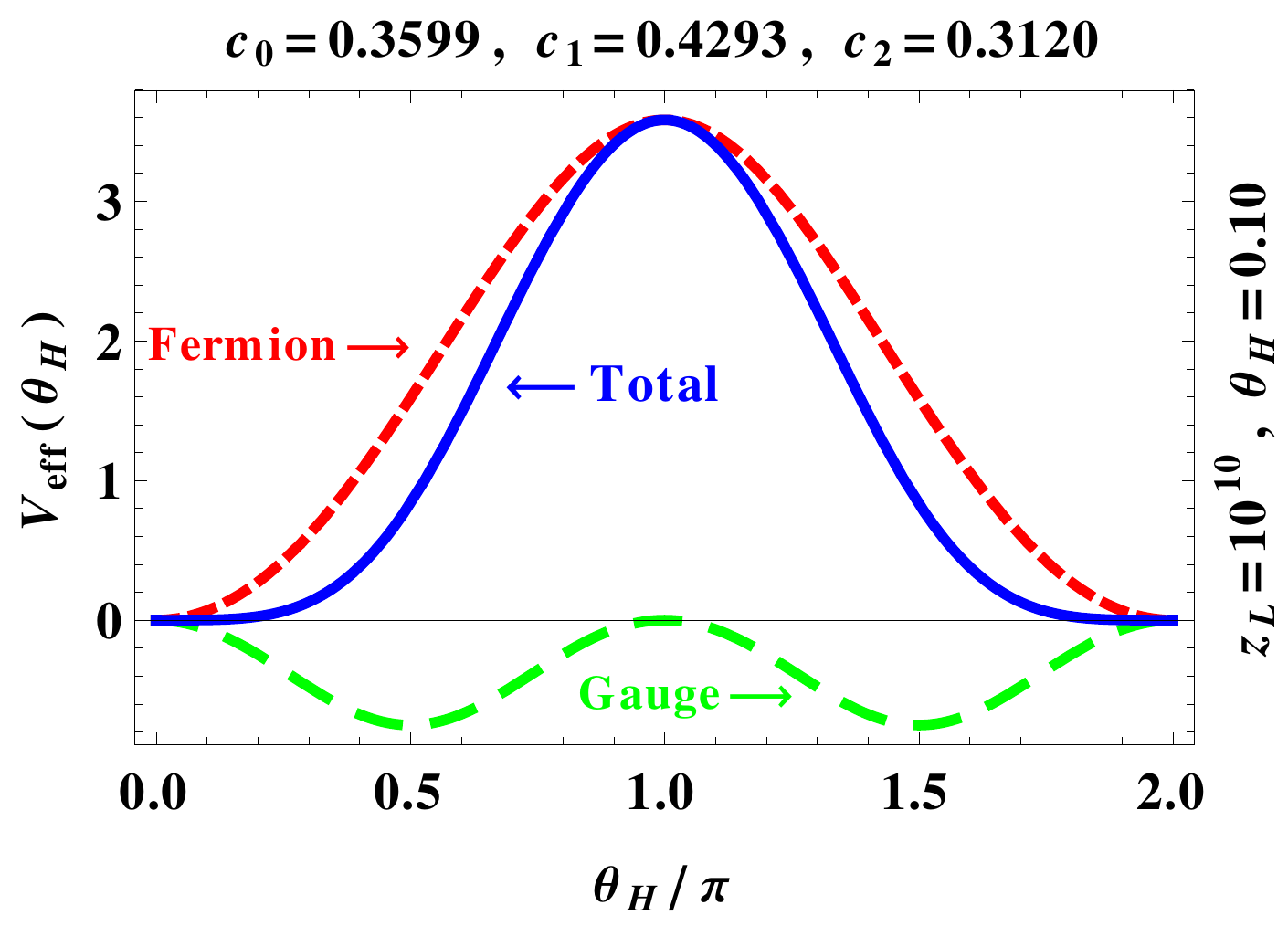}
\includegraphics[bb=0 0 427 266, height=5.4cm]{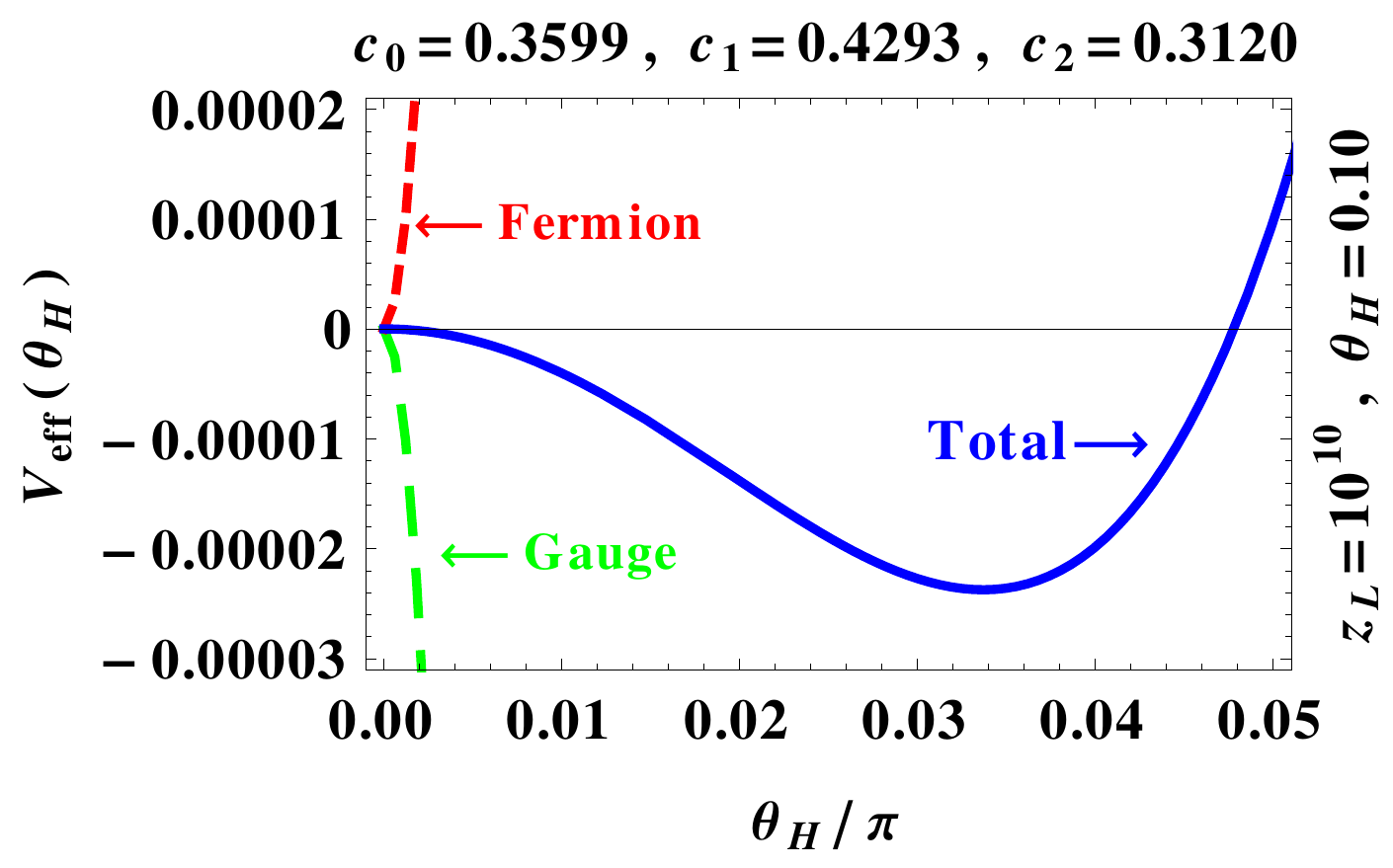}
\end{center}
\caption{The effective potential $V_{\rm eff}(\theta_H)$   for
$\theta_H^{\rm min} =0.10$, $z_L=10^{10}$  and $m_t=170\,$GeV.    
$V_{\rm eff}(\theta)/[(kz_L^{-1})^4/(4\pi)^2]$ has been plotted.
The bulk mass parameters are given by $(c_0=0.3599, c_1=0.4293, c_2=0.3120)$.
The bottom figure shows the behavior near the minimum.
The blue solid, the green dashed, and the red short dashed lines show
the effective potential containing the contributions from all the
$SO(11)$ bulk gauge boson and fermions,
only $SO(11)$ bulk gauge boson, and 
only $SO(11)$ bulk fermions, respectively.
}
\label{Figure:Sample-1}
\end{figure}

\begin{figure}[tbh]
\begin{center}
\includegraphics[bb=0 0 427 288, height=6.5cm]{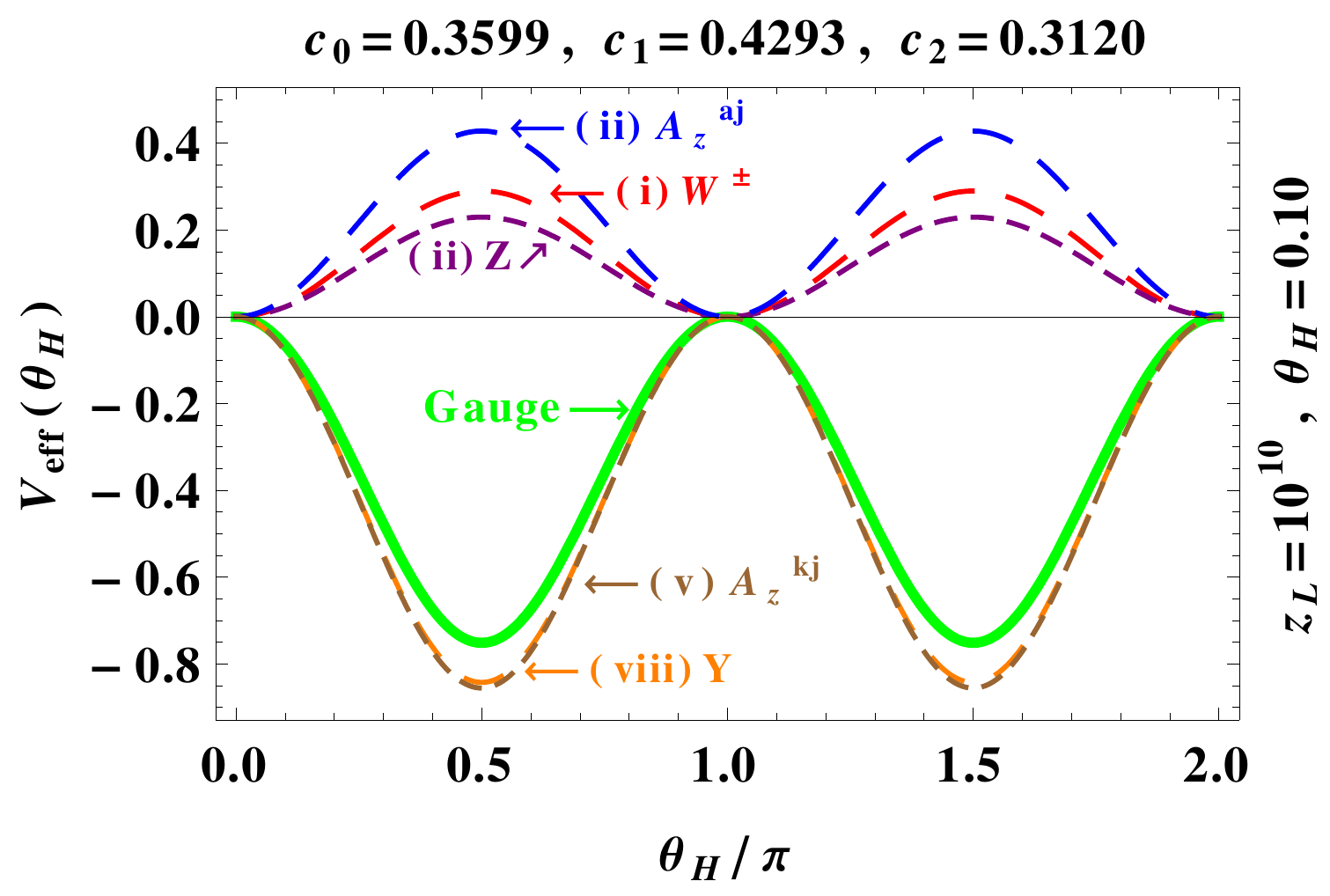}
\end{center}
\caption{Contributions of gauge fields to $V_{\rm eff}(\theta_H)$.
The input parameters are the same as in Figure \ref{Figure:Sample-1}.
The green solid line represents all gauge field
contributions for the effective potential, which is the same as 
the green dashed in Figure~\ref{Figure:Sample-1}.
The red dashed line is (i) $W^{\pm}$  contribution,
the purple short dashed line is (ii) $Z$   contribution,
the orange dashed line is (viii) $Y$  contribution,
the blue dashed line is (ii) $A_z^{a4},A_z^{a 11}(a=1,2,3)$  contribution, and
the brown short dashed line is (v) $A_z^{k4},A_z^{k11} (k=5 \sim 10)$  contribution.
}
\label{Figure:Sample-1-gauge}
\end{figure}

\begin{figure}[tbh]
\begin{center}
\includegraphics[bb=0 0 427 297, height=6.5cm]{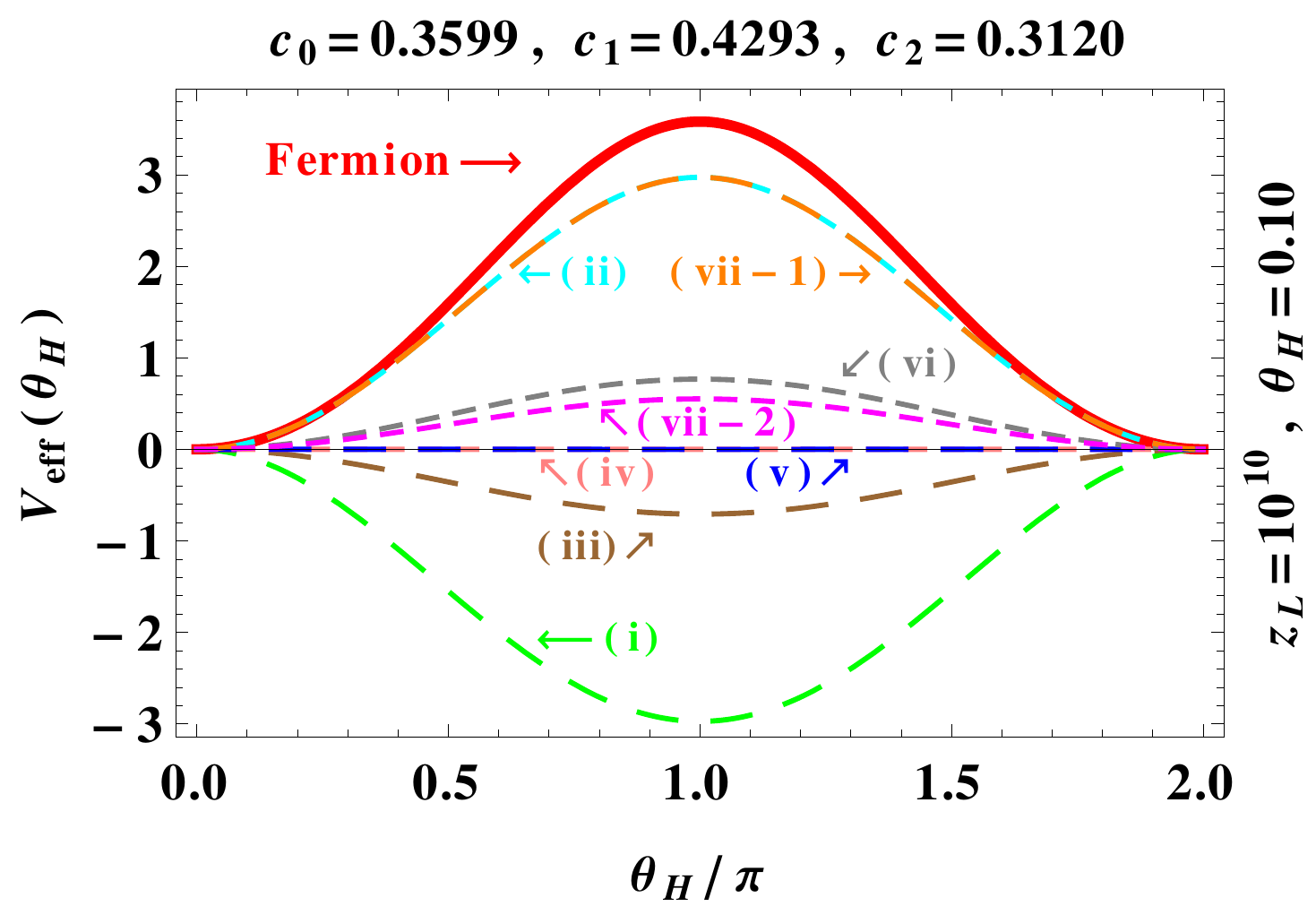}
\end{center}
\caption{Contributions of fermions to $V_{\rm eff}(\theta_H)$.
The parameter set is the same as  in Figure~\ref{Figure:Sample-1}.
The red solid line is the total fermion contribution for
effective potentials, which is the same as the red dashed in 
Figure~\ref{Figure:Sample-1}.
The green dashed line is (i) $Q_{em}=+\twothird$ fermion contribution,
the cyan short dashed line is (ii) $Q_{em}=-\twothird$ fermion contribution,
the brown dashed line is (iii) $Q_{em}=-\onethird$ fermion contribution,
the pink short dashed line is (iv) $Q_{em}=+\onethird$ fermion contribution,
the blue dashed line is (v) $Q_{em}=-1$ fermion contribution,
the gray short dashed line is (vi) $Q_{em}=+1$ fermion contribution,
the orange dashed line is (vii-1) $Q_{em}=0$ fermion contribution,
and 
the magenta short dashed line is (vii-2) $Q_{em}=0$ fermion
 contribution.
}
\label{Figure:Sample-1-fermion}
\end{figure}

The effective potential for $m_t=170\,$GeV is displayed in Figure~\ref{Figure:Sample-1}.
The global minimum is located at $\theta_H=0.10$, and the EW symmetry breaking 
takes place.  In Figure~\ref{Figure:Sample-1-gauge} and Figure~\ref{Figure:Sample-1-fermion}
contributions of gauge fields and fermions are plotted separately.

It is seen in Figure~\ref{Figure:Sample-1-gauge}, the contributions from 
 (v) $A_z^{k4},A_z^{k11}$  ($k=5 \sim 10$) and (viii) $Y$ bosons
 dominate over others in the gauge field sector.
 In the fermion sector there appears cancellation among contributions
 from various components.
 It is seen in Figure~\ref{Figure:Sample-1-fermion}, the contribution of
 (i) the top quark  is almost canceled by that of
 (ii) the $\hat u$-type $\hat t$ fermion.
 The bottom quark and $\hat d$-type $\hat b$ fermion contributions are not canceled out, but
each contribution is small.
The tau lepton and $\hat e$-type $\hat \tau$ fermion contributions are not canceled out, 
but each contribution is small. 
The four contributions from $b$, $\hat b$, $\tau$, and $\hat \tau$ add up almost zero.
The  contribution  from  neutral fermions is appreciable in the current model.

In the previous section we observed that there appear light exotic fermions
which should not exist in reality.  In this section we have observed that 
there appear cancellations among the contributions to $V_\eff (\theta_H)$
from fermions and their corresponding exotics.  These two seem to be related, 
and the too light Higgs boson mass  $m_H$ is inferred to be  a result of those cancellations.

\section{Conclusion and discussions}

In the present paper we have explored the $SO(11)$ gauge-Higgs grand unification in the RS space.
$SO(11)$ gauge symmetry is broken to $SO(4) \times SO(6)$ symmetry by the orbifold boundary conditions,
which is spontaneously broken to $SU(2)_L \times U(1)_Y \times SU(3)_C$ by
the brane scalar $\Phi_{\bf 16}$ on the Planck brane.  The EW $SU(2)_L \times U(1)_Y$  symmetry is 
dynamically broken to $U(1)_\EM$ by the Hosotani mechanism.
The Higgs boson appears as the four-dimensional fluctuation mode of the AB phase in the fifth 
dimension, or the zero mode of $A_y$.  Thus the gauge-Higgs unification is achieved.

Quark-lepton fermion multiplets are introduced in $\Psi_{\bf 32}$, $\Psi_{\bf 11}$ and $\Psi_{\bf 11}'$
in each generation.  Unlike the $SO(5) \times U(1)_X$ gauge-Higgs EW unification, one need not
introduce brane fermions on the Planck brane.  The quark-lepton masses are generated by the 
Hosotani mechanism with $\theta_H \not= 0$, supplemented with the brane interactions on the
Planck brane.  We have demonstrated that the quark-lepton mass spectrum can be reproduced
by adjusting the parameters of the brane interactions.

One of the interesting features of the model is that the proton decay is forbidden, in sharp 
contrast to the GUT models in four dimensions.  The quark-lepton number $N_{\Psi}$  is conserved by
the gauge interactions and brane interactions.

In the current model, however, there appear light exotic fermions associated with $\hat u$-type, $\hat d$-type
and $\hat e$-type fermions, which contradicts with the observation.
The Higgs boson mass $m_H$, which is predicted in the current gauge-Higgs  grand unification, 
turns out too small.   The small $m_H$ is a result of the partial cancellation among the contributions
of the quark-lepton component and the exotic fermion component to the effective potential $V_\eff (\theta_H)$.
In other words the exotic fermion problem  and the small $m_H$ problem seem to be related to each other.
The model need improvement in this regard.  We hope to report how to cure those problems in the near future.

\vskip 1.cm

\subsection*{Acknowledgements}
This work was supported in part by Japan Society for the Promotion of Science, Grants-in-Aid for Scientific Research, No. 23104009 and No.\ 15K05052.


\appendix

\section{$SO(11)$, $SO(10)$ and $SO(4)$}
The generators of $SO(11)$, $T_{jk} = - T_{kj} =T_{jk}^\dagger $ ($j,k=1 \sim 11$), satisfy
the algebra
\beeq
[T_{ij} , T_{kl} ] = i (\delta_{ik} T_{jl}  - \delta_{il} T_{jk} + \delta_{jl} T_{ik} - \delta_{jk} T_{il} ) ~.
\label{algebra1}
\eneq
In the adjoint representation
\beqn
&&\hskip -1.cm
(T_{ij})_{pq} = -i (\delta_{ip} \delta_{jq} - \delta_{iq} \delta_{jp} ) ~ ~, \cr
\noalign{\kern 10pt}
&&\hskip -1.cm
\Tr T_{jk} T_{lm} =2 (\delta_{jl} \delta_{km} - \delta_{jm} \delta_{kl} ) ~~,~~
\Tr (T_{jk})^2 = 2 ~.
\label{AdRep1}
\eeqn

As a basis of $SO(11)$ Clifford algebra, it is convenient to adopt
\beqn
&&\hskip -1.cm
\{ \Gamma_j , \Gamma_k \} = 2 \delta_{jk} \, I_{32}~, \cr
\noalign{\kern 5pt}
&&\hskip -1.cm
\Gamma_1 = \sigma^1 \otimes \sigma^1 \otimes \sigma^1 \otimes \sigma^1 \otimes \sigma^1 ~, \cr
&&\hskip -1.cm
\Gamma_2 = \sigma^2 \otimes \sigma^1 \otimes \sigma^1 \otimes \sigma^1 \otimes \sigma^1 ~, \cr
&&\hskip -1.cm
\Gamma_3 = \sigma^3 \otimes \sigma^1 \otimes \sigma^1 \otimes \sigma^1 \otimes \sigma^1 ~, \cr
&&\hskip -1.cm
\Gamma_4 = \sigma^0 \otimes \sigma^2 \otimes \sigma^1 \otimes \sigma^1 \otimes \sigma^1 ~, \cr
&&\hskip -1.cm
\Gamma_5 = \sigma^0 \otimes \sigma^3 \otimes \sigma^1 \otimes \sigma^1 \otimes \sigma^1 ~, \cr
&&\hskip -1.cm
\Gamma_6 = \sigma^0 \otimes \sigma^0 \otimes \sigma^2 \otimes \sigma^1 \otimes \sigma^1 ~, \cr
&&\hskip -1.cm
\Gamma_7 = \sigma^0 \otimes \sigma^0 \otimes \sigma^3 \otimes \sigma^1 \otimes \sigma^1 ~, \cr
&&\hskip -1.cm
\Gamma_8 = \sigma^0 \otimes \sigma^0 \otimes \sigma^0 \otimes \sigma^2 \otimes \sigma^1 ~, \cr
&&\hskip -1.cm
\Gamma_9 = \sigma^0 \otimes \sigma^0 \otimes \sigma^0 \otimes \sigma^3 \otimes \sigma^1 ~, \cr
&&\hskip -1.15cm
\Gamma_{10} = \sigma^0 \otimes \sigma^0 \otimes \sigma^0 \otimes \sigma^0 \otimes \sigma^2 ~, \cr
&&\hskip -1.15cm
\Gamma_{11} = \sigma^0 \otimes \sigma^0 \otimes \sigma^0 \otimes \sigma^0 \otimes \sigma^3  
~~ = -i \Gamma_1 \cdots \Gamma_{10}~, 
\label{Clifford1}
\eeqn
where $\sigma^0= I_2$ and $\{ \sigma^k \}$ are Pauli matrices.
In terms of $\Gamma_j$ the $SO(11)$ generators in the spinorial representation are given by
\beqn
&&\hskip -1.cm
T_{jk} = - \frac{i}{4} \, [ \Gamma_j , \Gamma_k] ~ 
\big( = - \frac{i}{2} \, \Gamma_j  \Gamma_k  ~~~ {\rm for~} j \not= k \big) ~, \cr
\noalign{\kern 10pt}
&&\hskip -1.cm
 (T_{jk})^2 = \frac{1}{4} \, I_{32} ~~,~~ \Tr (T_{jk})^2 = 8 ~.
 \label{SpinorRep1}
\eeqn

The orbifold boundary conditions $P_0, P_1$ in \eqref{BCP1} and \eqref{BCP2} break $SO(11)$
to $SO(4) \times SO(6)$.  The generators of the corresponding 
$SO(4) \simeq SU(2)_L \times SU(2)_R$ in the spinorial representation are given by
\beqn
&&\hskip -1.cm
\vec T_L = 
\onehalf \begin{pmatrix} T_{23} + T_{14} \cr T_{31} + T_{24} \cr T_{12} + T_{34} \end{pmatrix} 
= \onehalf \vec \sigma  \otimes \begin{pmatrix} 1\cr &0 \end{pmatrix}
 \otimes \sigma^0 \otimes \sigma^0 \otimes \sigma^0 ~, \cr
 \noalign{\kern 10pt}
&&\hskip -1.cm
\vec T_R =
\onehalf \begin{pmatrix} T_{23} - T_{14} \cr T_{31} - T_{24} \cr T_{12} - T_{34} \end{pmatrix} 
= \onehalf \vec \sigma  \otimes \begin{pmatrix} 0\cr &1 \end{pmatrix}
 \otimes \sigma^0 \otimes \sigma^0 \otimes \sigma^0 ~. 
\label{SPgenerator1}
\eeqn

The orbifold boundary condition $P_0$ at the Planck brane reduces $SO(11)$ to $SO(10)$,
whose generators are given by $T_{jk}$ ($j,k = 1 \sim 10$).  
In the representation \eqref{Clifford1} those generators become block-diagonal
$T_{jk}^{SO(10)} = [ \cdots ] \otimes (\sigma^0 ~{\rm or}~ \sigma^3)$ so that
a spinor {\bf 32} of $SO(11)$ splits into ${\bf 16} \oplus \overline{{\bf 16}}$ of $SO(10)$;
\beeq
\Psi_{\bf 32}= \begin{pmatrix} \Psi_{\bf 16} \cr 
\noalign{\kern 5pt} \Psi_{\overline{\bf 16}} \end{pmatrix} .
\label{Spinor1}
\eneq
With \eqref{Clifford1} one finds that
\beqn
&&\hskip -1.cm 
\Gamma_j^* = (-1)^{j+1} \, \Gamma_j ~, \cr
\noalign{\kern 5pt}
&&\hskip -1.cm
R\,  \Gamma_j \,  R = (-1)^j \, \Gamma_j ~,~~~
R\,  \Gamma_j^* \,  R = - \, \Gamma_j ~, \cr
\noalign{\kern 5pt}
&&\hskip -1.cm
R\,  T_{jk}^*  \,  R = - T_{jk} ~, \cr
\noalign{\kern 5pt}
&&\hskip -1.cm
R = \Gamma_2 \, \Gamma_4 \, \Gamma_6 \, \Gamma_8 \, \Gamma_{10}
= R^\dagger = R^{-1}  \cr
\noalign{\kern 5pt}
&&\hskip -.6cm 
= - \sigma^2 \otimes  \sigma^3 \otimes \sigma^2 \otimes \sigma^3 \otimes \sigma^2  \cr
\noalign{\kern 5pt}
&&\hskip -.6cm 
\equiv - \hat R \otimes \sigma^2 ~.
\label{Rtransform1}
\eeqn
It follows that for an $SO(11)$ spinor $\Psi_{\bf 32}$, the $R$-transformed one also transforms as {\bf 32}.
\beqn
&&\hskip -1.cm 
\tilde \Psi_{\bf 32} \equiv i R \, \Psi_{\bf 32}^* ~, \cr
\noalign{\kern 10pt}
&&\hskip -1.cm 
\Psi_{\bf 32}' = \Big( 1 + \frac{i}{2} \, \ep_{jk} T_{jk} \Big) \Psi_{\bf 32}
\quad \Rightarrow \quad
\tilde \Psi_{\bf 32}' = \Big( 1 + \frac{i}{2} \, \ep_{jk} T_{jk} \Big) \tilde \Psi_{\bf 32} ~.
\label{Rtransform2}
\eeqn
Its $SO(10)$ content is given by
\beeq
\tilde \Psi_{\bf 32}= 
\begin{pmatrix} \tilde \Psi_{\bf 16} \cr 
\noalign{\kern 5pt} \tilde \Psi_{\overline{\bf 16}} \end{pmatrix}
=\begin{pmatrix} - \hat R \, \Psi_{\overline{\bf 16}}^* \cr
\noalign{\kern 5pt}  +\hat R \, \Psi_{\bf 16}^* \end{pmatrix} ~.
\label{Rtransform3}
\eneq

\section{Basis functions in RS space}

Mode functions  of various fields in the RS spacetime are expressed  in terms of Bessel functions.  
We define, for gauge fields, 
\beqn
&&\hskip -1cm 
C(z;\lambda) = \frac{\pi}{2}\lambda   z z_L F_{1,0}(\lambda z, \lambda z_L) ~, \quad
C'(z;\lambda)= \frac{\pi}{2}\lambda^2 z z_L F_{0,0}(\lambda z, \lambda z_L) ~, \cr \noalign{\kern 5pt}
&&\hskip -1cm 
S(z;\lambda) = -\frac{\pi}{2}\lambda   z  F_{1,1}(\lambda z, \lambda z_L) ~, \quad
S'(z;\lambda)= -\frac{\pi}{2}\lambda^2 z F_{0,1}(\lambda z,  \lambda z_L) ~, 
\label{BesselF1}
\eeqn
where $F_{\alpha,\beta}(u,v) = J_\alpha (u) Y_\beta(v)-Y_\alpha(u) J_\beta(v)$.  
They satisfy
\beqn
&&\hskip -1.cm
z \frac{d}{dz} \frac{1}{z} \frac{d}{dz}
 \begin{pmatrix} C(z; \lambda) \cr S(z; \lambda) \end{pmatrix} = 
- \lambda^2 \begin{pmatrix} C(z; \lambda) \cr S(z; \lambda) \end{pmatrix} , \cr
\noalign{\kern 10pt}
&&\hskip -1cm 
C(z_L; \lambda) = z_L ~,~~ C' (z_L; \lambda) =0~, ~~ 
S(z_L; \lambda) = 0 ~,~~ S' (z_L; \lambda) =\lambda ~,  \cr \noalign{\kern 5pt}
&&\hskip -1cm 
C S' - S C' = \lambda z ~. 
\label{BesselF2}
\eeqn
It follows that 
\beqn
&&\hskip -1.cm
\frac{d}{dz} z \frac{d}{dz} \frac{1}{z}
 \begin{pmatrix} C'(z; \lambda) \cr S'(z; \lambda) \end{pmatrix} = 
- \lambda^2 \begin{pmatrix} C'(z; \lambda) \cr S'(z; \lambda) \end{pmatrix} , \cr
\noalign{\kern 10pt}
&&\hskip -1cm 
\frac{d}{dz} \Big\{  \frac{1}{z} S'(z; \lambda) \Big\} \Big|_{z=z_L} = 0 ~.
\label{BesselF3}
\eeqn

For fermions with a bulk mass parameter $c$ we define
\beqn
&&\hskip -1cm
\begin{pmatrix} C_L \cr S_L \end{pmatrix}  (z;\lambda, c)
= \pm \frac{\pi}{2} \lambda\sqrt{zz_L} F_{c+{1\over 2},c\mp{1\over 2}}  (\lambda z, \lambda z_L) ~, \cr
\noalign{\kern 10pt}
&&\hskip -1cm  
\begin{pmatrix} C_R \cr S_R \end{pmatrix}  (z;\lambda, c)
= \mp \frac{\pi}{2} \lambda\sqrt{zz_L} F_{c-{1\over 2},c\pm {1\over 2}} (\lambda z, \lambda z_L)~,
\label{BesselF4}
\eeqn
which satisfy
\beqn
&&\hskip -1.cm
D_+ (c)\begin{pmatrix} C_L \cr S_L \end{pmatrix} = \lambda \begin{pmatrix} S_R \cr C_R \end{pmatrix} ~,~~
D_- (c)\begin{pmatrix} C_R \cr S_R \end{pmatrix} = \lambda \begin{pmatrix} S_L \cr C_L \end{pmatrix} ~, \cr
\noalign{\kern 5pt}
&&\hskip 1.cm
D_\pm (c)  = \pm \frac{d}{dz} + \frac{c}{z}~, \cr
\noalign{\kern 5pt}
&&\hskip -1.cm
C_R=C_L =1 ~,~~  S_R=S_L=0 ~,   ~~{\rm at~} z= z_L ~, \cr 
\noalign{\kern 5pt}
&&\hskip -1.cm
C_L C_R - S_L S_R =1 ~.
\label{BesselF5}
\eeqn

We note that for $\lambda z_L \ll 1$ and $c \ge 0$
\beqn
&&\hskip -1.cm
C(1; \lambda) \sim z_L \cdot \big\{ 1 + O(\lambda^2 z_L^2) \big\} ~, \cr
&&\hskip -1.cm
C' (1; \lambda) \sim \lambda^2 z_L \ln z_L \cdot \big\{ 1 + O(\lambda^2 z_L^2) \big\} ~, \cr
&&\hskip -1.cm
S(1;\lambda) \sim - \onehalf \lambda z_L \cdot \big\{ 1 + O(\lambda^2 z_L^2) \big\} ~, \cr
&&\hskip -1.cm
S'(1;\lambda) \sim \lambda z_L^{-1}  \cdot \big\{ 1 + O(\lambda^2 z_L^2) \big\} ~, \cr
\noalign{\kern 10pt}
&&\hskip -1.cm
C_L(1; \lambda, c) \sim z_L^c \cdot \big\{ 1 + O(\lambda^2 z_L^2) \big\} ~, \cr
&&\hskip -1.cm
C_R(1; \lambda, c) \sim z_L^{-c} \cdot \big\{ 1 + O(\lambda^2 z_L^2) \big\} ~, \cr
\noalign{\kern 10pt}
&&\hskip -1.cm
S_L(1; \lambda, c) \sim - \frac{\lambda z_L^{c+1}}{2(c+\onehalf)} 
 \cdot \big\{ 1 + O(\lambda^2 z_L^2) \big\} ~, \cr
\noalign{\kern 10pt}
&&\hskip -1.cm
S_R(1; \lambda, c) \sim  
\begin{cases}
\myfrac{\lambda z_L^{c}}{2(c-\onehalf)} 
\cdot \big\{ 1 + O(\lambda^2 z_L^2) \big\} &{\rm for~} c > \onehalf \cr
\noalign{\kern 5pt}
\lambda z_L^{1/2} \ln z_L &{\rm for~} c = \onehalf \cr
\noalign{\kern 5pt}
\myfrac{ \lambda z_L^{1-c}}{2(\onehalf - c )} 
\cdot \big\{ 1 + O(\lambda^2 z_L^2) \big\} &{\rm for~} c < \onehalf 
 \end{cases} ~.
\label{LEspectrum1}
\eeqn
In particular,
\beeq
S_L (1; \lambda, c) S_R(1; \lambda, c) \sim  
\begin{cases}
- \myfrac{\lambda^2  z_L^{2c+1}}{4 c^2 -1} 
\cdot \big\{ 1 + O(\lambda^2 z_L^2) \big\} &{\rm for~} c > \onehalf \cr
\noalign{\kern 5pt}
- \onehalf \lambda^2 z_L^2 \ln z_L &{\rm for~} c = \onehalf \cr
\noalign{\kern 5pt}
- \myfrac{ \lambda^2 z_L^{2}}{1 - 4c^2} 
\cdot \big\{ 1 + O(\lambda^2 z_L^2) \big\} &{\rm for~} c < \onehalf 
 \end{cases} ~.
\label{LEspectrum2}
\eneq

\vskip .5cm

\ignore{
\renewenvironment{thebibliography}[1]
         {\begin{list}{[$\,$\arabic{enumi}$\,$]}  
         {\usecounter{enumi}\setlength{\parsep}{0pt}
          \setlength{\itemsep}{0pt}  \renewcommand{\baselinestretch}{1.2}
          \settowidth
         {\labelwidth}{#1 ~ ~}\sloppy}}{\end{list}}
}

\def\jnl#1#2#3#4{{#1}{\bf #2},  #3 (#4)}

\def\Zphys{{\em Z.\ Phys.} }
\def\jssc{{\em J.\ Solid State Chem.\ }}
\def\jpsJ{{\em J.\ Phys.\ Soc.\ Japan }}
\def\ptps{{\em Prog.\ Theoret.\ Phys.\ Suppl.\ }}
\def\PTP{{\em Prog.\ Theoret.\ Phys.\  }}
\def\PTEP{{\em Prog.\ Theoret.\ Exp.\  Phys.\  }}
\def\JMP{{\em J. Math.\ Phys.} }
\def\NPB{{\em Nucl.\ Phys.} B}
\def\NP{{\em Nucl.\ Phys.} }
\def\PLB{{\it Phys.\ Lett.} B}
\def\PL{{\em Phys.\ Lett.} }
\def\PRL{\em Phys.\ Rev.\ Lett. }
\def\PRB{{\em Phys.\ Rev.} B}
\def\PRD{{\em Phys.\ Rev.} D}
\def\PRe{{\em Phys.\ Rep.} }
\def\AP{{\em Ann.\ Phys.\ (N.Y.)} }
\def\RMP{{\em Rev.\ Mod.\ Phys.} }
\def\ZPC{{\em Z.\ Phys.} C}
\def\SCI{\em Science}
\def\CMP{\em Comm.\ Math.\ Phys. }
\def\MPLA{{\em Mod.\ Phys.\ Lett.} A}
\def\IJMPA{{\em Int.\ J.\ Mod.\ Phys.} A}
\def\IJMPB{{\em Int.\ J.\ Mod.\ Phys.} B}
\def\EPJC{{\em Eur.\ Phys.\ J.} C}
\def\PR{{\em Phys.\ Rev.} }
\def\JHEP{{\em JHEP} }
\def\JCAP{{\em JCAP} }
\def\cmp{{\em Com.\ Math.\ Phys.}}
\def\JPA{{\em J.\  Phys.} A}
\def\JPG{{\em J.\  Phys.} G}
\def\NJP{{\em New.\ J.\  Phys.} }
\def\PoS{{\em PoS} }
\def\CQG{\em Class.\ Quant.\ Grav. }
\def\ATMP{{\em Adv.\ Theoret.\ Math.\ Phys.} }
\def\ibid{{\em ibid.} }

\renewenvironment{thebibliography}[1]
         {\begin{list}{[$\,$\arabic{enumi}$\,$]}  
         {\usecounter{enumi}\setlength{\parsep}{0pt}
          \setlength{\itemsep}{0pt}  \renewcommand{\baselinestretch}{1.0}
          \settowidth
         {\labelwidth}{#1 ~ ~}\sloppy}}{\end{list}}

\def\reftitle#1{{\it ``#1,'' }}    


\end{document}